\documentclass[3p]{elsarticle}

\usepackage[american]{babel}
\usepackage{amsmath}
\usepackage{textcomp}
\usepackage{mathtools}
\usepackage{adjustbox}

\usepackage{silence}
\usepackage{changepage}

\usepackage{xfrac}
\usepackage{lmodern}
\usepackage[hidelinks]{hyperref}
\usepackage{cleveref}
\usepackage{microtype}
\usepackage{listings}
\usepackage[autostyle, english = american]{csquotes}
\usepackage{longtable}
\usepackage{subcaption}
\usepackage{booktabs}
\usepackage{gensymb}
\usepackage[normalem]{ulem}

\usepackage{comment}  % easy to comment things out
\usepackage{cleveref} % Referencing multiple tables
\usepackage{lipsum} % for dummy text only

\usepackage[]{siunitx}
\usepackage{color}

\usepackage[colorinlistoftodos]{todonotes}

\usepackage[section]{placeins}

\usepackage{natbib} %to use \citet, now mixed with [nr]
\usepackage{ecrc-ari}

\volume{00}

\firstpage{1}

\journalname{Applied Radiation and Isotopes}

\runauth{H. Ekeberg et al.}

\jid{ari}
\usepackage{amssymb}
\usepackage{amsthm}
\biboptions{sort&compress}

\usepackage[figuresright]{rotating}

\usepackage{fancyvrb}
\usepackage{color}
\usepackage[version=4]{mhchem}
\definecolor{mygreen}{rgb}{0,0.6,0}
\definecolor{mygray}{rgb}{0.5,0.5,0.5}
\definecolor{mymauve}{rgb}{0.58,0,0.82}

\newcommand{\natIr}{\ce{^{\textrm{nat}}Ir}}
\newcommand{\natFe}{\ce{^{nat}Fe}}
\newcommand{\natCu}{\ce{^{\textrm{nat}}Cu}}
\newcommand{\natNi}{\ce{^{\textrm{nat}}Ni}}

\newcommand{\Pt}{\ce{^{193\textrm{m}}Pt}}

\usepackage{multirow}
\usepackage{makecell} %centering tables
\usepackage{longtable}
\usepackage{tabu, booktabs}
\usepackage{multicol}


\begin{document}

\begin{frontmatter}


\title{Nuclear excitation functions for medical isotope production: targeted radionuclide therapy via  {\ce{^{nat}}Ir}$(d,x)${\ce{^{193m}Pt}}}

% \dochead{Short}

\author[uio,nnrc,ife]{H.~L.~O.~Ekeberg }

\author[ucb]{A.~S.~Voyles \corref{cor1}}
\ead{andrew.voyles@berkeley.edu}

\author[lbl]{M.~S.~Basunia}

\author[ucb]{J.~C.~Batchelder}

\author[ucb,lbl]{L.~A.~Bernstein}

\author[llnl]{D.~L.~Bleuel}

\author[uio]{K.~C.~W.~Li}

\author[uio,nnrc]{E.~M.~Martinsen}

\author[ucb]{E.~F.~Matthews}

\author[ucb]{J.~T.~Morrell}

\author[uio]{N.~I.~J.~Pettersen}

%\author[uio]{O.~A.~Ranum}

\author[uio]{S.~Siem}

\address[uio]{Department of Physics, University of Oslo, N-0316 Oslo, Norway}
\address[nnrc]{Norwegian Nuclear Research Centre, Norway}
\address[ife]{Tracer Technology Department, Institute for Energy Technology, IFE, Kjeller, Norway}
\address[ucb]{Department of Nuclear Engineering, University of California, Berkeley, Berkeley CA, 94720 USA}
% \address[uio]{Department of Physics, University of Oslo, N-0316 Oslo, Norway}
\address[lbl]{Lawrence Berkeley National Laboratory, Berkeley CA, 94720 USA}
\address[llnl]{Lawrence Livermore National Laboratory, Livermore, CA 94551 USA}


\cortext[cor1]{Corresponding author}

% \address[ucb]{Department of Nuclear Engineering, University of California, Berkeley, Etcheverry Hall, 2521 Hearst Ave, Berkeley, CA 94709}
% \address[lbl]{Lawrence Berkeley National Laboratory,  1 Cyclotron Rd, Berkeley, CA 94720}
% \address[llnl]{Lawrence Livermore National Laboratory, 7000 East Ave, Livermore, CA 94550}

\begin{abstract}

\Pt{} is an Auger emitting radionuclide which may have therapeutic potential, particularly when labeled to the chemotherapeutic drug cisplatin.
One challenge to broader explorations of its clinical potential is the need for production routes with high specific activity. 
As part of a larger campaign  to address gaps in reaction data for emerging medical radionuclides, this work seeks to characterize the \natIr(d,x) reactions as a potential production pathway for \Pt{}. 
A stacked target irradiation, consisting of natural iridium, iron, nickel, and copper foils, was performed using a 33 MeV deuteron beam at the Lawrence Berkeley National Laboratory 88-Inch Cyclotron. 
This measurement, along with previous experimental data, suggests an energy window between 11 to 18 MeV to maximize the production and radiopurity of \Pt{}. 
This experiment has yielded cross sections for 43 channels of deuteron-induced reactions from threshold to 30~MeV, including the first experimental results of \natIr(d,x)\ce{^{188m1+g,190m1+g}Ir} (cumulative), \natNi(d,x)\ce{^{56,57,58m,58g}Co} (independent), \natCu(d,x)\ce{^{61}Co} (cumulative) and \natFe(d,x)\ce{^{53}Fe},\ce{^{48}V} (cumulative).
The results were compared with literature data, the TENDL-2023 database, and default theoretical calculations from the \textsc{TALYS-2.04}, \textsc{CoH-3.6.0}, \textsc{EMPIRE-3.2.3}, and \textsc{ALICE-2020} reaction modeling codes. 
This work presents another example of the lack of predictive capabilities for this set of modern nuclear-reaction modeling codes, and highlights the unsatisfactory modeling of experimental cross sections. 
Experimental data are important to improve the codes in general, and new experimental results can be used to improve the models. 
Finally, this measurement has revealed the need for an updated evaluation of the  \natCu(d,x)\ce{^{63}Zn} deuteron monitor reaction.

\end{abstract}


\end{frontmatter}

% \linenumbers

% \listoftodos

\newpage
\section{Introduction}

Use of radioactive isotopes for \emph{in vivo} applications within the medical field has had a transformative impact on global health. 
Medical practitioners make use of single-photon emission computed tomography (SPECT) and positron emission tomography (PET) imaging agents---like \ce{^{99\textrm{m}}Tc} and \ce{^{18}F}---to routinely diagnose diseased tissue in millions of patients per year worldwide. 
Use of low-energy $\beta$-emitting radionuclides---such as \ce{^{177}Lu} (Lutathera\textsuperscript{\textregistered}), \ce{^{90}Y} (Zevalin\textsuperscript{\textregistered}), and \ce{^{131}I} (Bexxar\textsuperscript{\textregistered})---and $\alpha$-emitting radionuclides---like \ce{^{223}Ra} (Xofigo\textsuperscript{\textregistered})---as therapeutic agents have similar potential for positively redefining the medical field. 
Although it has yet to be fully realized, widespread use of radiotheranostic agents combining a matched pair of a therapeutic radionuclide with an imaging surrogate has potential for even larger impact than any singular diagnostic or therapeutic agent \cite{Filippi2020,Langbein2019}.
Unfortunately, large-scale production of emerging radionuclides, particularly those whose decay-properties enable  
therapeutic applications, has not been realized to date.
In selecting candidates for targeted radionuclide therapy, a vital figure-of-merit is the linear energy transfer (LET, typically reported in keV/$\micro$m) of decay radiation, which measures energy deposition per unit length. 
High-LET radiation produces a high density of ionization events, damaging cells and their DNA. 
As LET is inversely proportional to the radius over which this
energy is deposited, high-LET radionuclides are prized for therapeutic potential, as their decay radiation produces high cellular lethality over a narrow region \cite{Qaim201731}.

The future of radionuclide therapy appears to feature two major groups: those which decay by emission of an alpha particle (“alpha emitters”), and those which emit a cascade of Auger electrons in their decay (“Auger emitters”) \cite{Muller2017,bolcaen2023marshalling}. 
These are examples of the increasing trend in nuclear medicine towards using targeted therapies developed for individual cases, which require short-range, accurately delivered doses.
Alpha emitters feature high LET giving them a short range (on the order of a single cell) and significantly reducing the dose delivered to surrounding healthy tissue, and have demonstrated significant early clinical success, particularly for the case of \ce{^{225}Ac}  \cite{Kratochwil2016,Kratochwil2017,Kratochwil2018}. 
In a complementary fashion, Auger emitters produce a cascade of 5--40 low-energy (10~eV--10~keV) electrons with a LET of 5--25~keV/$\micro$m (which corresponds to a range of 2--500~nm in tissue), leading to a massive dose in a volume comparable to the nucleus of a single cell. 
This energy transfer induces double-strand breaks in the DNA of a cancerous cell from which it is nearly impossible for the cell to survive, while sparing surrounding healthy cells \cite{Howell2020}. 
In addition to the shorter range facilitating highly selective treatment modalities, Auger emitters avoid the common challenges of alpha emitters: a chain of multiple radioactive decays prior to reaching a stable nuclide (which will likely dissociate the  daughters away from the targeting site and increase non-useful patient dose), and the location of many alpha-emitters far from stability (which often limits production paths and constrains potential target options) \cite{Filosofov2021}.
However, this extreme dose localization has created challenges in matching radionuclides with suitable targeting biomolecules.

Despite tremendous potential, transitioning Auger-emitting therapeutics from the realm of R\&D to an actual drug used by medical practitioners has yet to be demonstrated  beyond scarce investigational and dosimetry studies.
One of the most daunting problems facing the development of Auger-based medical applications is associated with limited radionuclide access and production routes with low specific activity.
Auger therapy is far more sensitive to low specific activity than most $\alpha$- or $\beta^-$-radiopharmaceuticals, as much greater quantities of the keV-scale electrons per decay must be accumulated within a finite volume to induce the desired therapeutic dose \cite{Filosofov2021}.
One interesting class of  radiometals with high-spin isomeric states \cite{Muller2017}, such as \ce{^{117\textrm{m}}Sn} ($t_{1/2}$ = 14.00 $\pm$ 0.05 d \cite{Stevenson2015,Blachot2002a}) and \ce{^{193\textrm{m}}Pt} ($t_{1/2}$ = 4.33 $\pm$ 0.03 d \cite{Howell1986,ShamsuzzohaBasunia2017}), have gained increasing interest for use in conversion- and Auger-electron therapy, respectively. 
These long-lived isomers result from a large difference in angular momentum from their low-lying ground state,  inhibiting their ability to de-excite by isomeric transition or $\beta/\epsilon$ decay.

Radioplatinums, \ce{^{193\textrm{m}}Pt} in particular,  have significant potential for targeted radionuclide therapy \cite{Tarkanyi2006}, especially in conjunction with the platinum-based chemotherapeutic drug cisplatin.
Cisplatin has been commonly used since the 1970s for treatment of a wide variety of cancers, as it  binds to the DNA bases of rapidly-dividing cancer cells. 
Conventional cisplatin therapy is limited by its chemical toxicity, with the severe long-term side effects including irreversible hearing loss, and nerve and renal damage \cite{national1979cancer}. 
However, it has been suggested that for \Pt{}-labeled cisplatin,  the  added radiotoxicity 
may significantly reduce the  amount of cisplatin required, and the chemotoxicity limitations may thus be reduced \cite{Howell1986,Howell1994a,Azure1992}. 
The long half-life of the \Pt{} isomer allows for binding to the cellular DNA, facilitates the use of large biomolecules for targeting disease, and offers the possibility for large-scale regional production and shipment. 
The cascade of Auger electrons emitted in each decay are expected to have excellent clinical efficacy, with ranges less than the cell diameter, minimizing dose to surrounding tissue \cite{Howell1992}.

However, as with many Auger emitters, a dearth of production pathways with high specific activity has  prevented wider clinical investigations of \Pt{}.
As the isomer ($J^\Pi=\sfrac{13}{2}^+$) has a high spin state,  transfer of large amounts of angular momentum is needed for reaction pathways to preferentially populate the medically-valuable isomer instead of its long-lived ground state ($J^\Pi=\sfrac{1}{2}^-$) \cite{ShamsuzzohaBasunia2017}.
The typical production mechanism for  \Pt{} currently used is via the \ce{^{192}Os}($\alpha$,3n) reaction at $\approx$30~MeV, which yields high specific activity, but the \ce{^{192}Os} target is highly toxic, chemically volatile, and mechanically brittle \cite{Uddin2010a}. 
There are alternatives involving slightly heavier beams and potentially simpler targetry such as the \ce{^{186}W}(\ce{^{9}Be},2n) reaction, but these severely limit the number of accelerators for potential production. 
Proton-induced reactions on iridium have been considered for the production of radioplatinum  \cite{Tarkanyi2005,Obata2019}, but since (p,xn) reaction channels preferentially populate nuclear states closer in $J^\Pi$ to the target, they are less favorable for selective production of the \ce{^{193\textrm{m}}Pt} isomer \cite{Qaim2018a,Skobelev2015,Vandenbosch1960,Tarkanyi2015a}.
Likewise, reactor production is not expected to yield high specific activity \ce{^{193\textrm{m}}Pt}, due to the low spin transfer for thermal neutrons \cite{Toth1980a}.
Recent work by Tarkanyi  \emph{et al.} \cite{Tarkanyi2006,Tarkanyi2019} has proposed the use of Ir(d,xn) as a viable pathway for high specific activity \ce{^{193\textrm{m}}Pt}, with more preferential isomer population than Ir(p,xn), and without the need for 30~MeV $\alpha$-particles for \ce{^{192}Os}($\alpha$,3n).
The \natIr{}(d,x){\Pt{}}   production route explored in this work seeks to improve upon the earlier work by Tarkanyi  \emph{et al.}, resolve disagreements in  these previously reported channels, and provide the most complete characterization of reaction channels to date in this mass region. 
In this work, our stacked-target activation method was applied to measure multiple cross sections from threshold to 33~MeV. 
Finally, this measurement serves as the first extension of our variance minimization technique to deuteron beams.
\section{Experimental methods and materials}

Preliminary discussion of this measurement was reported in H. Ekeberg's Master thesis \cite{Lovise2020a}; here we present the final report of that work. 
Unless otherwise stated, all values are presented herein as mean $\pm$ SD (standard deviation), or as the calculated result $\pm$ half the width of a 68\% confidence interval.

\subsection{ Stacked-target design}
\label{subsec: Target stack}

In this work, we used the stacked target activation technique  to measure cross sections using an incident deuteron beam up to 33~MeV on a stack of well-characterized thin targets. 
This approach has been well-established in our recent work for the measurement of (p,x) reactions \cite{Voyles2018a,Morrell2020,Fox2021a,Fox2021,Voyles2021}; this work serves as our first extension of these methods for energetic deuteron beams.
% The method is well documented in the literature (\textcolor{red}{citations}). 
The stack consisted of ten \ce{^{\textrm{nat}}Ir} (99.9\%, lot \#LS510228) target foils, and ten \ce{^{\textrm{nat}}Ni} (99.9\%, lot \#LS471921), ten \ce{^{\textrm{nat}}Cu} (99.95\%, lot \#LS471698) and three \ce{^{\textrm{nat}}Fe} (99.5\%, lot \#LS470411) monitor foils, (all  nominal 25~$\micro$m thickness, from Goodfellow Corporation, Coraopolis, PA 15108, USA).

\begin{figure}[ht]
    \centering
    \includegraphics[width=0.5\textwidth]{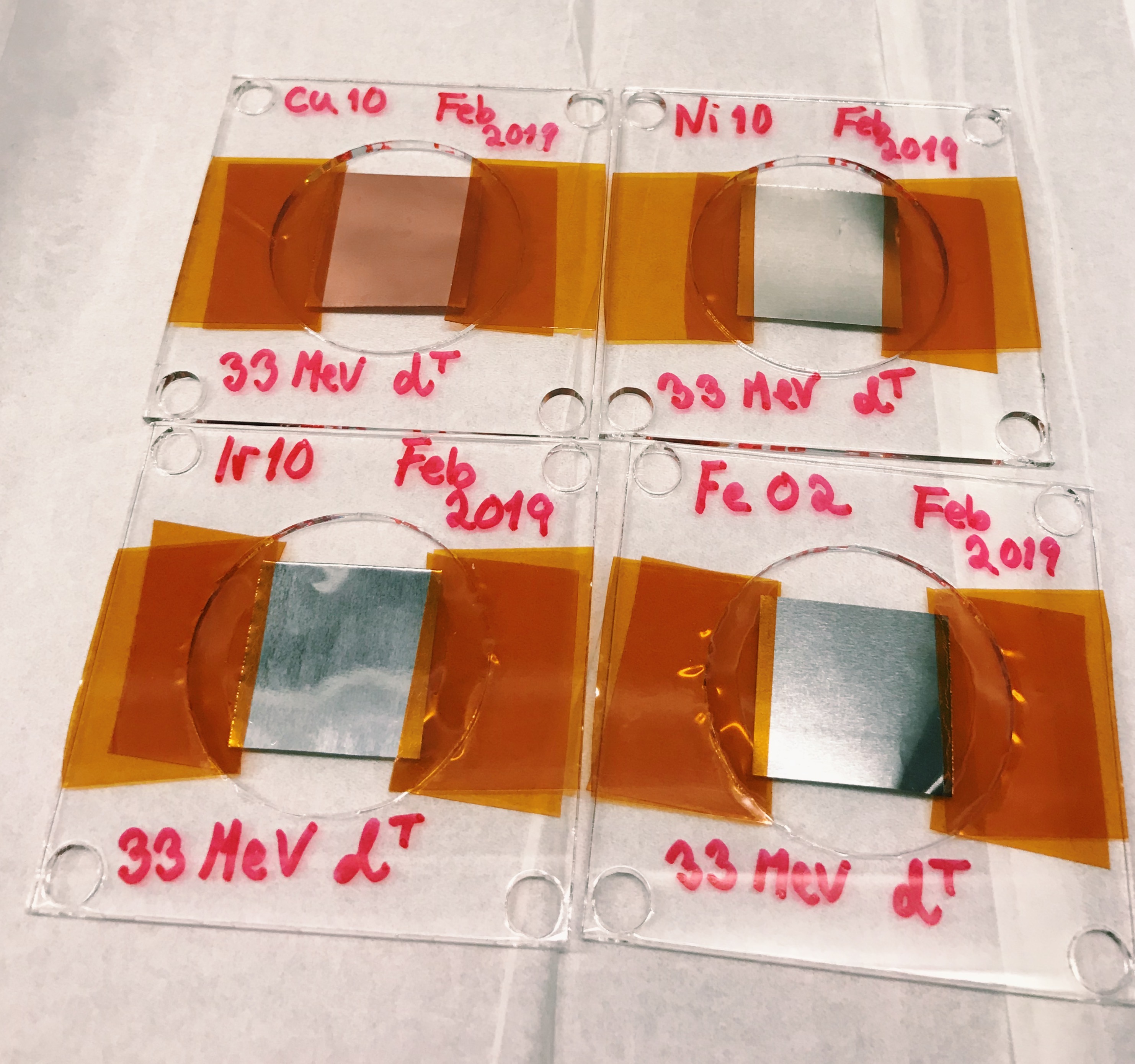}
    \caption{Each target material mounted on acrylic frames with a hollow center. Kapton tape is attached along the edges of the foils, outside of the beam strike area.}    \label{fig:targets_on_frame}
\end{figure}

The iridium foils were obtained as pre-cut in 25$\times$25~mm squares,  the nickel, copper, and iron foils were cut down to match, and all foils were cleaned with isopropanol. 
All foils were spatially characterized at four different locations using a digital caliper and micrometer (Mitutoyo America Corp.) and four mass measurements were performed using an analytical balance (Mettler Toledo) in order to determine their areal density.
Unlike in our previous work, no beam energy degraders were needed for this stack---due to the increased stopping power of deuterons (relative to protons), the target and monitor foils served to sufficiently degrade the beam to cover our energy range of interest.
Likewise, no Kapton tape was used for sealing the stack foils, due to a combination of significant ablation and scorch marks, as well as low levels of dispersible \ce{^{13}N} contamination, found previously for $E_d<$10~MeV  on Kapton tape.
Instead, the foils were  mounted by their edges over the hollow aperture of 1.5875 mm-thick acrylic frames, with all Kapton tape remaining out of the beam strike area. The mounted foils can be seen in \autoref{fig:targets_on_frame}.
The specifications of the target stack are detailed in Table \ref{tab:foilCharacterization} of \ref{ch_app:stack_design}.

The foils were placed in a target holder (made from 6061 aluminum alloy) with a hollow upstream aperture to provide an unobstructed path between the foils and the beam, as  in our prior work \cite{Voyles2018a,Morrell2020,Fox2021a,Fox2021,Voyles2021}.
The target stack was irradiated with 33~MeV deuterons at the  Lawrence Berkeley National Laboratory  (LBNL) 88-Inch Cyclotron \cite{7999622}, for one hour at a nominal current of 125\,nA,   measured using a current integrator on the electrically-isolated beamline. 
The beam current remained stable under these conditions for the duration of the irradiation.
The approximately 1.2~cm-diameter deuteron beam incident upon the stack's upstream stainless steel profile monitor had  an approximately 2\% energy width due to multi-turn extraction---this energy profile was used for all later analysis.
Following end of irradiation, the stack was removed from the beamline and disassembled, with each mounted foil individually sealed into plastic bags for containment.
All activated foils were transported to a counting lab for gamma spectrometry, which started approximately 15\,minutes following the end of the irradiation.

\subsection{Quantification of induced activities}

One ORTEC GMX Series (model \#GMX-50220-S) and six ORTEC IDM-200-V High-Purity Germanium (HPGe) detectors were used to determine the activities in each target. 
Samples were counted at fixed positions ranging 10--60 cm (5\% maximum permissible dead-time) from the front face of the detector. 
The foils were counted for 5 weeks following irradiation. 
% An example of one of the gamma-ray spectra collected is shown in Fig. 1. 
Net peak areas were fitted using the code FitzPeaks \cite{fitzgerald2009fitzpeaks}, which utilizes the SAMPO fitting algorithms for gamma-ray spectra \cite{Aarnio2001}.
Activities for all products observed in the collected $\gamma$-ray spectra, along with  all calculations of measured cross sections reported in this work, were calculated using half-lives and gamma-ray branching ratios  from the most recent edition of Nuclear Data Sheets for each mass chain \cite{ShamsuzzohaBasunia2017,Baglin2012a,Vanin2007,Kondev2018,Johnson2017,SINGH2003,Singh2006,Nesaraja2010,Junde2011, Burrows2006, Wang2017,Dong2015,Dong2014, JUNDE2008787,Bhat1998,Basunia2018,Browne2010,Browne2013,Singh2007,Zuber20151}, with the exception of \ce{^{61}Cu} \cite{hermanne2025critical}, with all utilized data tabulated in \ref{ch_app:tables}.
Corrections (typically \textless 0.2\%) for gamma-ray self-attenuation within each foil  were made, using photon attenuation coefficients from the XCOM photon cross sections database \cite{berger2011xcom}, and assuming attenuation through half the thickness of each foil. 
End-of-Beam (EoB) activities were determined by $\chi^2$-fitting of all observed decay gammas for a product to the decay curve. 
The total uncertainty in activity is the propagated sum of the uncertainty in fitted peak areas, uncertainty in detector efficiency calibration, uncertainty in the gamma-ray branching ratio data, and uncertainty in photon attenuation coefficients (taken as 5\%).
Where the decay of any precursors could be measured and the in-growth contribution separated, or where no decay precursors exist, independent cross sections for direct production of a nucleus are reported. 
Where the in-growth due to parent decay could not be deconvolved, due to timing or decay-property limitations, cumulative cross sections are reported.
Further details on this spectroscopy may be found in Ref. \cite{Lovise2020a, Voyles2018a}.

\subsection{Deuteron energy  and fluence assignments}

Thin \natNi{}, \natFe{}, and \natCu{}    foils were co-irradiated to measure beam current at each position within the stack. 
The International Atomic Energy Agency (IAEA)-recommended \natNi{}(d,x)\ce{^{61}Cu},\ce{^{56,58}Co}, \natFe{}(d,x)\ce{^{56}Co}, and \natCu{}(d,x)\ce{^{62,63,65}Zn}  monitor reactions were used  \cite{Hermanne2018}.
Using the formalism outlined in our previous work, the integral form of the well-known activation equation was used to  determine deuteron current ($I$),
in order to account for energy loss across each monitor foil \cite{Voyles2018a}.
The propagated uncertainty in beam current is calculated as the quadrature sum of (1) the uncertainty in quantified EOB activity, (2) uncertainty in the duration of irradiation, (conservatively estimated at 10 s, to account for any transient changes in beam current), (3) uncertainty in foil areal density, (4) uncertainty in monitor product half-life (included, but normally negligible), (5) uncertainty in IAEA recommended cross section (using values from the 2018 IAEA re-evaluation\,\cite{Hermanne2018}), and (6) uncertainty in differential deuteron fluence (from transport simulations).

\begin{figure}[ht]
    \centering
    \includegraphics[width=0.5\textwidth]{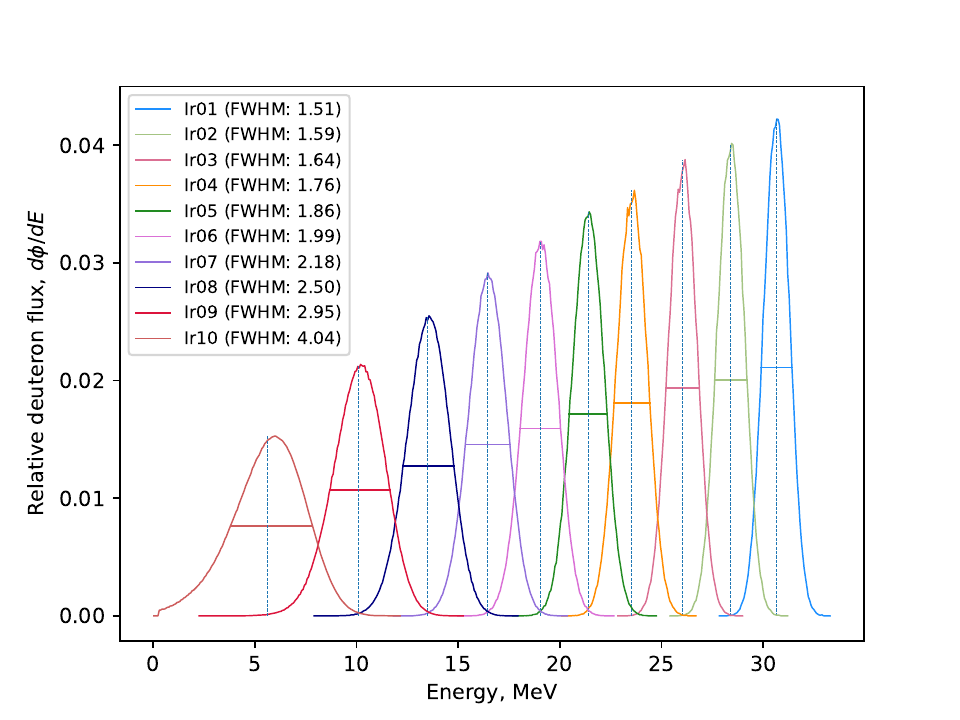}
    \caption{Visualization of the \texttt{NPAT}-calculated deuteron  energy spectrum for each of the 10 iridium foils. As the beam energy degrades, the distribution becomes progressively more skewed, and the full width half maximum (FWHM) becomes progressively larger.} 
    \label{fig:Ir_flux_distribution}
\end{figure}

% \begin{figure*}[ht] 
%     \centering
%     \subfloat[]{
%         \subfigimg[width=0.44\textwidth,trim={0 0 0 0},clip]{(a)}{chi_squared.pdf}{200}
%         \label{fig:VarianceMinimization_before}}
%     \hspace{0.01\textwidth} % juster for luft mellom bildene
%     \subfloat[]{
%         \subfigimg[width=0.46\textwidth,trim={0 0 0 0},clip]{(b)}{july25_compartment_beamcurrent_after_var_min.pdf}{200}
%         \label{fig:VarianceMinimization_after}}
%         \caption{(a) Result of $\chi^2$ analysis used in the variance minimization technique to determine the required adjustment (“Optimized”) to stopping power within the deuteron energy spectrum calculations. These corrections show marked improvement in monitor reaction agreement over the “Default” calculations.   (b) Monitor reaction beam currents following the “Optimized” stopping power adjustments.}
%      \label{fig:variance_minimization_fig}
% \end{figure*}

\begin{figure}[h!]
    \centering
    % Første figur (a)
    \begin{subfigure}[b]{0.45\textwidth}
        \centering
        \includegraphics[width=\textwidth]{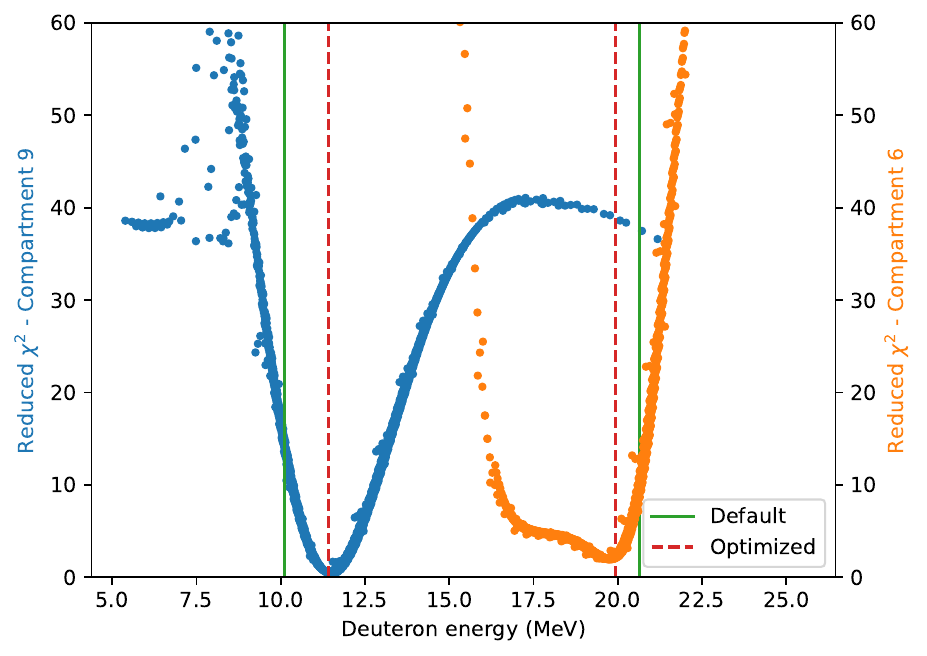}
        \subcaption*{(a)}
        \label{fig:a}
    \end{subfigure}
    \hfill
    % Andre figur (b)
    \begin{subfigure}[b]{0.45\textwidth}
        \centering
        \includegraphics[width=\textwidth]{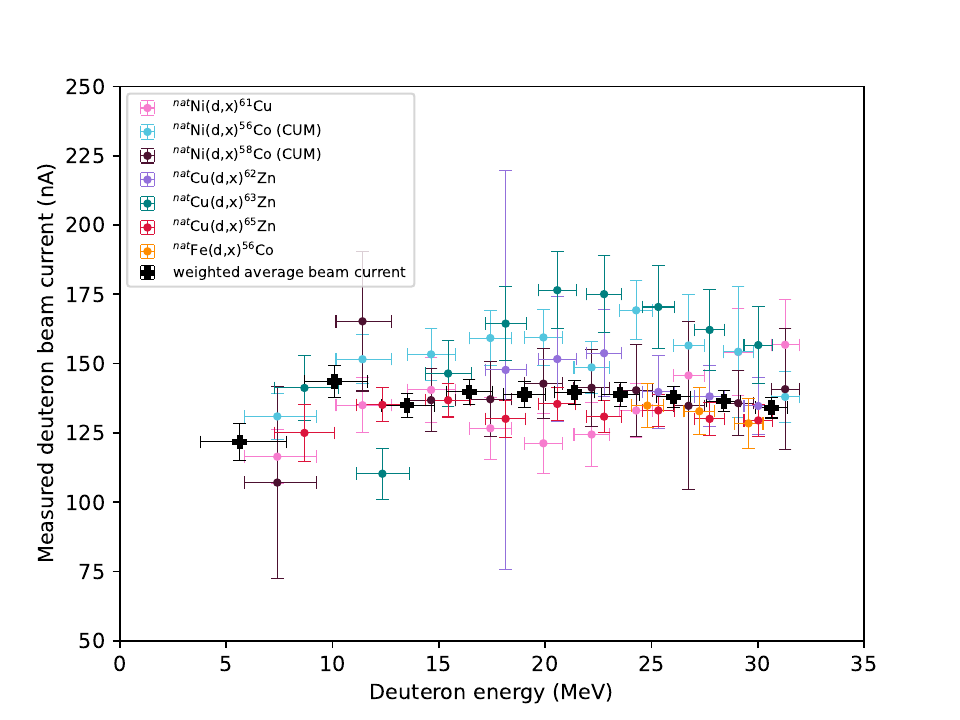}
        % \caption{}
        \subcaption*{(b)}
        \label{fig:b}
    \end{subfigure}
    
    \caption{(a) Result of $\chi^2$ analysis used in the variance minimization technique to determine the required adjustment (“Optimized”) to stopping power within the deuteron energy spectrum calculations. These corrections (with the red and green lines denoting the average energy of the beam entering each of the two listed compartments) show marked improvement in monitor reaction agreement (“Optimized” calculations) over the “Default” calculations. (b) Monitor reaction beam currents following the “Optimized” stopping power adjustments.}
    \label{fig:variance_minimization_fig}
\end{figure}

Particle transport calculations for the target stack were performed using  the  Andersen \& Ziegler stopping power formalism, as implemented in \texttt{NPAT} (now superseded by \texttt{Curie})
\cite{npat,curie_code}. 
The calculated energy spectrum for the iridium foils in the target stack is shown in \autoref{fig:Ir_flux_distribution} as an example.
As detailed in our previous work \cite{Voyles2018a,Morrell2020,Fox2021a,Fox2021,Voyles2021}, the “variance minimization” technique is used here to calculate  the  deuteron energy spectrum and current  in each foil, as well as reduce the overall systematic uncertainties in these calculations.
Our earlier work has established that a poor characterization of particle transport for the degraded beam leads to compounding disagreement in beam current towards the rear of the stack, and that the variance minimization technique is successful as a corrective tool for this discrepancy.
Most of our prior work has used the effective density of energy degrader foils as a free parameter in variance minimization, as they account for the overwhelming majority of stopping power for the stack.
However, when no degraders are present, or if the stack foils themselves constitute the majority of the total stopping power, a global scaling of the effective areal density of all stack components and a minor (\textless5\%) scaling of the incident beam energy   serve as better free parameters for this technique \cite{Morrell2020}.
Applying changes in the areal density and beam energy does not imply that the nominal values are incorrect, but act as surrogate corrections for improperly characterized deuteron stopping power and systematic uncertainties in beam characterization to best reproduce the IAEA monitor cross section data \cite{Morrell2020,Voyles2021}. 
In this work, the stack components' effective areal densities were
varied uniformly in the stopping power simulations by a factor
of up to $\pm$25\% of nominal values, and the incident beam energy was simultaneously varied up to $\pm$5\% of the nominal 33~MeV.
The resulting reduced $\chi^2$ for all monitor channels in a compartment (e.g., Ni06/Ir06/Cu06) was minimized simultaneously for compartments 6 and 9.
Compartments further back in the stack, or right above the energetic threshold for a monitor reaction, provide the greatest sensitivity to incorrect stopping powers during variance minimization---\#6 was chosen as the first compartment above the \natCu{}(d,x)\ce{^{62}Zn} threshold, and \#9 was chosen as the first compartment above the \natCu{}(d,x)\ce{^{63}Zn} threshold.

The results of the minimization technique, seen in \autoref{fig:variance_minimization_fig}, indicate a clear minimum in the deuteron current $\chi^2$ values, for both compartments 6 and 9.
This minimum occurs for a 2\% increase in incident beam energy (33.7 MeV) and a 4.25\% increase in effective areal density.
This implies that the monitor data are best reproduced if the average energy from the cyclotron tune (which includes multi-turn extraction) was slightly higher than 33 MeV, or if the deuteron stopping power calculations were underestimated.
Since the positive enhancements to both beam energy and the areal density have compensating effects on the transport calculation, the net change in stopping power is well within the range seen in our earlier work \cite{Voyles2018a,Morrell2020,Fox2021a,Fox2021,Voyles2021}. 
The variance minimization led to 
a clear improvement in agreement for the monitor beam currents over the default transport  calculations,
where the \natNi(d,x)\ce{^{56}Co} and \natCu(d,x)\ce{^{62}Zn} monitor reactions  improved the most. 
\natCu(d,x)\ce{^{63}Zn} and, to a lesser degree, \natNi(d,x)\ce{^{61}Cu}  yield   beam currents which were completely insensitive to variance minimization and consistently $\approx 1.5-2 \sigma$ larger in magnitude than the average beam current above approximately 15 MeV, discussed further in \autoref{section:discussion_monitors}. 
The changes in the beam current for the remaining monitor reactions were minor 
as a result of the variance minimization, confirming that the overall physical impact of applying this technique on the transport calculations was minimal. 

In each compartment, the correlated uncertainty-weighted mean  for the  \ce{^{nat}Cu}(d,x), \ce{^{nat}Fe}(d,x), and  \ce{^{nat}Ni}(d,x) monitor channels were calculated at each energy position, to determine the final fluence assignments for each monitor foil, respectively. The correlated uncertainty-weighted mean  for all observed monitor channels was used to determine the final fluence assignments for the Ir foils.
As in our recent work \cite{Voyles2021}, this correlated uncertainty-weighted mean beam current was calculated as:

\begin{equation} \label{eq:weighted_average_beamcurrent}
    \langle I \rangle  = \frac{ \sum_{i,j} I_j  \left( \mathbf{V}^{-1}_{ij}  \right) }{\sum_{i,j}  \left( \mathbf{V}^{-1}_{ij}  \right)}.
\end{equation}

Uncertainty in each final deuteron fluence is calculated by error propagation of the individual monitor channel fluence values at each energy position:

\begin{equation}
    \label{eq:corrWeight}
    \delta_{\langle I \rangle} = \sqrt{\frac{ 1}{\sum_{i,j}  \left( \mathbf{V}^{-1}_{ij}  \right)}}. 
\end{equation}
Each element $ij$ of the covariance matrix, $V_{ij}$ is calculated using the robust
sandwich estimator  \cite{huber1967behavior}:
\begin{equation}
    \label{eq:Vij}
    \mathbf{V}_{ij} = \mathrm{Cov}[I_i, I_j] = \sum_{\beta} \frac{\partial I}{\partial \beta_i} \delta_{\beta_i} \mathrm{Corr}[\beta_i, \beta_j] \delta_{\beta_j} \frac{\partial I}{\partial \beta_j},
\end{equation}
where $i$ and $j$ are the monitor reactions in a given compartment, and $\beta \in [A_0$, $\rho \Delta r$, $\lambda$, $\Delta t_{irr}$, $\int \sigma(E) \frac{d\phi}{dE}dE$]. Here, $A_0$ is the EoB-activity for the given product, $\rho\Delta r$ is the foil areal density, $\lambda$ is the decay constant for the given product, $\Delta t_{irr}$ is the duration of irradiation, and $\int \sigma(E) \frac{d\phi}{dE}dE$ is the flux-averaged monitor cross section for the given reaction.
% $\Sigma_{\beta_{ij}}$ is the correlation matrix for each $\beta$. 
$A_0$ is assumed to be 30\% correlated for all reactions, $\rho \Delta r$ is assumed to be 100\% correlated for reactions from the same monitor foil, $\lambda$ is assumed to be uncorrelated, $\Delta t_{irr}$ is assumed to be 100\% correlated, and $\int \sigma(E) \frac{d\phi}{dE}dE$ is assumed to be 30\% correlated for reactions from the same monitor foil.  

\subsection{Cross sections}

As in our recent work \cite{Voyles2021}, this variance-minimized deuteron current $\langle I \rangle$ in each foil is used to calculate cross sections.
The reported asymmetric uncertainty in beam energy corresponds to the full width at half maximum (FWHM) for the flux distribution in each foil.
Using the weighted average beam current over each compartment determined from the monitor cross sections and fluence assignments, the reported cross sections were calculated as:
\begin{equation}
    \sigma = \frac{A_0}{\rho\Delta r \cdot \langle I \rangle \cdot (1-e^{-\lambda \Delta t_\text{irr}})}.
\end{equation}

The propagated uncertainty in cross sections are reported as the quadrature sum of the  uncertainty in (1) EoB-activities, (2) beam current, (3) foil areal density, and (4) the decay constant of the product. 

The measured data are compared to previous experimental data from the EXFOR database \cite{Zerkin2020}, recommended data from the TENDL-2023 library \cite{Koning2019}, along with calculations from the nuclear reaction modeling codes TALYS-2.04, ALICE-2020, CoH-3.6.0, and EMPIRE-3.2.3 
\cite{Herman2007, KAWANO2010, Blann1996, Koning2012}. 
The codes were run similarly to  \cite{Voyles2021}, using all default parameters, and energies from 0-40 MeV.

\section{Results and discussion} \label{ch:results}

This work presents the first extension of our variance minimization methods, performed for a deuteron beam irradiation, where small changes in the incident deuteron energy and the monitor foils were applied rather than to the beam degraders. 
A total of 50 reaction channels  (including 7 measurements in the monitor foils) for deuteron-induced reactions on natural iridium, iron, nickel, and copper are reported in this work. 
This includes the first experimental measurements of \natIr(d,x)\ce{^{188m1+g}Ir} (independent), \natIr(d,x)\ce{^{190m1+g}Ir} (cumulative), \natFe(d,x)\ce{^{48}V} (cumulative),\ce{^{53}Fe} (cumulative), \natNi(d,x)\ce{^{59}Fe} (cumulative), \natNi(d,x)\ce{^{56, 57, 58m, 58g}Co} (independent). 
In addition, the measurement of \natIr(d,x)\ce{^{189}Ir} revealed a potential error in the absolute intensity  of the decay gammas. 
The decay data used in analysis are listed in \ref{ch_app:tables}. 
Plots of the excitation functions for reactions in the monitor foils are attached in \ref{excitationfunctions}.
All reported cross section data are tabulated in \ref{chapter:cross_sections_tabulated}. 

The measured \natIr(d,x) cross sections in this work are compared to prior experimental data from Tárkányi \emph{et. al.} \cite{Tarkanyi2006, Tarkanyi2019} and Obata \emph{et. al.} \cite{Obata2019} who have performed similar
experiments with deuteron energies in the ranges 1.7–38.0 MeV, 17.09–49.50 MeV, and 40.3–23.8 MeV, respectively.

\subsection{Excitation function for \natIr(d,x)\ce{^{193m}Pt}}
\begin{figure}
    \centering
    \includegraphics[width=0.8\textwidth]{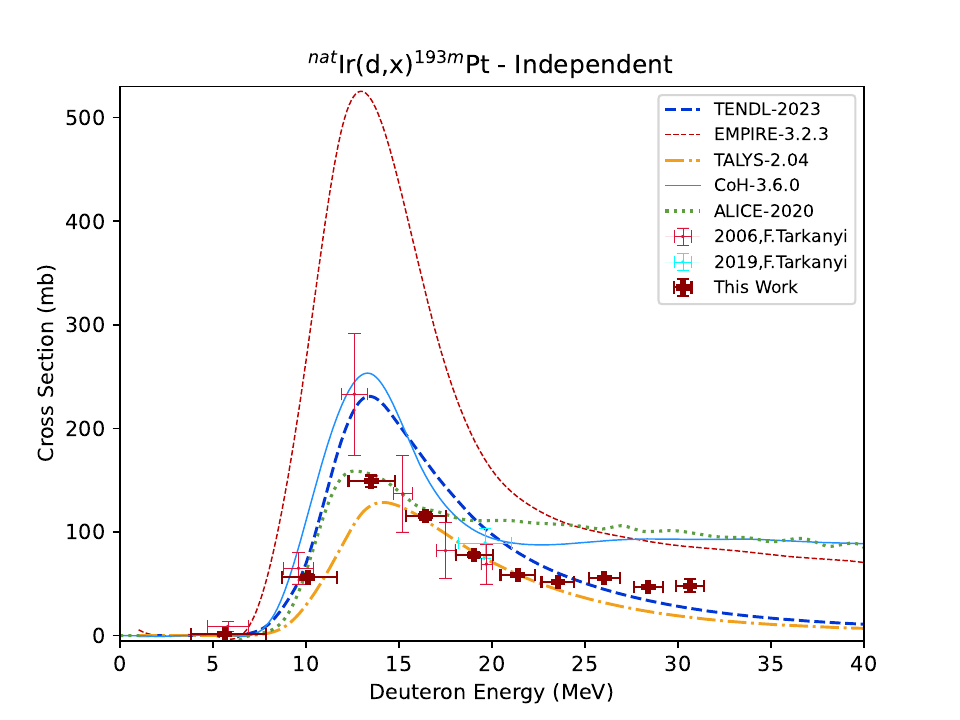}
    \caption{Excitation function for \natIr(d,x)\ce{^{193m}Pt}, visibly peaking between approximately  11-13 MeV. }
    \label{fig:Ir_193mPt_i}
\end{figure}

\ce{^{193m}Pt} was observed in all foils (\autoref{fig:Ir_193mPt_i}).
The lack of strong gamma lines makes it challenging to make precise measurements. 
\ce{^{193m}Pt} was first identified exclusively using the weakly fed gamma-line ($E_\gamma$=135.5 keV, intensity=0.11\%). Including the characteristic X-ray after internal conversion (K$_{\alpha1}$=66.831 keV, intensity=7.21\%) improved the statistics significantly. This work presents a far more precise measurement than prior work, likely due to this inclusion.
As shown in the figure, there is good agreement with the experimental data, with a predicted peak in the excitation function between 12--14 MeV. 
Additional future measurements near this compound peak would be useful to resolve its sharpness, as well as the amplitude. 
TENDL-2023 and CoH-3.6.0 match the experimental data around the peak. 
TALYS-2.04 and TENDL-2023 best reproduce the pre-equilibrium tail. 
ALICE-2020 underpredicts the peak, while overpredicting the pre-equilibrium tail. 
EMPIRE-3.2.3 overpredicts the entire excitation function. 

\subsection{Excitation function for \natIr(d,x)\ce{^{188,189,191}Pt}}

\begin{figure}
    \centering
    \includegraphics[width=0.8\textwidth]{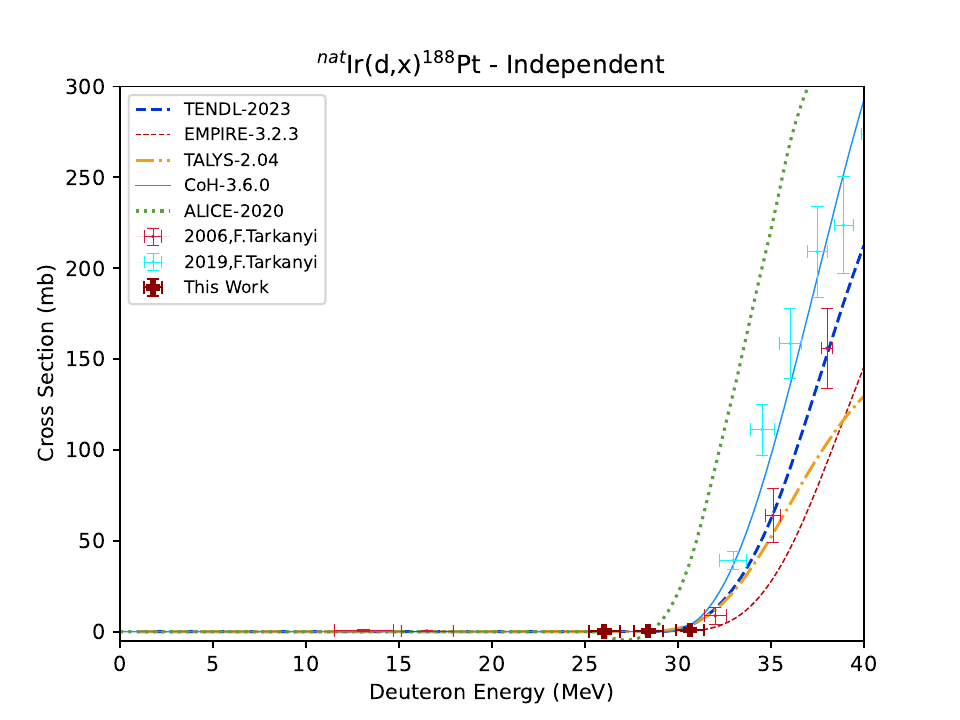}
    \caption{Excitation function for \natIr(d,x)\ce{^{188}Pt}. The \ce{^{191}Ir}(d, 5n)\ce{^{188}Pt} channel opens at approximately $E_d$=26 MeV.}
    \label{fig:Ir_188Pt_i}
\end{figure}

\begin{figure}
    \centering
    \includegraphics[width=0.8\textwidth]{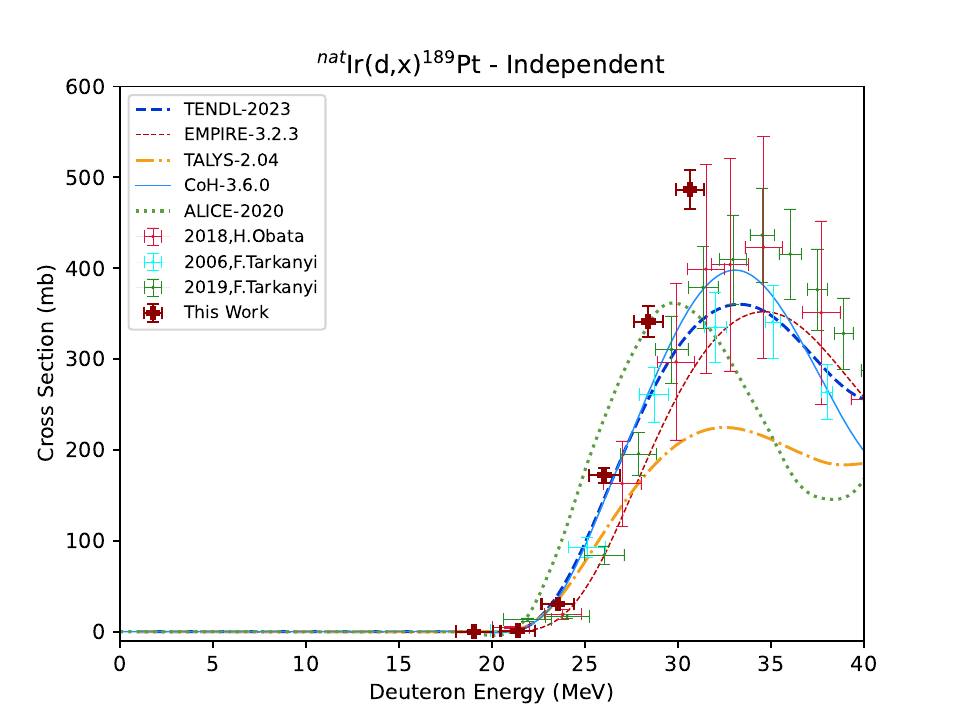}
    \caption{Excitation function for \natIr(d,x)\ce{^{189}Pt}. Reaction \ce{^{191}Ir}(d, 4n)\ce{^{189}Pt} opens at approximately $E_d$=19 MeV.}
    \label{fig:Ir_189Pt_i}
\end{figure}

\begin{figure}
    \centering
    \includegraphics[width=0.8\textwidth]{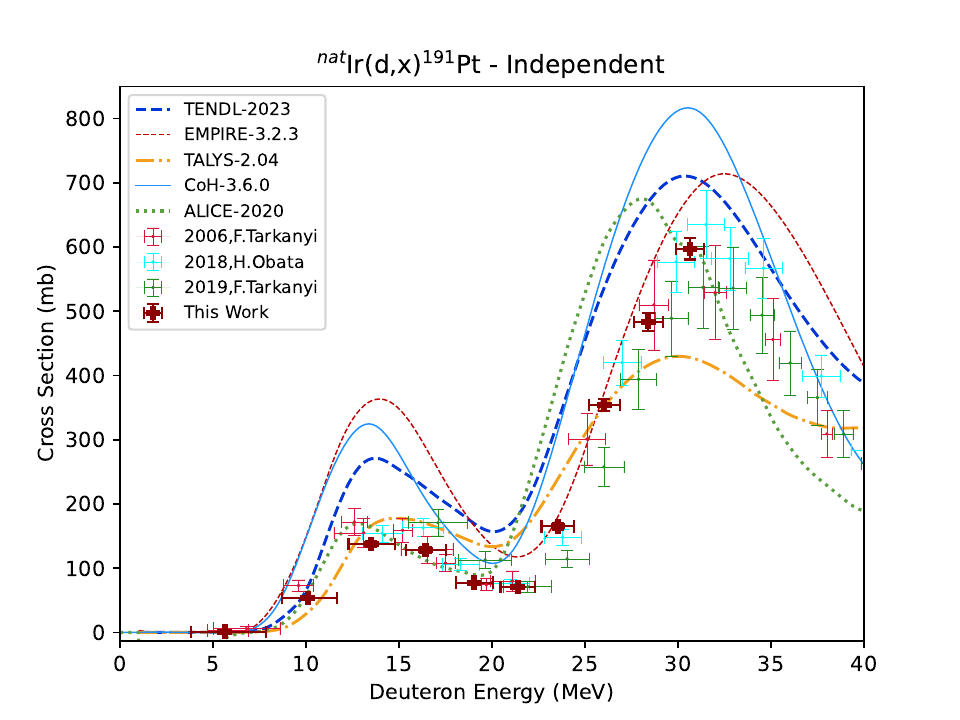}
    \caption{Excitation function for \natIr(d,x)\ce{^{191}Pt}. Both reaction channels contributions via \ce{^{191}Ir}(d,2n) and \ce{^{193}Ir}(d,4n) can be seen, where the former  peaks at approximately 12 MeV and the latter  peaks at approximately 30 MeV.}
    \label{fig:Ir_191Pt_i}
\end{figure}

\ce{^{188}Pt} (\autoref{fig:Ir_188Pt_i}) is produced by \ce{^{191}Ir}(d,5n), and is energetically possible with deuterons of approximately 26 MeV and higher. 
The measured cross sections in this work are near threshold, and, beyond our measured energies, we can see a rapid increase in this channel as the deuteron energy increases. 
As the prior data were measured above the energies of this work, it is difficult to make a direct comparison, though the trend appears to be consistent.
TENDL-2023 and CoH-3.6.0 appear to agree best with the experimental data for this channel.
ALICE-2020 predicts a steeper increase. 
\ce{^{189}Pt} (\autoref{fig:Ir_189Pt_i}) is produced through \ce{^{191}Ir}(d,4n), opening at approximately 19 MeV. 
EMPIRE-3.2.3, CoH-3.6.0 and TENDL-2023 all agree with the previously measured data. 
The excitation function predicted by ALICE-2020 is shifted slightly toward a lower deuteron energy. 
TALYS-2.04 predicts the peak location, but disagrees in magnitude. The results obtained in this analysis tend to estimate slightly higher cross sections than previous experimental data,
in particular on the rise of the compound peak. The data however agree within uncertainties, in addition to the shape of the excitation function.
Given that our results agree in magnitude with previous experimental data in all other reaction channels, systematic factors such as beam current determination likely do not explain the minor discrepancies seen in this one channel, particularly for the two highest energy datapoints. A more likely explanations is the difference in adopted nuclear data for gamma branching ratios in this channel. This can be seen in the difference in magnitude between the 2006 and 2019 Tárkányi \emph{et. al.} measurements \cite{Tarkanyi2006, Tarkanyi2019}, whose difference in branching ratios lead to the difference in magnitude in their results. As both Obata \cite{Obata2019} and Tárkányi measurements use the two most intense gammas for \ce{^{189}Pt}, whereas we use an average of five of its decay gammas, this may explain the minor differences observed. 
\ce{^{191}Pt} (\autoref{fig:Ir_191Pt_i}) is produced from both stable \ce{^{191}Ir} and \ce{^{193}Ir} isotopes, and the excitation function visibly peaks around 12 MeV and 30 MeV for both reactions, respectively. 
The results are in good agreement with previous experimental data. 
The reaction modeling codes all generally show good performance in predicting the shape. 
For the first peak, ALICE-2020 best predicts the magnitude of the experimental data, whereas TENDL-2023 and CoH-3.6.0 best predict its shape. 
For the second peak, there is some disagreement of peak location, but overall, the codes predict the experimental data fairly well.

\subsection{Excitation function for \natIr(d,x)\ce{^{188}Ir} - independent and cumulative}
\begin{figure}
    \centering
    \includegraphics[width=0.8\textwidth]{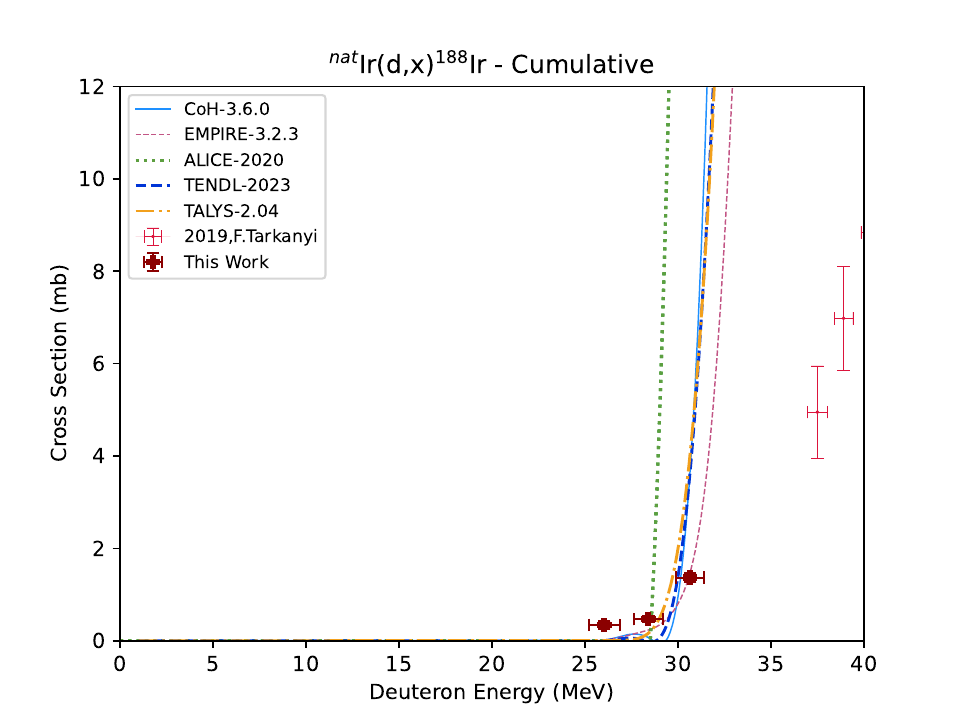}
    \caption{Excitation function for  \natIr(d,x)\ce{^{188}Ir}. This is the total cumulative cross section, including feeding from \ce{^{188}Pt}. }
    \label{fig:Ir_188Ir_c}
\end{figure}

\begin{figure}
    \centering
    \includegraphics[width=0.8\textwidth]{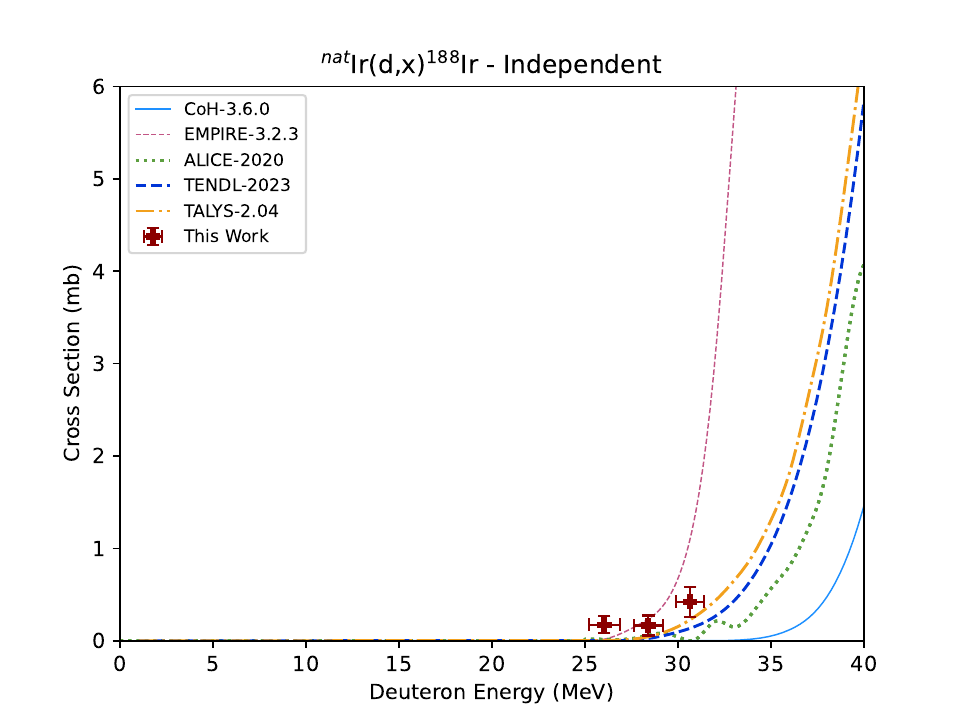}
    \caption{Excitation function for \natIr(d,x)\ce{^{188}Ir}. The green points show the total cumulative cross section with feeding from \ce{^{188}Pt}, while the red points show the independent measurements of \ce{^{188}Ir} production. }
    \label{fig:Ir_188Ir_i}
\end{figure}

\ce{^{188}Ir} is produced directly via \ce{^{191}Ir}(d,p4n) and via $\beta^+$ decay from \ce{^{188}Pt} ($\epsilon: 100\%$). 
Both total cumulative (\autoref{fig:Ir_188Ir_c}), and the independent (\autoref{fig:Ir_188Ir_i}) are reported, where the independent measurements were obtained by subtracting the independent contribution from \ce{^{188}Pt}. 
The energetic threshold for this reaction is approximately 16 MeV, and there is a rapid increase when the reaction channel for \ce{^{188}Pt} opens at approximately 26 MeV. 
For the total cumulative cross sections, the experimental data in this work suggest a more rapid increase above threshold, which is generally matched in shape  by EMPIRE-3.2.3, TALYS-2.04, and TENDL-2023.
However, given the paucity of existing experimental data, as well as the weakly-fed nature of this channel, it is no surprise that all codes seem to be performing poorly here, with additional measurements needed to better tune predictive capabilities.
However, our independent results for this channel are well-reproduced by TALYS-2.04 and TENDL-2023, suggesting that the overly sharp rise of the cumulative prediction lies with overestimation of the feeding of \ce{^{188}Pt}, as seen previously in \autoref{fig:Ir_188Pt_i}. 

\subsection{Excitation function for \natIr(d,x)\ce{^{189}Ir}}
\begin{figure}
    \centering
    \includegraphics[width=0.8\textwidth]{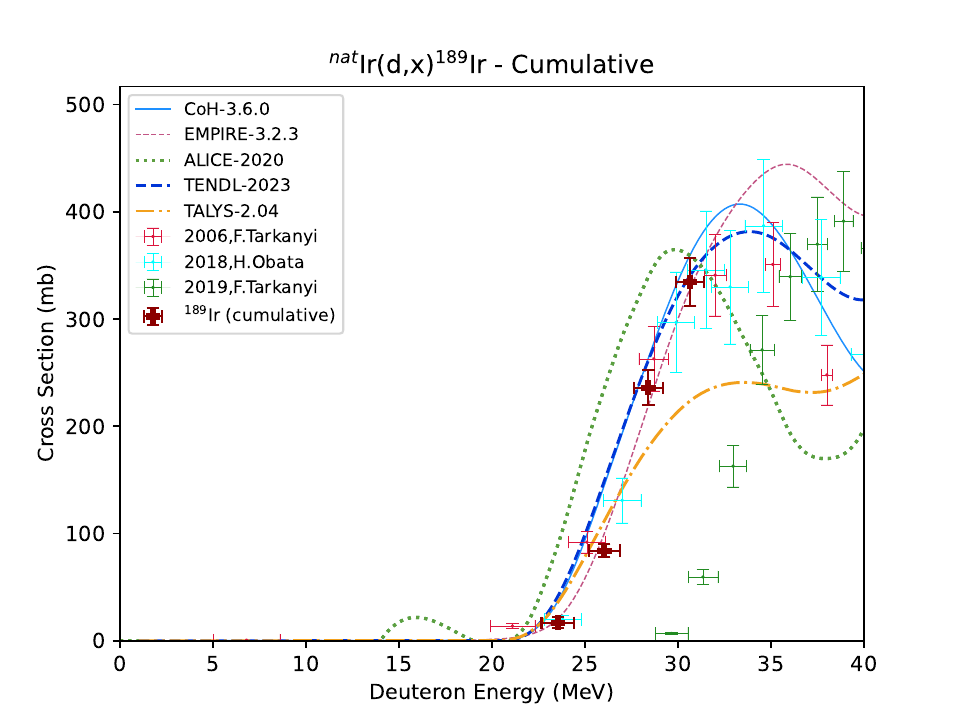}
    \caption{Excitation function for \natIr(d,x)\ce{^{189}Ir}.}
    \label{fig:Ir_189Ir_c}
\end{figure}

\ce{^{189}Ir} (\autoref{fig:Ir_189Ir_c}) is produced directly via \ce{^{191}Ir}(d,p3n) and via $\beta^+$ decay ($\epsilon: 100\%$) from \ce{^{189}Pt}. 
The total cumulative cross section for \ce{^{189}Ir} should therefore be greater than the independent measurements of \ce{^{189}Pt}. 
As shown in \autoref{tab:Iridium_Cross_sections}, the independent measurements of \ce{^{189}Pt} provide higher cross section measurements, resulting in a nonphysical behavior. 
This becomes evident when attempting to subtract feeding from \ce{^{189}Pt}, resulting in negative measurements of the independent cross sections for \ce{^{189}Ir}. 
This may imply that the absolute intensity normalization of the decay-gammas may be wrong in the current mass chain evaluation of A=189 \cite{Johnson2017}. 
Therefore, to obtain the cumulative \ce{^{189}Ir} cross sections, spectra taken more than 4-5 days after EoB were used, assuming that \ce{^{189}Pt} had decayed completely after ten half-lives. 
There is disagreement with the measured cross sections of Tárkányi \emph{et. al.} (2019) \cite{Tarkanyi2019}, suggesting a 10 MeV shift toward higher energy for the compound peak. 
However, the measured cross sections in this work agree with the experimental data of Tárkányi \emph{et. al.} (2006) \cite{Tarkanyi2006} and Obata \emph{et. al.} (2018) \cite{Obata2019}, as well as the reaction modeling codes build confidence in the location of the peak. 
While no clear cause for this discrepancy in their 2006 results can be found, neither do the authors present an explanation in their more recent 2019 manuscript.
This particular reaction requires further investigation.

\subsection{Excitation functions for \natIr(d,x)\ce{^{190, 190m1+g, 190m2}Ir} }
\begin{figure}
    \centering
    \includegraphics[width=0.8\textwidth]{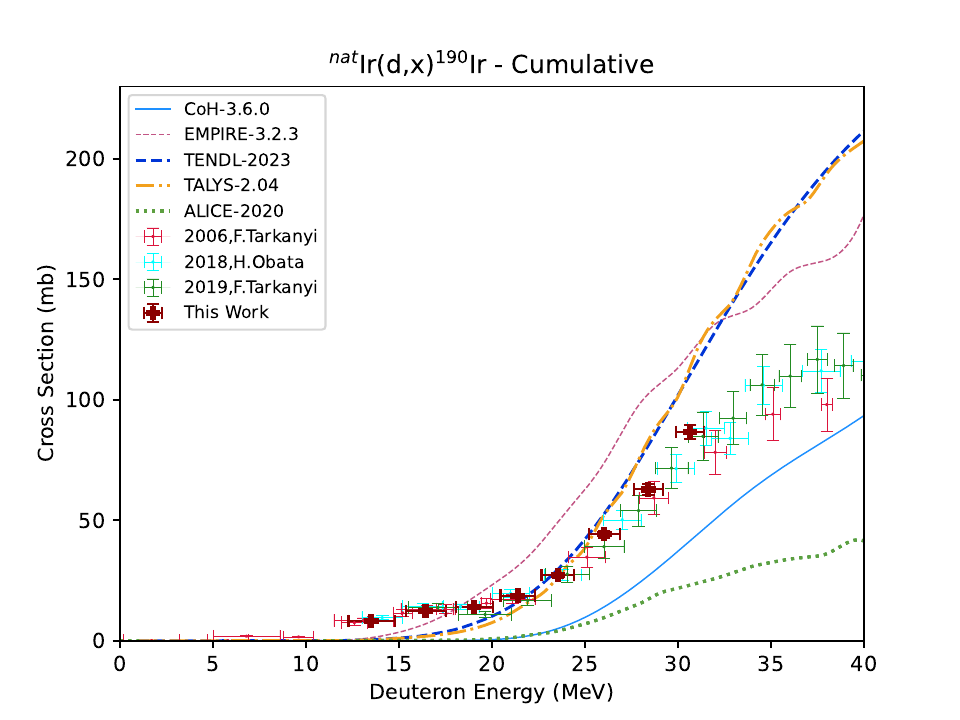}
    \caption{Excitation function for \ce{^{190}Ir} (cumulative), including feeding of the ground state and from decay of its two isomers. }
    \label{fig:Ir_190Ir_c}
\end{figure}

\begin{figure}
    \centering
    \includegraphics[width=0.8\textwidth]{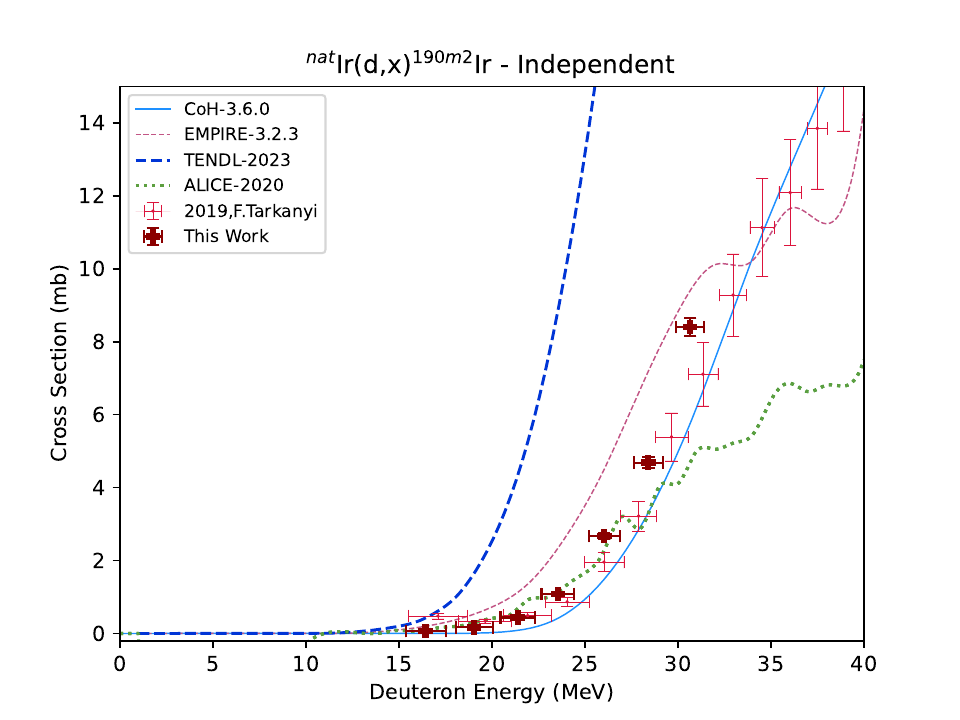}
    \caption{ Excitation function for \ce{^{190m2}Ir}.}
    \label{fig:Ir_190m2Ir_i}
\end{figure}

\begin{figure}
    \centering
    \includegraphics[width=0.8\textwidth]{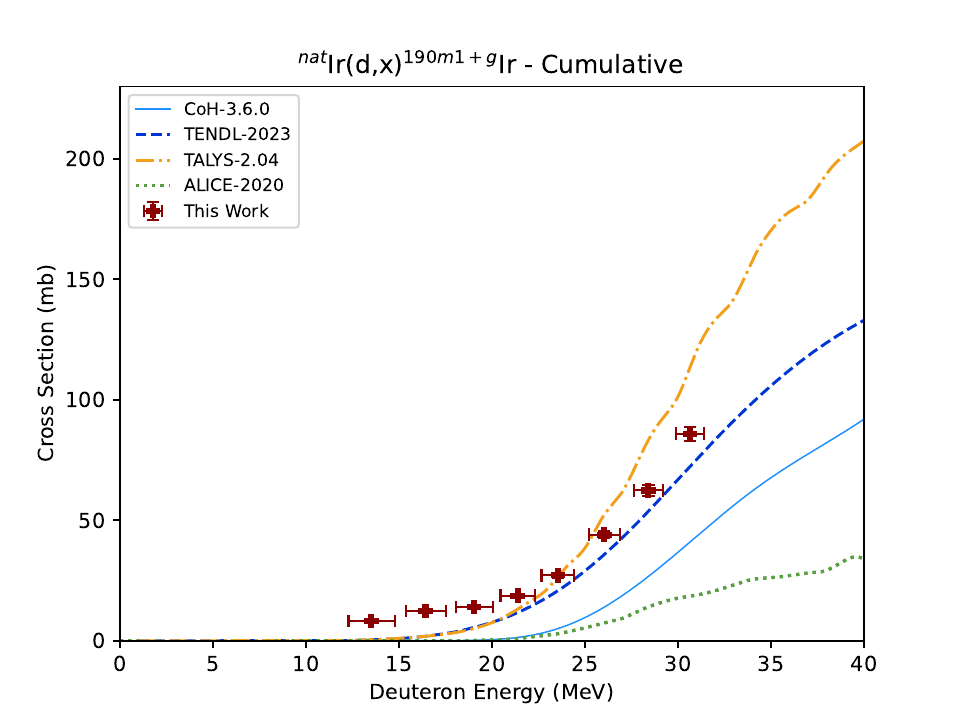}
    \caption{Excitation function for \ce{^{190m1+g}Ir}.} 
    % The green points represents the total cumulative cross section of the ground state and isomers. The red points show the subtracted contribution from the m2 isomer.  }
    \label{fig:Ir_190m1+gIr_c}
\end{figure}

\ce{^{190}Ir} has two isomers, both of which decay by isomeric transition (m2: 8.6\%, m1: 100\%). A total cumulative cross section (\autoref{fig:Ir_190Ir_c}), an independent measurement of the m2 isomer (\autoref{fig:Ir_190m2Ir_i}),
and the subtracted measurement of \ce{^{190m1+g}Ir} (\autoref{fig:Ir_190m1+gIr_c}) are reported. The total cumulative cross section is consistent with the existing experimental data. TENDL-2023 and TALYS-2.04 agree with the 
experimental data from approximately 20 MeV to 30 MeV. CoH-3.6.0 and ALICE-2020 consistently underestimate the excitation function, while EMPIRE-3.2.3 consistently overestimates the shape above 20 MeV. The contribution from the m2 isomer to the 
ground state is minimal, given that cross sections and the branching ratio are low. CoH-3.6.0, ALICE-2020 and EMPIRE-3.2.3 reproduce the shape of the excitation function of \ce{^{190m2}Ir}, where CoH-3.6.3 in particular follows the shape of 
the experimental data accurately. TENDL-2023 predicts a lower effective energetic threshold, as well as a more rapid rise of the excitation function, compared to both experimental data as well as the other codes. For \ce{^{190m1+g}Ir}, TENDL-2023 and TALYS-2.04 reproduce the shape above 20 MeV. 
ALICE-2020 and CoH-3.6.0, as for the total cumulative cross section, consistently underestimate the excitation function. It was not possible to model \ce{^{190m1+g}Ir} with EMPIRE-3.2.3 or \ce{^{190m2}Ir} with TALYS-2.04, as both codes lacked these isomers in their structure tables.

\subsection{Excitation functions for \natIr(d,x)\ce{^{192}Ir} }

\begin{figure}
    \centering
    \includegraphics[width=0.8\textwidth]{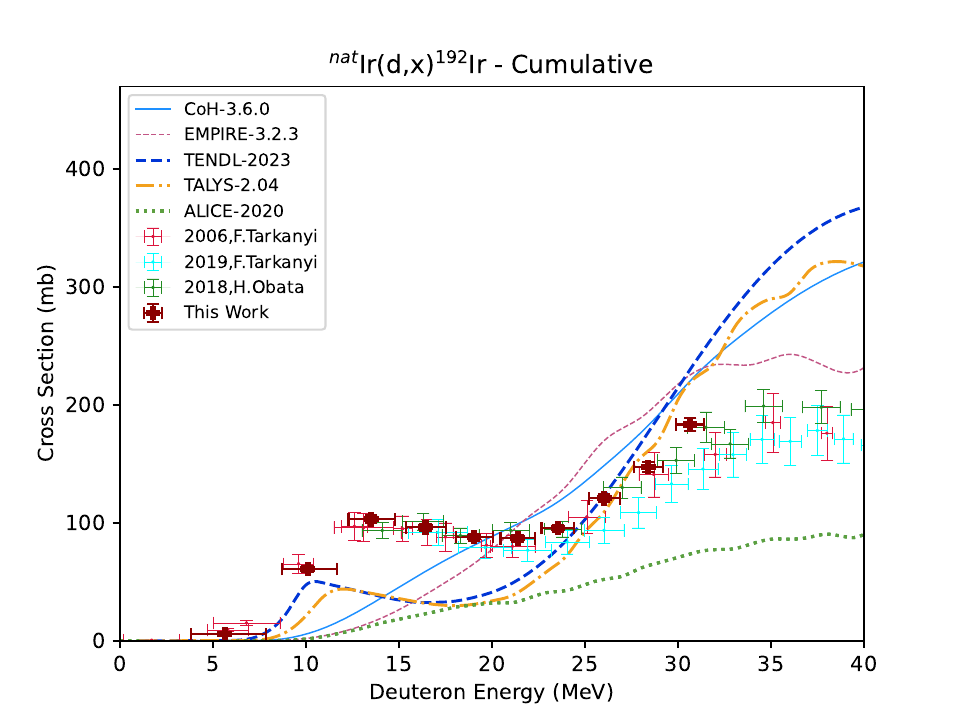}
    \caption{Excitation function for \natIr(d,x)\ce{^{192}Ir}. Both reaction channels via \ce{^{191}Ir}(d,p) and \ce{^{193}Ir}(d,2np) can be seen, where the former  peaks at approximately 14 MeV and the latter begins to peak at approximately 35-40 MeV.}
    \label{fig:Ir_192Ir_c}
\end{figure}

\ce{^{192}Ir} (\autoref{fig:Ir_192Ir_c}) is produced via both target isotopes, and the excitation function  visibly peaks around 10 MeV, increasing towards the second peak near 40 MeV. 
In addition, \ce{^{192}Ir} has two isomers, both of which decay by isomeric transition (m2: 100\%, m1: 99.9825\%). 
Neither were able to be directly measured in this work, due to their half-lives.
The experimental data in this work are consistent with previous experimental data, though slightly higher. 
ALICE-2020 underestimates the excitation function. 
TENDL-2023 and TALYS-2.04 accurately predict the shape, but underestimate in magnitude. 
Neither EMPIRE-3.2.3 and CoH-3.6.0 reproduce the shape of the excitation function well. 

\subsection{Excitation functions for \natIr(d,x)\ce{^{194m1+g}Ir}}
\begin{figure}
    \centering
    \includegraphics[width=0.8\textwidth]{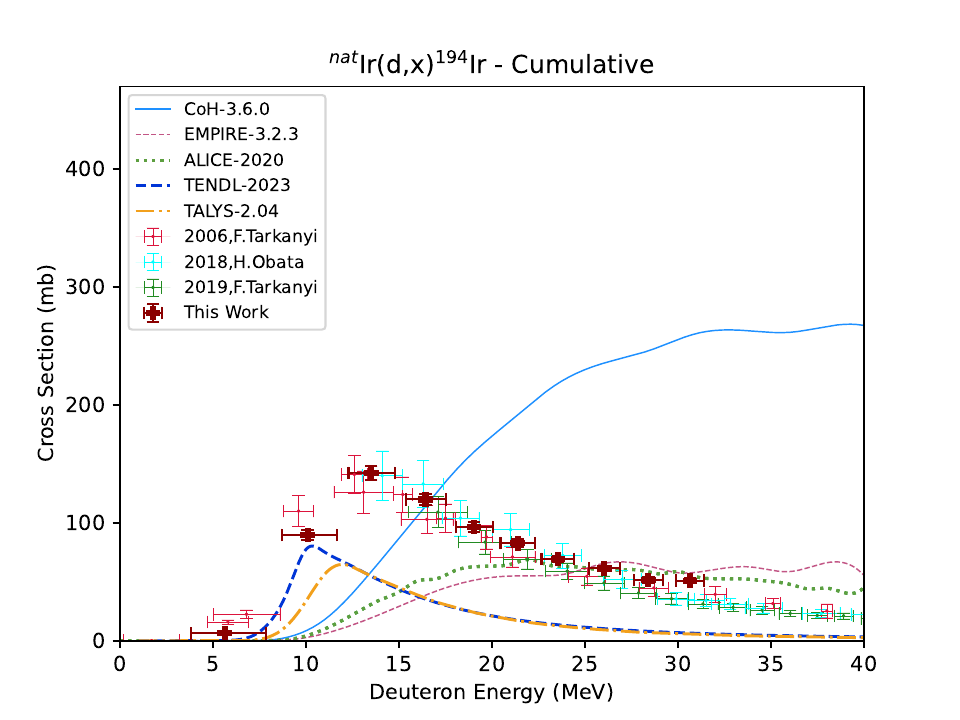}
    \caption{Excitation function for \natIr(d,x)\ce{^{194}Ir}.}
    \label{fig:Ir_194Ir_c}
\end{figure}

The cumulative cross sections for the ground state and m1 isomer  (\autoref{fig:Ir_194Ir_c}) are reported.  
The measurements agree well with previous experimental data. 
Both TALYS-2.04 and TENDL-2023 predict the shape of the excitation function, but suggest a peak maximum shifted slightly towards a lower deuteron energy. 
CoH-3.6.0, ALICE-2020 and EMPIRE-3.2.3 disagree in shape and magnitude of the excitation function.

\subsection{Physical thick target yield} 
\label{subsection:monitorandthicktargetyield}

In this work, physical yields were calculated for the observed iridium products.
The physical thick target yield is defined here as the yield at the beginning of the irradiation (at $\Delta t_\text{irr}=0$) per unit current (MBq/C) \cite{Otuka2015}:

\begin{equation} \label{eq:physicalYield}
    \alpha_\text{phys} = \lambda \int_{E_L}^{E_0} - \left( \frac{1}{\rho} \frac{dE}{dx} \right) ^{-1} \frac{\sigma(E)}{Z\cdot e} dE,
\end{equation}

where $\lambda$ represents the decay constant for a product with cross section $\sigma(E)$. The deuteron energy window is integrated from incident beam energy, $E_0$, to the energy at the desired depth in the material $E_L$ (typically to E=0 or threshold of the desired reaction).
$\rho$ is the mass density of the target, $dE/dx$ is the beam stopping power in the target, and $Z\cdot e$ is the charge of the particle beam, here +1 for deuterons. \\

\begin{figure}
    \centering
    \includegraphics[width=0.8\textwidth]{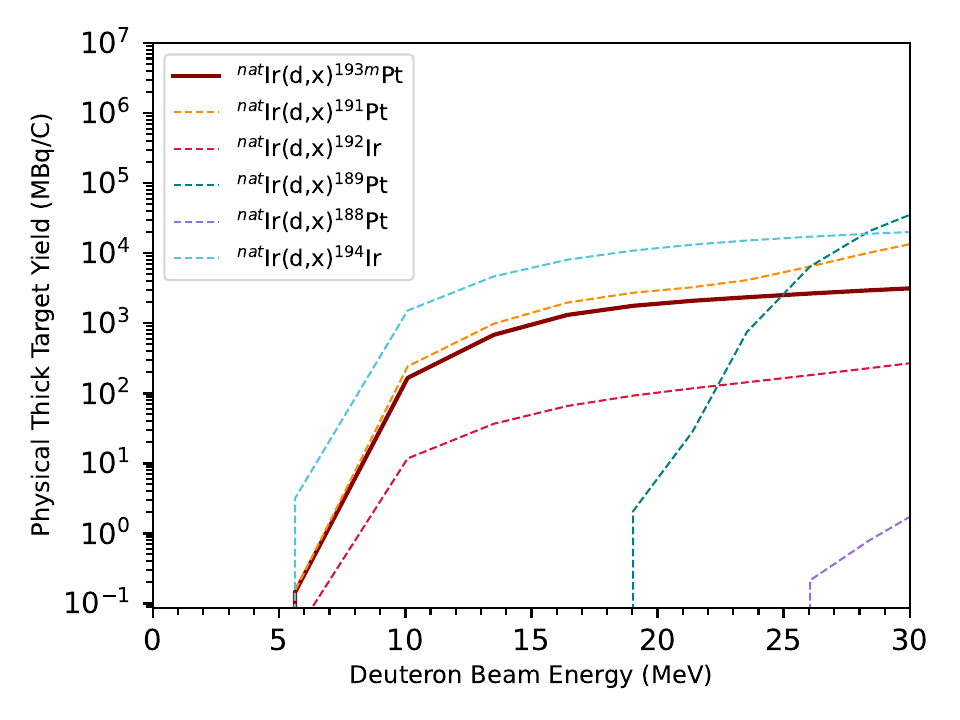}
    \caption{The physical thick target yield for various competing reactions for deuterons on natural iridium. }
    \label{fig:physicalyield}
\end{figure}

For medical applications, the primary competing measurable reaction channels for \ce{^{193m}Pt} in the energy region are from \ce{^{191}Ir}(d,2n)\ce{^{191}Pt}, \ce{^{191}Ir}(d,p)\ce{^{192}Ir} and \ce{^{193}Ir}(d,p)\ce{^{194}Ir}, as shown in \autoref{fig:physicalyield}. Radiocontamination from \ce{^{188, 189}Pt} appears above deuteron energies of approximately 19 MeV, well above the threshold for \ce{^{193m}Pt} production.
Contamination from other radioplatinums is not desired due to the inability to separate them via radiochemical methods. 
For a natural Ir target, the energy window yielding the highest cross sections is between 8--18 MeV, where the only competing radioplatinum is formed via \ce{^{191}Ir}(d,2n)\ce{^{191}Pt}.
Beyond 18 MeV, not only does \ce{^{189}Pt} (along with \ce{^{191}Pt} from  \ce{^{193}Ir}(d,4n)) rise rapidly, but very little additional \ce{^{193m}Pt} is produced, making higher energies useless for production.
By using an enriched target of \ce{^{193}Ir}, contaminants of radioplatinums can likely be significantly reduced. 
In addition, production of stable \ce{^{192}Pt} will be reduced, which is preferable as production of \enquote{cold} platinums will compete with \ce{^{193m}Pt} during labeling, reducing its specific activity and, thus, clinical potency. 
The stable/long-lived platinum contaminants \ce{^{193, 194}Pt} will  be produced in this energy window, which will reduce the specific activity of the final product, though \ce{^{193m}Pt} production should still dominate on a mass basis. 
Additionally, both \ce{^{192}Ir} and \ce{^{194}Ir} decay to stable \ce{^{192}Pt} and \ce{^{194}Pt}. 
To reduce further increase of stable platinum contaminants, separation of \ce{^{193m}Pt} should take place shortly after end of beam, to minimize this stable platinum feeding from \ce{^{194}Ir} in particular, with a half-life of 19.28 h. 
Further measurements of both cross sections and derived thick-target yields should be performed using an enriched \ce{^{193}Ir} target, to further optimize production conditions if \ce{^{193m}Pt} gains widespread clinical interest. \\

\begin{figure}[h!]
    \centering
    % Første figur (a)
    \begin{subfigure}[b]{0.45\textwidth}
        \centering
        \includegraphics[width=\textwidth]{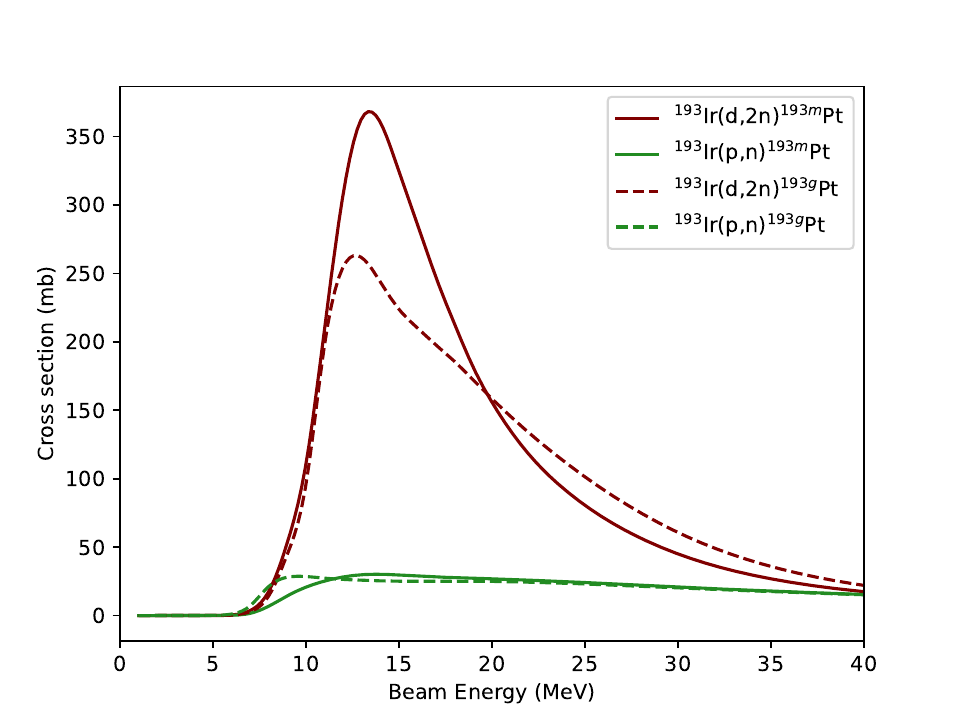}
        % \caption{}
        % \subcaption*{(a)}
        \subcaption{}
        \label{fig:tendl_protonsvsdeuterons}
    \end{subfigure}
    \hfill
    % Andre figur (b)
    \begin{subfigure}[b]{0.45\textwidth}
        \centering
        \includegraphics[width=\textwidth]{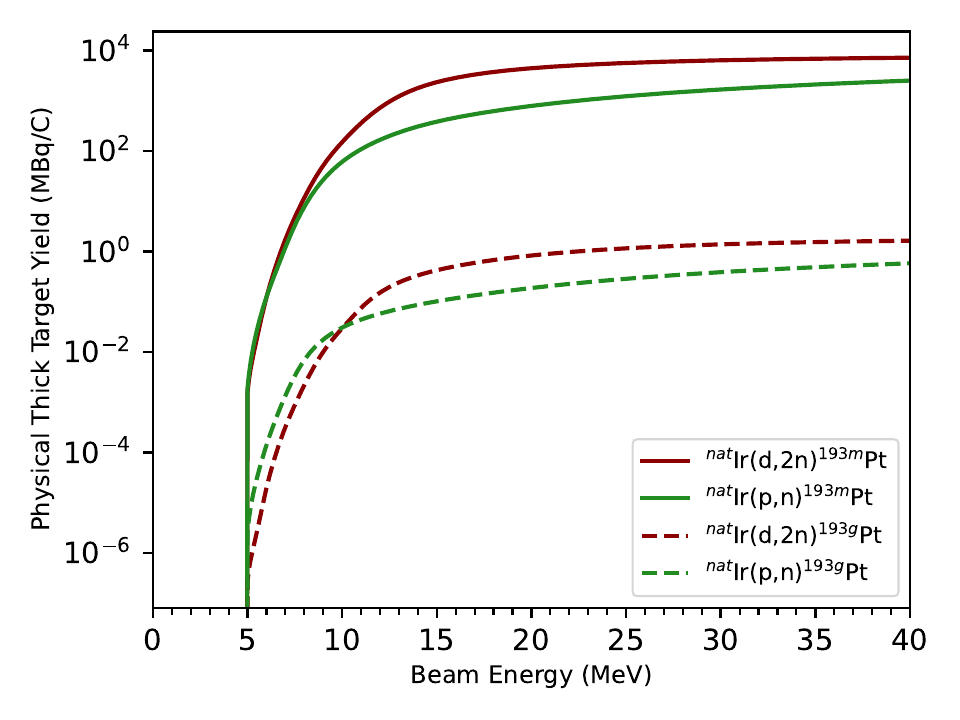}
        % \caption{}
        % \subcaption*{(b)}
        \subcaption{}
        \label{fig:tty_protonsvsdeuterons}
    \end{subfigure}
    
    \caption{(a) The excitation functions of the ground state and isomer production of \ce{^{193}Pt}, produced with protons and deuterons. Deuteron production yields higher cross sections than protons. (b) Thick target yield calculated using the TENDL-2023 cross sections. This indicates that the population of the isomer using deuterons will provide the higher yield. For both routes, the isomer yield is significantly larger than the yields of the \ce{^{193}Pt} ground state due to both larger cross sections as well as a significantly shorter half life. }
    \label{fig:protons_vs_deuterons_fig}
\end{figure}

As mentioned in the introduction, deuterons are expected to more strongly populate \ce{^{193m}Pt} due to their increased spin transfer, relative to protons. As there are currently no experimental data on \natIr(p,x)\ce{^{193m}Pt}, TENDL-2023 was used to compare the production cross sections of protons and deuterons for both the ground state and isomer (\autoref{fig:tendl_protonsvsdeuterons}), as well as their thick target yields (\autoref{fig:tty_protonsvsdeuterons}). We can see that the production cross section for \ce{^{193m}Pt} is about 12 times higher for deuterons than protons. The thick target yield emphasizes this. We can clearly see that both the isomer to ground state ratio is highest using deuterons, and that deuterons will provide a higher physical thick target yield than protons.

% \begin{figure*}[ht] 
%     \centering
%     \subfloat{
%         \centering
%         \subfigimg[width=0.49\textwidth,trim={0 0 0 0},clip]{}{tendl_protonsvsdeuterons.pdf}{190}
%          \label{fig:tendl_protonsvsdeuterons}}
%      \subfloat{
%         \centering
%         \subfigimg[width=0.45\textwidth,trim={0 0 0 0},clip]{}{physicalyield_deuteronsvsproton.pdf}{190}
%          \label{fig:tty_protonsvsdeuterons}}
%     \caption{(a, left) The excitation functions of the ground state and isomer production of \ce{^{193}Pt}, produced with protons and deuterons. Deuteron production yields higher cross sections than protons. (b, right) Thick target yield calculated using the TENDL-2023 cross sections. This indicates that the population of the isomer using deuterons will provide the higher yield. For both routes, the isomer yield is significantly larger than the yields of the \ce{^{193}Pt} ground state due to both larger cross sections as well as a significantly shorter half life.}
%      \label{fig:protons_vs_deuterons_fig}
% \end{figure*}

\subsection{Status of deuteron monitor reactions}\label{section:discussion_monitors}
\begin{figure}
    \centering
    \includegraphics[width=0.8\textwidth]{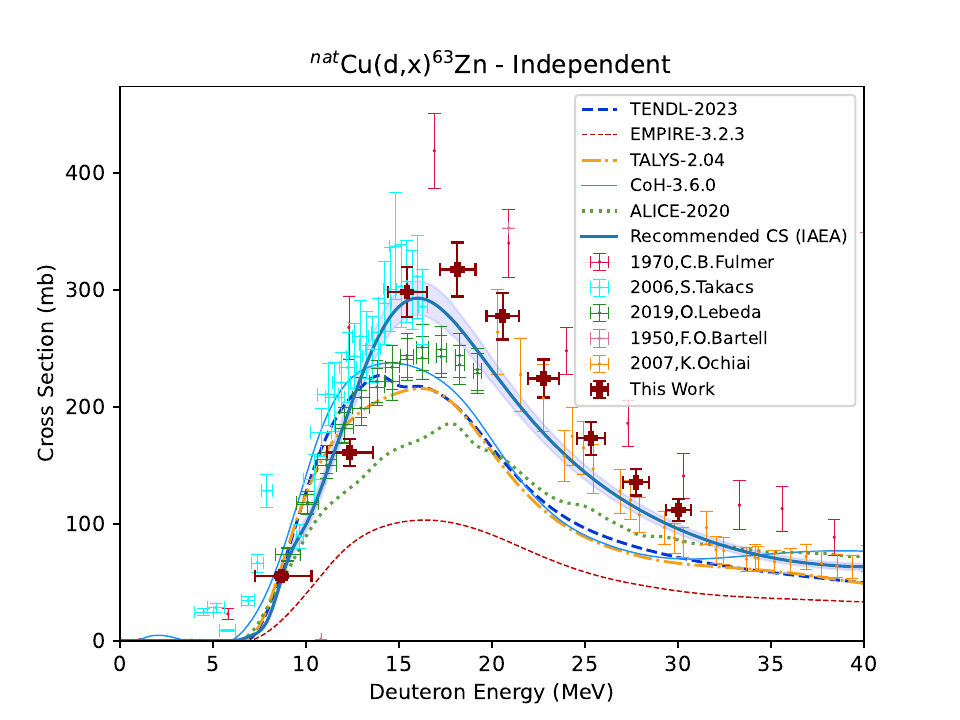}
    \caption{The excitation function of \natCu(d,x)\ce{^{63}Zn}. Experimental data are retrieved from the EXFOR database \cite{Fulmer1970, Takacs2006a, Lebeda2019, Bartell1950, Ochiai2007}.}
    \label{fig:Cu_63Zn_i}
\end{figure}

The current state of datasets used to obtain the IAEA-recommended cross section data for deuteron-induced monitor reactions is significantly more limited compared to proton-induced reactions. 
For instance, the reaction \natCu(p,x)\ce{^{63}Zn} has 3.7 times more experimental datasets than \natCu(d,x)\ce{^{63}Zn}. 
More experimental data will improve the monitor reactions, yielding more precise beam currents, and better cross section results. 
The particular monitor reaction of \natCu(d,x)\ce{^{63}Zn} (\autoref{fig:Cu_63Zn_i}) is a good example of opportunities for improvement in deuteron monitor reactions. 
While our apparent \natCu(d,x)\ce{^{63}Zn} cross sections (calculated using the final beam currents in the same fashion as with all other reactions on copper), agree with the IAEA evaluation up to approximately 15 MeV, we yield higher values beyond this energy and in the pre-equilibrium tail than the recommended IAEA data. 
Notably, relatively few measurements (with significant spread) exist in this region, with the current evaluation essentially fitting the Ochiai data \cite{Ochiai2007} beyond 30 MeV, and approximately interpolating between the average of the lower-energy Lebeda data \cite{Lebeda2019} and the high-energy Ochiai data from 20--30 MeV.
This is also apparent in the beam current plot from earlier (\autoref{fig:variance_minimization_fig}), where this particular reaction consistently yielded higher beam currents, without significant improvement after variance minimization. 
If the beam current solely relied upon this monitor reaction, all the measured cross sections would have resulted in systematically being offset above 15 MeV, paving a cautionary tale for those measuring cross sections without the use of multiple monitor reactions to build confidence. 
For this reaction, we recommend to perform additional measurements, in particular for deuteron energies from 15-30 MeV, to build confidence in the location of the compound peak and the pre-equilibrium tail, and to perform an updated evaluation based on this future work. 
Comparisons for the remaining monitor reactions are plotted in \autoref{fig:MonitorReactions} of Appendix C.

\begin{figure}
    \centering
    \includegraphics[width=0.8\textwidth]{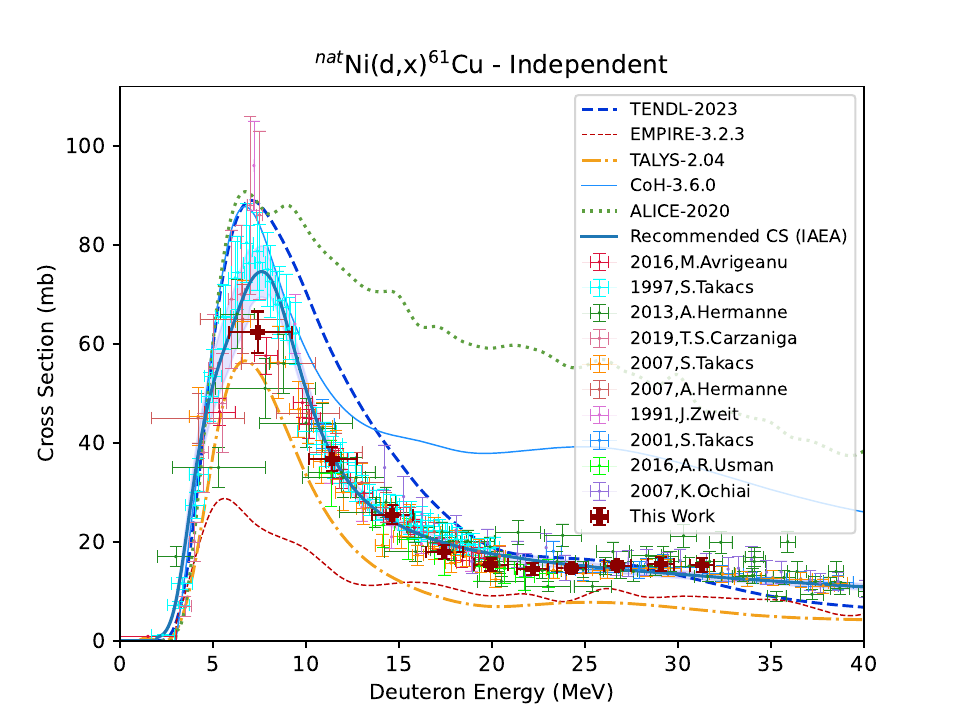}
    \caption{The excitation function of \natNi(d,x)\ce{^{61}Cu}. Experimental data are retrieved from the EXFOR database \cite{Avrigeanu2016, Takacs1997, Hermanne2013, Carzaniga2019, Takacs2007, hermanne2007activation, Zweit1991, Takacs2001, Usman2016a, Ochiai2007}.}
    \label{fig:Ni_61Cu_i}
\end{figure}

Similarly, in light of our recently-reported relative $\gamma$-ray intensities for the decay of \ce{^{61}Cu}
\cite{Bleuel2021}, we have adopted the decay data from the IAEA-proposed re-evaluation based upon our measurements \cite{hermanne2025critical}.
Using these decay data, we report here new independent cross sections for the \natNi(d,x)\ce{^{61}Cu} monitor reaction, seen in \autoref{fig:Ni_61Cu_i}.
The re-evaluation uses these decay data to report corrected cross sections for the \natNi(d,x)\ce{^{61}Cu} monitor reaction, but as the pointwise data are not yet publicly available, we can only compare to the existing evaluation.
Overall, the changes in decay data have only minor impact, with  less than approximately 5\% changes in cross section seen as a result. 
These changes also are consistent with the systematics seen in the re-evaluation, particularly for the inflection point seen near 20 MeV \cite{hermanne2025critical}.
While the impact in these results is minor/modest, due to our use of multiple gamma lines in spectroscopy, these new results will be valuable for future evaluations of this monitor reaction.

This observation indicates the need for the development and evaluation of new monitor reactions for deuteron-induced reactions. A good monitor reaction product should have the following properties: high intensity, distinct gamma lines without background contamination, sufficiently long half-life to allow for multiple observations with good statistics, no production possible through secondary neutrons, and preferably no decay feeding by other radionuclides. 
In addition, monitor targets should be available and inexpensive. 
Currently, the only deuteron monitor reaction for a \ce{^{nat}Fe} target is the \natFe(d,x)\ce{^{56}Co} reaction. Both \natFe(d,x)\ce{^{57}Co} and \natFe(d,x)\ce{^{58}Co} are candidate reactions that should perhaps be considered in the future, given the fairly modest amount of experimental data currently available for both reactions

\natNi(d,x) has a limited number of monitor reactions that meet the criteria, as the current reactions \natNi(d,x)\ce{^{56,58}Co} are both subject to decay feeding and can be produced through secondary neutrons.
\natNi(d,x)\ce{^{64}Cu} may be considered, but its lack of intense gamma lines make it challenging to use as a monitor channel. 
\natIr(d,x) is not practically suitable as a monitor foil due to the high cost of the target material. 

\section{Conclusions}

This work presents 43 cross-section measurements, in the beam-energy range between threshold and 33 MeV, for the \natIr(d,x), \natFe(d,x), \natNi(d,x), and
\natCu(d,x) reactions. Most of these results provide the best precision for the respective reactions to date. We report the first measurements of experimental measurements of \natIr(d,x)\ce{^{188m1+g}Ir} (independent), \natIr(d,x)\ce{^{190m1+g}Ir} (cumulative), \natFe(d,x)\ce{^{48}V},\ce{^{53}Fe} (cumulative), \natFe(d,x)\ce{^{61}Co} (cumulative), \natNi(d,x)\ce{^{59}Fe} (cumulative), \natNi(d,x)\ce{^{56, 57, 58m, 58g}Co} (independent).
In this work, the energy window which optimizes the production of \ce{^{193m}Pt} was investigated.
In addition, other medically valuable radionuclides such as \ce{^{191}Pt} and \ce{^{192,194}Ir} were also investigated. Physical thick target yield proves an optimized production window from threshold to 18 MeV. The physical thick target yield of protons and deuterons were compared both for the ground state and the isomer of \ce{^{193}Pt} using TENDL-2023. This emphasized the isomer-to-ground state ratio being higher using deuterons, as well as a higher yield for \ce{^{193m}Pt} in total.

Modeling of deuteron-induced reactions on iridium, in particular, has provided yet another example of the poor predictive capability of modern reaction codes, particularly for the range of high-spin isomers populated in these reactions. Developing a framework for parameter optimization similar to that in our recent work for proton-induced reactions \cite{Fox2021,Fox2021a} is an obvious next step, given the performance seen here. However, such an optimization would benefit from reaction data collected for deuteron irradiations of a naturally monoisotopic or enriched target, to significantly constrain the parameter space involved. Likewise, choosing a spherical or nearly spherical target nucleus as a first step will provide an important probe to optmize the nuclear level density model. In particular, our results suggest that these codes would benefit from improvements in the deuteron optical model, the role of angular momentum transfer in these reactions, and the competition between compound reaction mechanisms and deuteron breakup. While such an optimization is beyond the scope of the current work, these results nonetheless play a valuable role in highlighting the need for future work in proving the predictive performance of deuteron-induced reaction modeling. 
This work has also identified a potential error in the branching ratio of \ce{^{189}Ir}. Finally, this work has revealed weaknesses with the \natCu(d,x)\ce{^{63}Zn} and \natNi(d,x)\ce{^{61}Cu} monitor reaction evaluations, further emphasizing the importance of high-fidelity experimental data, and the dangers of reliance upon a single monitor reaction, in particular for deuterons.

\section*{Acknowledgments}

The authors would like to particularly acknowledge the assistance and support of Brien Ninemire, Scott Small, Nick Brickner, and all the rest of the operations, research, and facilities staff of the LBNL 88-Inch Cyclotron. We also wish to thank Amanda Lewis and Toshihiko Kawano, who provided assistance with reaction modeling, as well as Ian Kolaja, Oline A. Ranum, Daniel Murphy, Jesus Rios, and M. Shuza Uddin, who participated in these experiments. This research was supported by the U.S. Department of Energy Isotope Program, managed by the Office of Science for Isotope R\&D and Production, and carried out under Lawrence Berkeley National Laboratory (Contract No. DE-AC02-05CH11231) and Lawrence Livermore National Laboratory (Contract No. DE-AC52-07NA27344).
This research was supported by the Research Council of Norway through the Norwegian Nuclear Research Center (Project No. 341985) and through INTPART (Project No 310094).

\newpage

\appendix

\section{Stack design} \label{ch_app:stack_design}
The stack design is tabulated in Table \ref{tab:foilCharacterization}.

\begin{table}[]
    \centering
    \footnotesize
    \caption{Specifications of the  target stack design in the present work. The deuteron beam enters the stack upstream of the SS1  profile monitor, and travels through the stack in the order presented here. A 316 stainless steel foil is inserted at both the front and rear of the target stack as a monitor of the beam’s spatial profile by developing radiochromic film (Gafchromic EBT3) after end-of-bombardment. Unlike in our recent work with proton irradiations, no degrader foils or Kapton tape were included between the foils listed here, due to the increased stopping power of deuterons.}
    \label{tab:foilCharacterization}
    \begin{tabular}{lll}
\makecell{Foil} &  \makecell{Measured thickness ($\micro$m)} &  \makecell{Measured areal \\ density (mg/cm$^2$)}\\ 
\hline

\makecell{SS1}  & \makecell{124.5$\pm$1.0}                                              & \makecell{100.199$\pm$0.091}                                    \\ \hline
\makecell{Ni01} & \makecell{28.5$\pm$1.0}                             & \makecell{22.77$\pm$0.14}                                         \\
\makecell{Ir01} & \makecell{29.5$\pm$1.3}                             & \makecell{55.174$\pm$0.053}                                         \\
\makecell{Cu01} & \makecell{34.1$\pm$3.0}                             & \makecell{22.338$\pm$0.048}                                         \\
\makecell{Fe01} & \makecell{27.8$\pm$1.0}                             & \makecell{20.03$\pm$0.11}                                         \\ \hline
\makecell{Ni02} & \makecell{29.5$\pm$1.0}                             & \makecell{23.118$\pm$0.096}                                         \\
\makecell{Ir02} & \makecell{27.8$\pm$1.5}                             & \makecell{55.60$\pm$0.24}                                         \\
\makecell{Cu02} & \makecell{34.8$\pm$2.0}                             & \makecell{22.325$\pm$0.028}                                         \\
\makecell{Fe02} & \makecell{27.4$\pm$4.0}                             & \makecell{20.017$\pm$0.034}                                         \\ \hline
\makecell{Ni03} & \makecell{27.0$\pm$1.0}                             & \makecell{22.338$\pm$0.066}                                         \\
\makecell{Ir03} & \makecell{24.3$\pm$1.7}                             & \makecell{55.64$\pm$0.12}                                         \\
\makecell{Cu03} & \makecell{34.3$\pm$3.0}                             & \makecell{22.313$\pm$0.043}                                         \\
\makecell{Fe03} & \makecell{31.0$\pm$4.0}                             & \makecell{19.95$\pm$0.11}                                         \\ \hline
\makecell{Ni04} & \makecell{27.3$\pm$1.0}                             & \makecell{20.704$\pm$0.068}                                         \\
\makecell{Ir04} & \makecell{26.1$\pm$2.9}                             & \makecell{56.00$\pm$0.11}                                         \\
\makecell{Cu04} & \makecell{33.3$\pm$1.0}                             & \makecell{22.284$\pm$0.027}                                         \\ \hline
\makecell{Ni05} & \makecell{26.3$\pm$1.0}                             & \makecell{21.768$\pm$0.045}                                         \\
\makecell{Ir05} & \makecell{25.6$\pm$2.4}                             & \makecell{55.161$\pm$0.081}                                         \\
\makecell{Cu05} & \makecell{33.4$\pm$3.0}                             & \makecell{22.443$\pm$0.028}                                         \\ \hline
\makecell{Ni06} & \makecell{28.5$\pm$2.0}                             & \makecell{22.86$\pm$0.12}                                         \\
\makecell{Ir06} & \makecell{24.0$\pm$1.2}                             & \makecell{55.731$\pm$0.088}                                         \\
\makecell{Cu06} & \makecell{34.0$\pm$2.0}                             & \makecell{22.396$\pm$0.012}                                         \\ \hline
\makecell{Ni07} & \makecell{26.8$\pm$2.0}                             & \makecell{23.092$\pm$0.078}                                         \\
\makecell{Ir07} & \makecell{27.8$\pm$2.2}                             & \makecell{56.685$\pm$0.085}                                         \\
\makecell{Cu07} & \makecell{32.6$\pm$2.0}                             & \makecell{22.320$\pm$0.014}                                         \\ \hline
\makecell{Ni08} & \makecell{25.6$\pm$3.0}                             & \makecell{22.41$\pm$0.12}                                         \\
\makecell{Ir08} & \makecell{28.1$\pm$2.2}                             & \makecell{58.03$\pm$0.13}                                         \\
\makecell{Cu08} & \makecell{33.3$\pm$3.0}                             & \makecell{22.401$\pm$0.033}                                         \\ \hline
\makecell{Ni09} & \makecell{25.7$\pm$4.0}                             & \makecell{21.741$\pm$0.073}                                         \\
\makecell{Ir09} & \makecell{27.3$\pm$1.0}                             & \makecell{56.669$\pm$0.043}                                         \\
\makecell{Cu09} & \makecell{33.1$\pm$2.0}                             & \makecell{22.425$\pm$0.041}                                         \\ \hline
\makecell{Ni10} & \makecell{27.1$\pm$3.0}                             & \makecell{23.093$\pm$0.024}                                         \\
\makecell{Ir10} & \makecell{27.0$\pm$1.6}                             & \makecell{55.065$\pm$0.055}                                         \\
\makecell{Cu10} & \makecell{35.5$\pm$1.0}                             & \makecell{22.314$\pm$0.047}                                         \\ \hline
\makecell{SS2}  & \makecell{125.2$\pm$1.0}                                              & \makecell{100.865$\pm$0.097}                                    \\ 
\hline
    \end{tabular}
\end{table}

\newpage

\section{Tabulated nuclear and reaction data} \label{ch_app:tables}
Accepted decay data used to calculate activities for the monitor products are listed in \autoref{tab:monitor_reactions}, and \cref{tab:Products_Ir,tab:Products_Fe,tab:Products_Ni,tab:Products_Cu} for nuclei produced from iridium, iron, nickel and copper, respectively. Decay data is retrieved from the NuDat-3.0 database.

\clearpage
\footnotesize
\centering
\setlength{\tabcolsep}{3pt} % Default value: 6pt
\renewcommand{\arraystretch}{1} % Default value: 1
    \begin{longtable}{ccccc} 
    \caption{The table shows an overview of the gamma-lines used to calculate monitor reaction activities in each foil at the end of beam. Nuclear data from: \cite{Junde2011,Nesaraja2010,NICHOLS2012973,ERJUN2001,Browne_2010_A65,hermanne2025critical}. }
    \label{tab:monitor_reactions} \\ \hline
    \makecell{Monitor \\ reaction} & \makecell{Half-life} & \makecell{$E_\gamma$ \\ (keV)} & \makecell{$I_\gamma$(\%)} & \makecell{Useful\\ beam-energies \\ (MeV)\cite{Hermanne2018}}  \\
    \hline
    \makecell{\natFe(d,x)$^{56}$Co} & \makecell{77.236 d} & \makecell[t]{263.434 \\ 486.55 \\ 733.514 \\ 787.743 \\ 846.770 \\ 852.732 \\ 896.510 \\ 977.372 \\ 996.948 \\ 1037.843 \\ 1140.368 \\ 1159.944 \\ 1175.101 \\ 1198.888 \\ 1238.288 \\ 1335.40 \\ 1360.212 \\ 1771.357 \\ 1963.741 \\ 2015.215 \\ 2034.791 \\ 2212.944 \\ 2276.131 \\ 2598.500} & \makecell[t]{0.0220 $\pm$ 0.0003 \\ 0.0540 $\pm$ 0.002 \\ 0.191 $\pm$ 0.003 \\ 0.311 $\pm$ 0.003 \\ 99.9399 $\pm$ 0.0023 \\ 0.049 $\pm$ 0.003 \\ 0.073 $\pm$ 0.003 \\ 1.421 $\pm$ 0.006 \\ 0.111 $\pm$ 0.004 \\ 14.05 $\pm$ 0.04 \\ 0.132 $\pm$ 0.003 \\ 0.094 $\pm$ 0.006 \\2.252 $\pm$ 0.006 \\ 0.049 $\pm$ 0.005 \\ 66.46 $\pm$ 0.12 \\ 0.1224 $\pm$ 0.0012 \\ 4.283 $\pm$ 0.012 \\ 15.41 $\pm$ 0.06 \\ 0.707 $\pm$ 0.004 \\ 3.016 $\pm$ 0.012 \\ 7.77 $\pm$ 0.03 \\ 0.388 $\pm$ 0.004 \\ 0.118 $\pm$ 0.004 \\ 16.97 $\pm$ 0.04 }  & \makecell{10-50}\\ 
    \hline
    \makecell{\natNi(d,x)$^{56}$Co\\ (cumulative)} & \makecell{77.236 d}  & \makecell[t]{787.743 \\846.770\\ 977.372 \\ 1175.101 \\ 1963.741 \\ 2015.215 \\ 2034.791} & \makecell[t]{0.3111 $\pm$ 0.003 \\ 99.9399 $\pm$ 0.0023 \\ 1.421 $\pm$ 0.006 \\ 2.252 $\pm$ 0.006 \\ 0.707 $\pm$ 0.004 \\ 3.016 $\pm$ 0.012 \\ 7.77 $\pm$ 0.03} & \makecell{5-50} \\
    \hline
    \makecell{\natNi(d,x)$^{58}$Co\\ (cumulative)} & \makecell{70.86 d} & \makecell[t]{810.7593\\ 863.951 \\1674.725 } & \makecell[t]{99.450 $\pm$ 0.01 \\ 0.686 $\pm$ 0.01 \\ 0.517 $\pm$ 0.01 } & \makecell{5-50}  \\
    \hline
    \makecell{\natNi(d,x)$^{61}$Cu \\ \cite{hermanne2025critical}} & \makecell{3.339 h} & \makecell[t]{282.956 \\ 373.050 \\ 529.169\\ 588.605 \\ 625.605 \\ 656.008 \\ 816.692 \\ 841.211 \\ 902.294 \\ 1032.162 \\ 1073.465 \\ 1132.351 \\ 1185.234 \\ 1446.492} & \makecell[t]{12.7 $\pm$ 2.0 \\ 2.14 $\pm$ 0.33 \\ 0.37 $\pm$ 0.06 \\ 1.17 $\pm$ 0.18 \\ 0.044 $\pm$ 0.008 \\ 10.07 $\pm$ 1.68 \\ 0.35 $\pm$ 0.05 \\ 0.24 $\pm$ 0.04 \\ 0.087 $\pm$ 0.013 \\ 0.055 $\pm$ 0.012 \\ 0.042 $\pm$ 0.012 \\ 0.091 $\pm$ 0.014 \\ 3.6 $\pm$ 0.6 \\ 0.046 $\pm$ 0.007} & \makecell{3-50}  \\
    \hline
    \makecell{\natCu(d,x)$^{62}$Zn} & \makecell{9.193 h} & \makecell[t]{40.85 \\ 243.36 \\ 246.95 \\ 260.43 \\ 304.88 \\ 394.03 \\ 548.35 \\ 596.56 \\ 637.41} &  \makecell[t]{25.5  $\pm$ 2.4 \\ 2.52 $\pm$ 0.23 \\ 1.90 $\pm$ 1.8 \\ 1.35 $\pm$ 1.3 \\ 0.29 $\pm$ 0.03 \\ 2.24 $\pm$ 0.17 \\ 15.3 $\pm$ 1.4 \\ 26.0 $\pm$ 2.0 \\ 0.25 $\pm$ 0.03} & \makecell{15-50}  \\
    \hline
    \makecell{\natCu(d,x)$^{63}$Zn} & \makecell{38.47 min} & \makecell[t]{449.93 \\ 669.62 \\ 962.06} & \makecell[t]{0.236 $\pm$ 0.019 \\ 8.2 $\pm$ 0.3 \\6.5 $\pm$ 0.4 } & \makecell{8-50}  \\
    \hline
    \makecell{\natCu(d,x)$^{65}$Zn} & \makecell{243.93 d} &\makecell[t]{1115.539} & \makecell[t]{50.04 $\pm$ 0.1} & \makecell{5-50} \\
    \hline
    
    \end{longtable}

\newpage
\clearpage
\newpage

\footnotesize
\centering
\setlength{\tabcolsep}{3pt} % Default value: 6pt
\renewcommand{\arraystretch}{1} % Default value: 1
    \begin{longtable}{ccccccc}
    \caption{Products observed for reactions on Iridium foils. Iridium has two stable isotopes: \ce{^{191}Ir} (37.3\%) and \ce{^{193}Ir} (62.7\%). Nuclear data from: \cite{Kondev2018, Johnson2017, SINGH2003, Vanin2007, Baglin2012, ShamsuzzohaBasunia2017a, Singh2006}.   } 
    \label{tab:Products_Ir} \\ \hline
        \thead{Nuclide\\ level (keV)} & \thead{Half-life} & \thead{Decay \\ mode} & \thead{Reaction\\  route} & \thead{Q value \\ (keV)} & \thead{$E_{\gamma}$ (keV)} & \thead{${I_\gamma}$ (\%)}  \\
        \hline
        
        \makecell[t]{\ce{^{188}Ir}\\$\quad$(0.0)} & \makecell[t]{41.5 h} & \makecell[t]{$\epsilon=100\%$} & \makecell[t]{\ce{^{191}Ir}(d,4np) } & \makecell[t]{-24802.0} & \makecell[t]{1209.80 \\ 1715.67 \\ 2059.65} & \makecell[t]{6.9 $\pm$ 0.7 \\6.2 $\pm$ 0.6 \\ 7.0 $\pm$ 0.6} \\ \hline
        
        \makecell[t]{\ce{^{189}Ir}\\$\quad$(0.0)} & \makecell[t]{13.2 d} & \makecell[t]{$\epsilon=100\%$} & \makecell[t]{\ce{^{191}Ir}(d,3np) \\ \ce{^{193}Ir}(d,5np)} & \makecell[t]{-16626.0 \\ -30596.0} & \makecell[t]{ 233.5 \\ 245.1} & \makecell[t]{0.30 $\pm$ 0.03\\6.0 $\pm$ 0.6} \\ \hline
        
        \makecell[t]{\ce{^{190}Ir}\\$\quad$(0.0)} & \makecell[t]{11.78 d} & \makecell[t]{$\epsilon=100\%$} & \makecell[t]{\ce{^{191}Ir}(d,2np)\\ \ce{^{193}Ir}(d,4np)} & \makecell[t]{-10251.1\\-24221.2} & \makecell[t]{294.75\\380.03 \\1036.05} & \makecell[t]{6.6 $\pm$ 0.7\\ 2.03 $\pm$ 0.11 \\2.42 $\pm$ 0.16} \\ \hline
        
        \makecell[t]{\ce{^{190m2}Ir}\\$\quad$(376.4)} & \makecell[t]{3.087 h} & \makecell[t]{IT$=8.6\%$ \\ $\epsilon=91.4\%$} & \makecell[t]{\ce{^{191}Ir}(d,2np)\\ \ce{^{193}Ir}(d,4np)} & \makecell[t]{-10627.5 \\ -24597.6} & \makecell[t]{361.2 \\ 502.5 } & \makecell[t]{86.72 $\pm$ 0.21\\89.38  $\pm$ 0.2} \\ \hline
        
        \makecell[t]{\ce{^{192}Ir}\\$\quad$(0.0)} & \makecell[t]{73.829 d} & \makecell[t]{$\epsilon=4.76\%$ \\ $\beta^-=95.24\%$} & \makecell[t]{\ce{^{191}Ir}(d,p) \\ \ce{^{193}Ir}(d,2np)} & \makecell[t]{3973.55 \\-9996.6} & \makecell[t]{201.3112  \\ 374.4852 \\ 416.4688 \\ 468.06885 \\ 489.06 \\ 612.46215 \\ 1061.49 } & \makecell[t]{0.471 $\pm$ 0.006 \\ 0.727 $\pm$ 0.007 \\ 0.670 $\pm$ 0.021 \\ 47.84 $\pm$ 0.03 \\0.438 $\pm$ 0.015 \\ 5.34 $\pm$ 0.07 \\ 0.0531 $\pm$ 0.0006} \\ \hline
        
        \makecell[t]{\ce{^{194}Ir}\\$\quad$(0.0)} & \makecell[t]{19.28 h} & \makecell[t]{$\beta^-=100\%$} & \makecell[t]{\ce{^{193}Ir}(d,p)} & \makecell[t]{3842.22} & \makecell[t]{293.541  \\ 938.69  \\ 1468.91} & \makecell[t]{2.5 $\pm$ 0.3\\ 0.60 $\pm$ 0.08 \\  0.19 $\pm$ 0.03} \\ \hline
        
        \makecell[t]{\ce{^{188}Pt}\\$\quad$(0.0)} & \makecell[t]{10.16 d} & \makecell[t]{$\epsilon=99.999974\%$  \\ $\alpha=0.000026\%$} & \makecell[t]{\ce{^{191}Ir}(d,5n) } & \makecell[t]{-26109.0 } & \makecell[t]{195.05 \\ 381.43} & \makecell[t]{18.336 $\pm$ 0.955 \\ 7.3535 $\pm$ 0.382 } \\ \hline
        
        \makecell[t]{\ce{^{189}Pt}\\$\quad$(0.0)} & \makecell[t]{10.87 h} & \makecell[t]{$\epsilon=100\%$} & \makecell[t]{\ce{^{191}Ir}(d,4n) } & \makecell[t]{-19389.0} & \makecell[t]{94.34 \\ 113.82 \\ 243.50 \\ 317.65 \\ 721.38} & \makecell[t]{6.5$\pm$1.8  \\ 2.5$\pm$0.7 \\ 5.9 $\pm$ 1.8 \\ 2.8 $\pm$ 0.7 \\ 7.9 $\pm$ 2.1 } \\ \hline
        
        \makecell[t]{\ce{^{191}Pt}\\$\quad$(0.0)} & \makecell[t]{2.802 d} & \makecell[t]{$\epsilon=100\%$} & \makecell[t]{\ce{^{191}Ir}(d,2n) \\ \ce{^{193}Ir}(d,4n) } & \makecell[t]{-4017.0 \\ -17988.0} & \makecell[t]{178.96 \\351.17 \\ 409.44 \\ 456.47 \\ 538.87 \\ 624.06} & \makecell[t]{0.99  $\pm$ 0.17 \\3.4 $\pm$ 0.6 \\ 7.9 $\pm$ 1.4 \\ 3.3  $\pm$ 0.6  \\ 14.3 $\pm$ 2.4  \\  1.46 $\pm$ 0.25} \\ \hline
        
        \makecell[t]{\ce{^{193m}Pt}\\$\quad$(149.783)} & \makecell[t]{4.33 d} & \makecell[t]{IT$=100\%$} & \makecell[t]{\ce{^{193}Ir}(d,2n)} & \makecell[t]{-3063.5}  & \makecell[t]{66.831 \\ 135.5 } & \makecell[t]{7.21  $\pm$ 0.15\\ 0.1145475 $\pm$ 0.03} \\ \hline
    \end{longtable}

\newpage

\footnotesize
\centering
\setlength{\tabcolsep}{3pt} % Default value: 6pt
\renewcommand{\arraystretch}{1} % Default value: 1
    \begin{longtable}{ccccccc}
    \caption{Products observed for reactions on Iron foils. Iron has four stable isotopes: $^{54}$Fe (5.845\%), $^{56}$Fe (91.754\%), $^{57}$Fe (2.119\%) and  $^{58}$Fe (0.282\%). Nuclear data from: \cite{Burrows2006, Wang2017, Dong2015, Dong2014, Junde2011, Junde2009, Basunia2018, Junde2008, Bhat1998, Nesaraja2010}.  } 
    \label{tab:Products_Fe} \\         \hline
        \thead{Nuclide\\ level (keV)} & \thead{Half-life} & \thead{Decay \\ mode} & \thead{Reaction\\  route} & \thead{Q value \\ (keV)} & \thead{$E_{\gamma}$ (keV)} & \thead{${I_\gamma}$ (\%)}  \\
        \hline
        \makecell[t]{$^{48}$V \\$\quad$(0.0)} & \makecell[t]{15.9735 d} & \makecell[t]{$\epsilon=100\%$} & \makecell[t]{$^{54}$Fe(d,2$\alpha$) \\ $^{56}$Fe(d,2n2$\alpha$) \\ $^{57}$Fe(d,3n2$\alpha$)} & \makecell[t]{-3490.9 \\ -23986.1 \\ -31632.2} & \makecell[t]{944.130 \\ 983.525 \\ 1312.106} & \makecell[t]{7.870 $\pm$ 0.007 \\ 99.98 $\pm$ 0.04 \\98.2 $\pm$ 0.3} \\ \hline
        
        \makecell[t]{$^{51}$Cr\\$\quad$(0.0) } & \makecell[t]{27.704 d} & \makecell[t]{$\epsilon=100\%$} & \makecell[t]{$^{54}$Fe(d,p$\alpha$) \\ $^{56}$Fe(d,2np$\alpha$) \\ $^{57}$Fe(d,3np$\alpha$) \\ $^{58}$Fe(d,4np$\alpha$) } & \makecell[t]{-1381.3 \\ -21876.5 \\ -29522.6 \\ -39567.2 } & \makecell[t]{320.0824} & \makecell[t]{9.910 $\pm$ 0.01} \\ \hline
        
        \makecell[t]{$^{52}$Mn\\$\quad$(0.0)} & \makecell[t]{5.591 d} & \makecell[t]{$\epsilon=100\%$} & \makecell[t]{$^{54}$Fe(d,$\alpha$) \\ $^{54}$Fe(d,2n2p)  \\ $^{56}$Fe(d,2n$\alpha$) \\ $^{57}$Fe(d,3n$\alpha$)} & \makecell[t]{5163.6 \\ -23132.1 \\ -15331.6 \\ -22977.7 } & \makecell[t]{346.02 \\ 744.233 \\ 848.18 \\ 935.544 \\ 1246.278 \\ 1333.649 \\ 1434.092} & \makecell[t]{0.980 $\pm$0.014  \\ 90.0$\pm$1.2  \\ 3.32 $\pm$ 0.4 \\ 94.5 $\pm$ 1.3 \\ 4.21 $\pm$ 0.07 \\ 5.07 $\pm$ 0.07 \\ 100.0 $\pm$ 14.0 } \\ \hline        
        
        \makecell[t]{$^{54}$Mn\\$\quad$(0.0)} & \makecell[t]{312.20 d} & \makecell[t]{$\epsilon=100\%$} & \makecell[t]{$^{54}$Fe(d,2p)  \\ $^{56}$Fe(d,$\alpha$) \\ $^{57}$Fe(d,n$\alpha$) \\ $^{58}$Fe(d,2n$\alpha$)} & \makecell[t]{-2139.1 \\ 5661.4 \\ -1984.7 \\ -12029.3} & \makecell[t]{834.8480} & \makecell[t]{99.9760 $\pm$ 0.001} \\ \hline
        
        \makecell[t]{$^{56}$Mn\\$\quad$(0.0)} & \makecell[t]{2.5789 h} & \makecell[t]{$\beta^-=100\%$} & \makecell[t]{$^{56}$Fe(d,2p) \\ $^{57}$Fe(d,n2p) \\ $^{58}$Fe(d,$\alpha$)} & \makecell[t]{-5137.8 \\ -12783.8 \\ 5467.2} & \makecell[t]{846.7638 \\ 1810.726 \\ 2113.092} & \makecell[t]{98.86 $\pm$ 0.03 \\ 26.9 $\pm$ 0.4 \\ 14.2 $\pm$ 0.3} \\ \hline
        
        \makecell[t]{$^{53}$Fe\\$\quad$(0.0)} & \makecell[t]{8.51 min} & \makecell[t]{$\epsilon=100\%$} & \makecell[t]{$^{54}$Fe(d,2np)  \\$^{56}$Fe(d,4np) } & \makecell[t]{-15602.9 \\ -36098.1} & \makecell[t]{377.9} & \makecell[t]{42.0 $\pm$ 3.0} \\ \hline
        
        \makecell[t]{$^{59}$Fe\\$\quad$(0.0)} & \makecell[t]{44.490 d} & \makecell[t]{$\beta^-=100\%$} & \makecell[t]{$^{58}$Fe(d,p)} & \makecell[t]{4356.44} & \makecell[t]{1099.245 \\ 1291.590} & \makecell[t]{56.5 $\pm$ 0.9\\43.2 $\pm$ 0.9} \\ \hline
        
        \makecell[t]{$^{55}$Co\\$\quad$(0.0)} & \makecell[t]{17.53 h} & \makecell[t]{$\epsilon=100\%$} & \makecell[t]{$^{54}$Fe(d,n) \\ $^{56}$Fe(d,3n) \\ $^{57}$Fe(d,4n)} & \makecell[t]{2839.8 \\ -17655.4 \\ -25301.5} & \makecell[t]{91.9 \\ 477.2 \\ 803.7 \\827.0\\ 931.1 \\ 1316.6 \\ 1370.0 \\ 1408.5  \\ 2177.6 \\ 2872.4 \\ 2938.9 } & \makecell[t]{1.16 $\pm$ 0.09 \\ 20.2 $\pm$ 1.7 \\ 1.87 $\pm$ 0.15 \\ 0.21 $\pm$ 0.06 \\ 75 $\pm$ 4.0 \\ 7.1 $\pm$ 0.3 \\ 2.9 $\pm$ 0.3 \\ 16.9 $\pm$ 0.8 \\ 0.29 $\pm$ 0.04 \\ 0.118 $\pm$ 0.008 \\ 0.057 $\pm$ 0.01 } \\ \hline
            
        \makecell[t]{$^{57}$Co\\$\quad$(0.0)} & \makecell[t]{271.74 d} & \makecell[t]{$\epsilon=100\%$} & \makecell[t]{$^{56}$Fe(d,n) \\ $^{57}$Fe(d,2n) \\ $^{58}$Fe(d,3n)} & \makecell[t]{3802.9 \\ -3843.2 \\ -13887.8 } & \makecell[t]{122.06065 \\ 136.47356} & \makecell[t]{85.60 $\pm$ 0.17 \\ 10.68 $\pm$ 0.08 } \\ \hline
        
        \makecell[t]{$^{58}$Co\\$\quad$(0.0)} & \makecell[t]{70.86 d} & \makecell[t]{$\epsilon=100\%$} & \makecell[t]{$^{57}$Fe(d,n) \\ $^{58}$Fe(d,2n)} & \makecell[t]{4729.7 \\ -5314.9} & \makecell[t]{810.7593} & \makecell[t]{99.450 $\pm$ 0.01} \\ \hline 
    \end{longtable}

\newpage
\footnotesize
\setlength{\tabcolsep}{3pt} % Default value: 6pt
\renewcommand{\arraystretch}{1} % Default value: 1
    \centering
    \begin{longtable}{ccccccc}
    \caption{Products observed for reactions on Nickel foils. Nickel has five stable isotopes: $^{58}$Ni (68.077\%), $^{60}$Ni (26.223\%), $^{61}$Ni (1.1399\%), $^{62}$Ni (3.6346\%) and $^{64}$Ni (0.9255\%). Nuclear data from: \cite{Dong2015, Dong2014, Basunia2018, Junde2008, Junde2011, Bhat1998, Nesaraja2010, Browne2013, Browne2010, Singh2007}.} 
    \label{tab:Products_Ni} \\              \hline
    \small
        \thead{Nuclide\\ level (keV)} & \thead{Half-life} & \thead{Decay \\ mode} & \thead{Reaction\\  route} & \thead{Q value \\ (keV)} & \thead{$E_{\gamma}$ (keV)} & \thead{${I_\gamma}$ (\%)}  \\
        \hline
        \makecell[t]{$^{52}$Mn\\ $\quad$(0.0) } &\makecell[t]{5.591 d} & \makecell[t]{$\epsilon=100\%$ } & \makecell[t]{$^{58}$Ni(d,2$\alpha$) \\ $^{60}$Ni(d,2n2$\alpha$) \\ $^{61}$Ni(d,3n2$\alpha$)}   & \makecell[t]{-1235.6 \\ -21622.6 \\ -29442.7} & \makecell[t]{744.233 \\ 935.544 \\ 1246.278 \\ 1434.092} & \makecell[t]{90.0 $\pm$ 1.2 \\ 94.5 $\pm$ 1.3\\4.21 $\pm$ 0.07 \\100.0 $\pm$ 1.4} \\
        \hline

        \makecell[t]{$^{54}$Mn \\ $\quad$(0.0) } & \makecell[t]{312.20 d} & \makecell[t]{$\epsilon=100\%$} & \makecell[t]{$^{58}$Ni(d,2p$\alpha$) \\ $^{60}$Ni(d,2$\alpha$) \\  $^{61}$Ni(d,n2$\alpha$) \\ $^{62}$Ni(d,2n2$\alpha$)} & \makecell[t]{ -8538.3 \\ -629.6 \\ -8449.7 \\ -19045.4 }& 834.848 & 99.9760 $\pm$ 0.001 \\
        \hline 
        
        \makecell[t]{$^{59}$Fe\\ $\quad$(0.0)} & \makecell[t]{44.490 d} & \makecell[t]{$\beta^-=100\%$} & \makecell[t]{$^{60}$Ni(d,3p) \\ $^{61}$Ni(d,n3p) \\ $^{62}$Ni(d,p$\alpha$) \\ $^{64}$Ni(d,2np$\alpha$) } & \makecell[t]{-12539.5 \\ -20359.6 \\ -2659.7 \\ -19154.9}   & 1291.590 & 43.2 $\pm$ 0.9 \\
        \hline
        
        \makecell[t]{$^{55}$Co\\ $\quad$(0.0)} & \makecell[t]{17.53 h} & \makecell[t]{$\epsilon=100\%$ } & \makecell[t]{$^{58}$Ni(d,n$\alpha$) \\ $^{58}$Ni(d,3n2p) \\ $^{60}$Ni(d,3n$\alpha$) \\ $^{61}$Ni(d,4n$\alpha$) } & \makecell[t]{-3559.4 \\ -31855.0 \\ -23946.4 \\ -31766.5} & \makecell[t]{ 385.4 \\ 520.0 \\ 803.7 \\ 931.1 \\1212.8 \\ 1316.6 \\ 1370.0 \\  2177.6 } & \makecell[t]{0.54 $\pm$ 0.05 \\  0.83 $\pm$ 0.08 \\ 1.87 $\pm$ 0.15 \\ 75 $\pm$ 4.0 \\ 0.26 $\pm$ 0.03 \\ 7.1 $\pm$ 0.3 \\ 2.9 $\pm$ 0.3 \\ 0.29 $\pm$ 0.04 } \\
        \hline
        
        \makecell[t]{$^{57}$Co \\$\quad$(0.0)} & \makecell[t]{271.74 d} & \makecell[t]{$\epsilon$=100\%} & \makecell[t]{$^{58}$Ni(d,n2p) \\ $^{60}$Ni(d,n$\alpha$) \\ $^{60}$Ni(d,3n2p) \\ $^{61}$Ni(d,2n$\alpha$) \\ $^{62}$Ni(d,3n$\alpha$) \\}  & \makecell[t]{-10396.7 \\ -2488.1 \\ -30783.8	 \\ -10308.2 \\ -20903.9} & \makecell[t]{122.06065 \\ 136.47365} & \makecell[t]{85.60 $\pm$ 0.17\\ 10.68 $\pm$ 0.08} \\         \hline

        \makecell[t]{$^{58m}$Co \\ $\quad$(24.88921) } & \makecell[t]{9.10 h} & \makecell[t]{IT$=100\%$} & \makecell[t]{$^{58}$Ni(d,2n) \\ $^{60}$Ni(d,$\alpha$) \\ $^{61}$Ni(d,n$\alpha$) \\ $^{62}$Ni(d,2n$\alpha$) \\ $^{64}$Ni(d,4n$\alpha$)}  & \makecell[t]{-1848.7 \\ 6060.0 \\-1760.2 \\-12355.9\\ -28851.1}   & \makecell[t]{- } & \makecell[t]{-}\\
        
        \hline
        
        \makecell[t]{$^{60}$Co \\ $\quad$(0.0)} & \makecell[t]{1925.28 d} & \makecell[t]{$\beta^-=100\%$} & \makecell[t]{$^{60}$Ni(d,2p)  \\ $^{61}$Ni(d,n2p) \\ $^{62}$Ni(d,$\alpha$) \\ $^{64}$Ni(d,2n$\alpha$)}  & \makecell[t]{-4265.0 \\ -12085.1 \\ 5614.8 \\ -10880.4 } &\makecell[t]{1173.228 \\ 1332.492} & \makecell[t]{99.85 $\pm$ 0.03 \\ 99.9826 $\pm$ 0.0006} \\
        \hline
        
        \makecell[t]{$^{56}$Ni\\$\quad$(0.0)} & \makecell[t]{6.075 d} & \makecell[t]{$\epsilon=100\%$} & \makecell[t]{$^{58}$Ni(d,3np)} & \makecell[t]{-24688.4} & \makecell[t]{158.38 \\ 480.44 \\ 749.95 \\ 811.85 \\ 1561.80} & \makecell[t]{98.8$\pm$1.0  \\ 36.5 $\pm$ 0.8 \\ 49.5 $\pm$ 1.2 \\ 86.0 $\pm$ 0.9 \\ 14.0 $\pm$ 0.6 } \\ 
        \hline

        \makecell[t]{$^{57}$Ni\\ $\quad$(0.0)} & \makecell[t]{35.60 h} & \makecell[t]{$\epsilon=100\%$} & \makecell[t]{$^{58}$Ni(d,2np) \\ $^{60}$Ni(d,4np)} & \makecell[t]{-14440.8 \\ -34827.8} & \makecell[t]{1757.55 \\1919.52 \\ 2804.20 \\ } & \makecell[t]{5.75 $\pm$ 0.2 \\12.3 $\pm$ 0.4 \\ 0.098 $\pm$ 0.004 } \\
        \hline
         
         \makecell[t]{$^{65}$Ni \\ $\quad$(0.0)} & \makecell[t]{2.51719 h} & \makecell[t]{$\beta^-$= 100\% } & \makecell[t]{$^{64}$Ni(d,p)} & \makecell[t]{3873.51} &  \makecell[t]{366.27 \\ 1481.84 \\ 1623.42 \\ 1724.92} & \makecell[t]{4.81 $\pm$ 0.06 \\ 23.59 $\pm$ 0.0014 \\ 0.498 $\pm$ 0.014 \\ 0.399 $\pm$ 0.012 } \\
         \hline
         
         \makecell[t]{$^{60}$Cu\\ $\quad$(0.0)} & \makecell[t]{23.7 min } & \makecell[t]{$\epsilon=100\%$} & \makecell[t]{$^{60}$Ni(d,2n) \\ $^{61}$Ni(d,3n) \\ $^{62}$Ni(d,4n) } & \makecell[t]{-9134.9 \\ -16955.0 \\ -27550.7 } & \makecell[t]{467.3 \\ 497.9 \\ 643.2 \\ 952.4 \\ 1035.2 \\ 1110.5 \\ 1293.7 \\ 1791.6 \\ 1861.6 \\ 1936.9 \\ 2061.0 \\ 2158.9 \\ 2403.3 \\ 2687.9 \\ 2746.1} & \makecell[t]{3.52 $\pm$ 0.18  \\ 1.67 $\pm$ 0.09 \\ 0.97 $\pm$ 0.05 \\ 2.73 $\pm$ 0.18 \\3.70 $\pm$ 0.18 \\ 1.06 $\pm$ 0.18 \\ 1.85 $\pm$ 0.18 \\ 45.4 $\pm$ 2.3 \\ 4.8 $\pm$ 0.3 \\ 2.20 $\pm$ 0.09 \\ 0.79 $\pm$ 0.04 \\ 3.34 $\pm$ 0.18 \\ 0.77 $\pm$ 0.08 \\ 0.44 $\pm$ 0.07 \\ 1.06 $\pm$ 0.09}    \\
         \hline
    
        \makecell[t]{$^{64}$Cu\\ $\quad$(0.0)} & \makecell[t]{12.701 h} & \makecell[t]{$\epsilon=61.5\%$ \\ $\beta^-=38.5\%$} & \makecell[t]{$^{64}$Ni(d,2n)} & \makecell[t]{-4681.3} & \makecell[t]{1345.77} & \makecell[t]{0.475 $\pm$ 0.011}   \\
        \hline 
    \end{longtable}

\newpage

\centering
\footnotesize
\setlength{\tabcolsep}{3pt} % Default value: 6pt
\renewcommand{\arraystretch}{1} % Default value: 1
    \begin{longtable}{ccccccc}
    \caption{Products observed on for reactions on Copper foils. Copper has two stable isotopes: $^{63}$Cu (69.15\%) and $^{65}$Cu (30.85\%). Nuclear data from: \cite{Basunia2018, Browne2013, Browne2010, Singh2015a, Singh2007}. } 
    \label{tab:Products_Cu} \\         \hline
        \thead{Nuclide\\ level (keV)} & \thead{Half-life} & \thead{Decay \\ mode} & \thead{Reaction\\  route} & \thead{Q value \\ (keV)} & \thead{$E_{\gamma}$ (keV)} & \thead{${I_\gamma}$ (\%)}  \\
        \hline
        
        \makecell[t]{$^{59}$Fe\\$\quad$(0.0)} & \makecell[t]{44.490 d} & \makecell[t]{$\beta^-$= 100\%} & \makecell[t]{$^{63}$Cu(d,2p$\alpha$) \\ $^{65}$Cu(d,2$\alpha$)} & \makecell[t]{-8782.1 \\ 1687.0} & \makecell[t]{1099.245 \\ 1291.590} & \makecell[t]{56.5 $\pm$ 1.8 \\43.2 $\pm$ 1.4 } \\ \hline
        
        \makecell[t]{$^{60}$Co\\$\quad$(0.0)} & \makecell[t]{1925.28 d} & \makecell[t]{$\beta^-$=100\%} & \makecell[t]{$^{63}$Cu(d,p$\alpha$) \\ $^{65}$Cu(d,2np$\alpha$)} & \makecell[t]{-507.6 \\ -18334.1} & \makecell[t]{1173.228 \\ 1332.492} & \makecell[t]{99.85 $\pm$ 3.0 \\ 99.9826 $\pm$ 0.06 } \\ \hline
        
        \makecell[t]{$^{61}$Co\\$\quad$(0.0)} & \makecell[t]{1.649 h} & \makecell[t]{$\beta^-$=100\%} & \makecell[t]{$^{63}$Cu(d,n3p) \\ $^{65}$Cu(d,np$\alpha$)} & \makecell[t]{-19484.2 \\ -9015.1 } & \makecell[t]{67.412} & \makecell[t]{84.7} \\ \hline
        
        \makecell[t]{$^{65}$Ni\\$\quad$(0.0)} & \makecell[t]{2.51719 h} & \makecell[t]{$\beta^-$=100\%} & \makecell[t]{$^{65}$Cu(d,2p)} & \makecell[t]{-3580.2} & \makecell[t]{1481.84} & \makecell[t]{23.59 $\pm$ 0.14} \\ \hline
        
        \makecell[t]{$^{61}$Cu \cite{hermanne2025critical} \\$\quad$(0.0)} & \makecell[t]{3.339 h} & \makecell[t]{$\epsilon=100\%$} & \makecell[t]{$^{63}$Cu(d,3np) \\ $^{65}$Cu(d,5np)} & \makecell[t]{-21962.9 \\-39789.4} & \makecell[t]{282.956 \\ 656.008 \\ 1185.234} & \makecell[t]{12.7  $\pm$ 2.0 \\10.07 $\pm$ 1.68 \\3.6 $\pm$ 0.6} \\ \hline
        
        \makecell[t]{$^{64}$Cu\\$\quad$(0.0)} & \makecell[t]{12.701 h} & \makecell[t]{$\epsilon=61.5\%$ \\ $\beta^-= 38.5$} & \makecell[t]{$^{63}$Cu(d,p)\\ $^{65}$Cu(d,2np)} & \makecell[t]{5691.54 \\ -12135.0} & \makecell[t]{1345.77} & \makecell[t]{0.475 $\pm$ 0.011} \\ \hline        
    \end{longtable}
    
\normalsize

\newpage
\section{Excitation functions}\label{excitationfunctions}
\begin{figure}[h!!]
    \begin{adjustwidth}{-1.5cm}{}
  \begin{minipage}[b]{0.6\textwidth}
    \includegraphics[width=\textwidth]{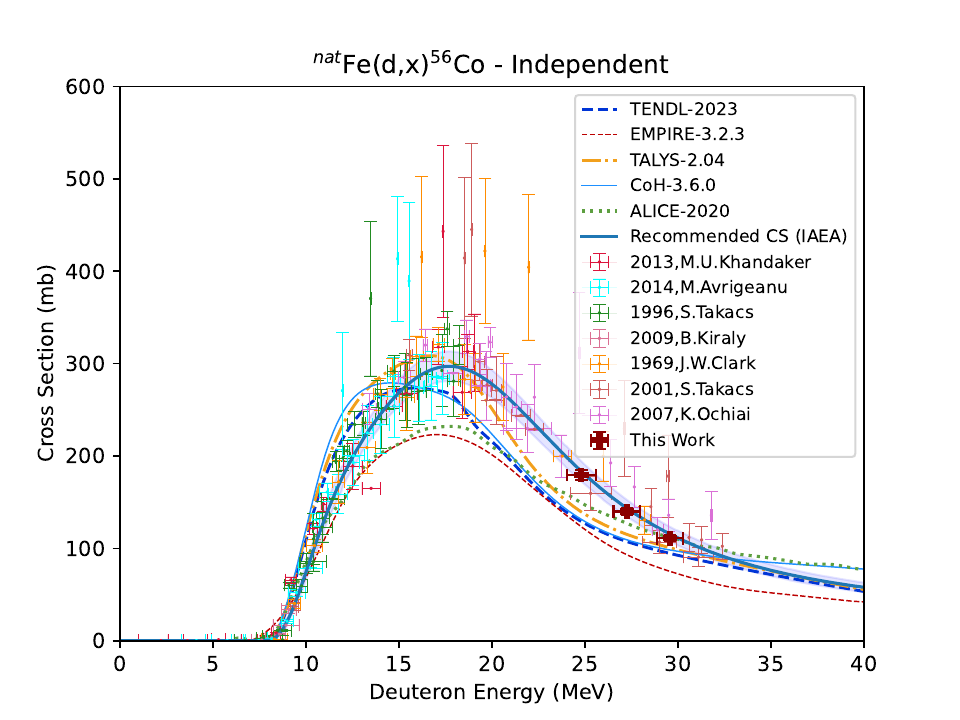}
    % \caption{}
    % \label{fig:Fe_56Co_i}
  \end{minipage} 
  \begin{minipage}[b]{0.6\textwidth}
    \includegraphics[width=\textwidth]{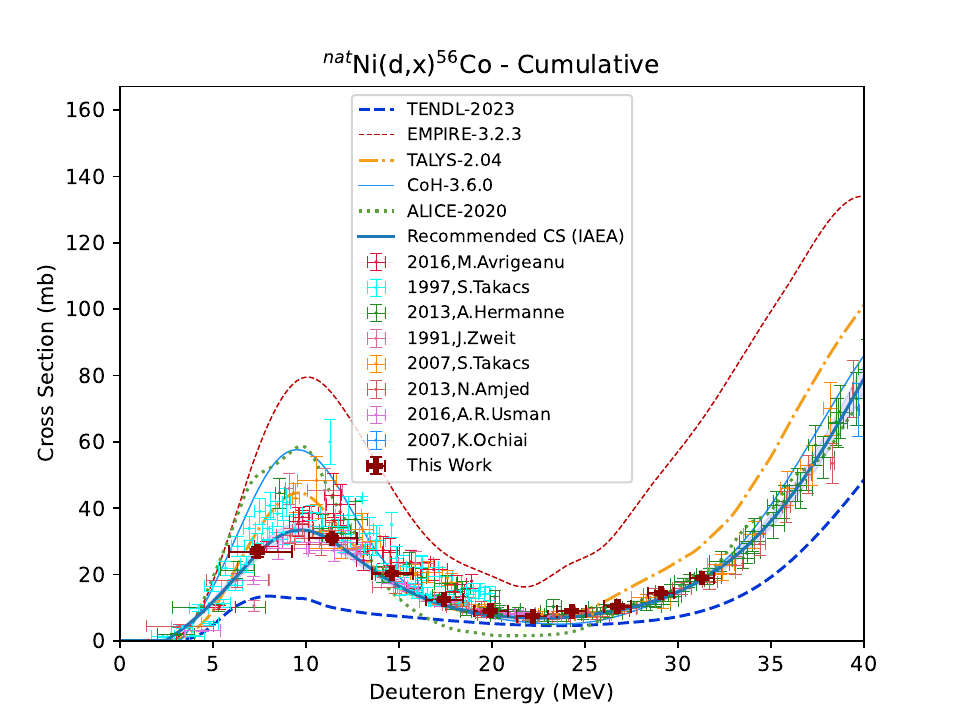}
    % \caption{}
    % \label{fig:Ni_56Co}
  \end{minipage} 
  \begin{minipage}[b]{0.6\textwidth}
    \includegraphics[width=\textwidth]{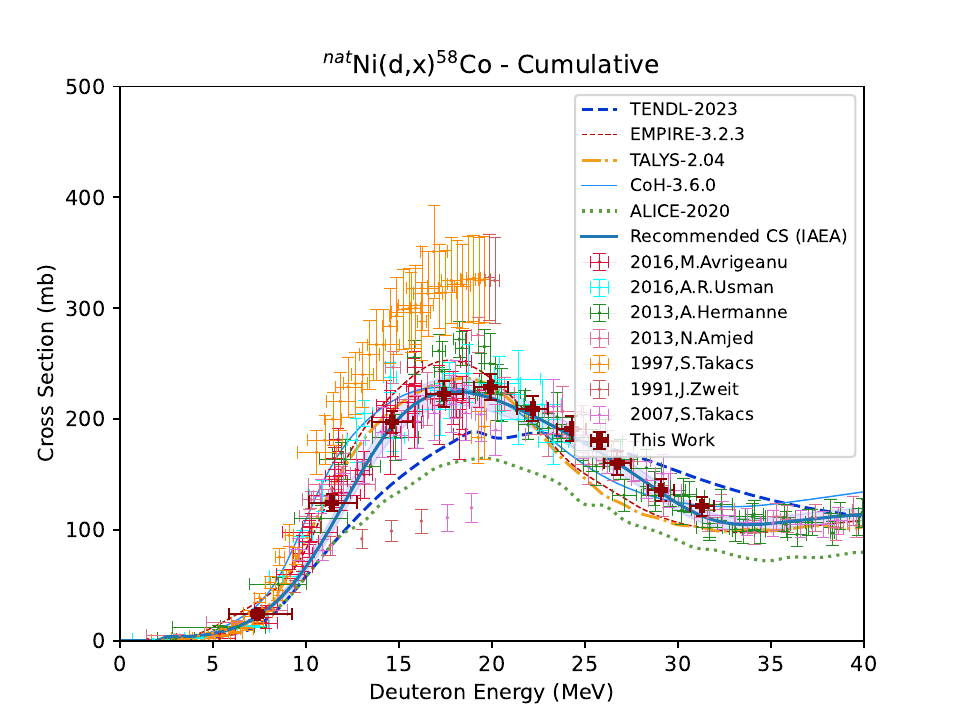}
    % \caption{}
    % \label{fig:Ni_58Co}
  \end{minipage}%\hspace*{-0.2em}
  \hfill
  %a\hskip 0pt plus .35 fill b\hskip 0pt plus .65 fill 
  % \begin{minipage}[b]{0.6\textwidth}
  %   \includegraphics[width=\textwidth]{Ni_61Cu_i.png}
  %   % \caption{}
  %   % \label{fig:Ni_61Cu}
  % \end{minipage}
  % \vfill
  \begin{minipage}[b]{0.6\textwidth}
    \includegraphics[width=\textwidth]{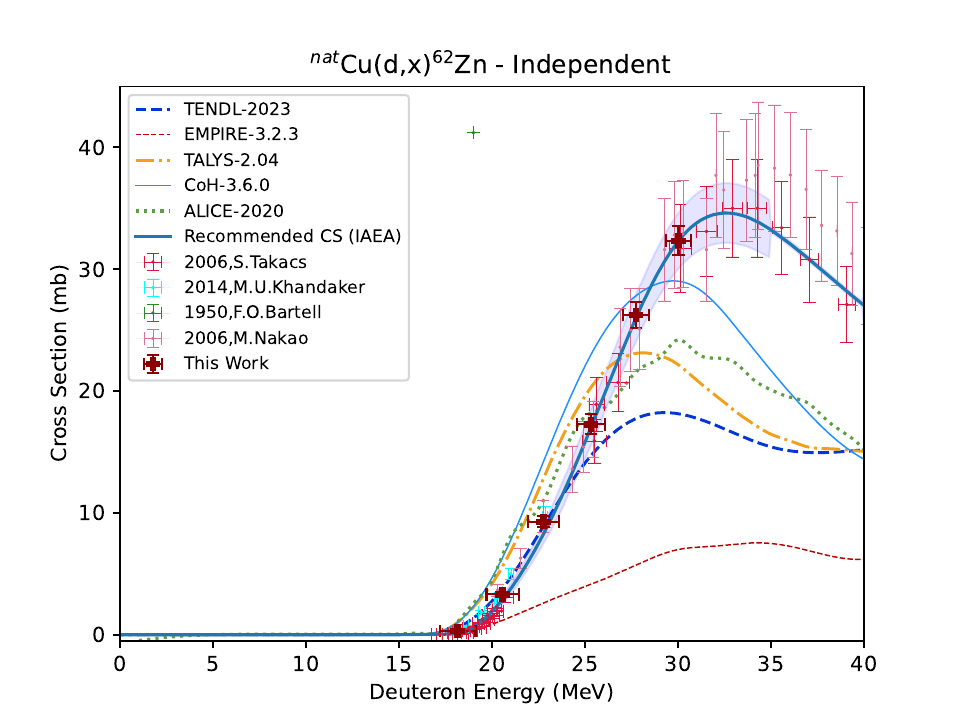}
    % \caption{}
    % \label{fig:Cu_62Zn}
  \end{minipage}  
   \begin{minipage}[b]{0.6\textwidth}
    \includegraphics[width=\textwidth]{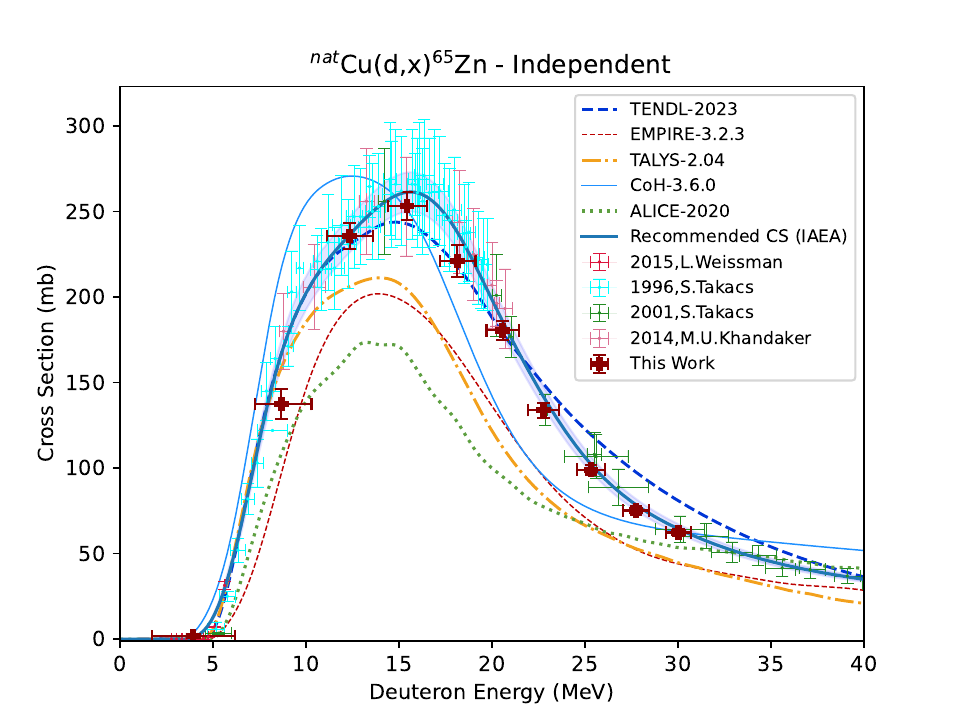}
    % \caption{}
    % \label{fig:Cu_65Zn}
  \end{minipage} 
  \vfill
  \caption{Excitation functions for the monitor reactions (except for \natCu(d,x)\ce{^{63}Zn} and \natNi(d,x)\ce{^{61}Cu}, seen in \autoref{fig:Cu_63Zn_i} and \autoref{fig:Ni_61Cu_i}). }
  \label{fig:MonitorReactions}
 \end{adjustwidth}
\end{figure}

\begin{figure}[h!!]
    \begin{adjustwidth}{-1.5cm}{}
  \begin{minipage}[b]{0.6\textwidth}
    \includegraphics[width=\textwidth]{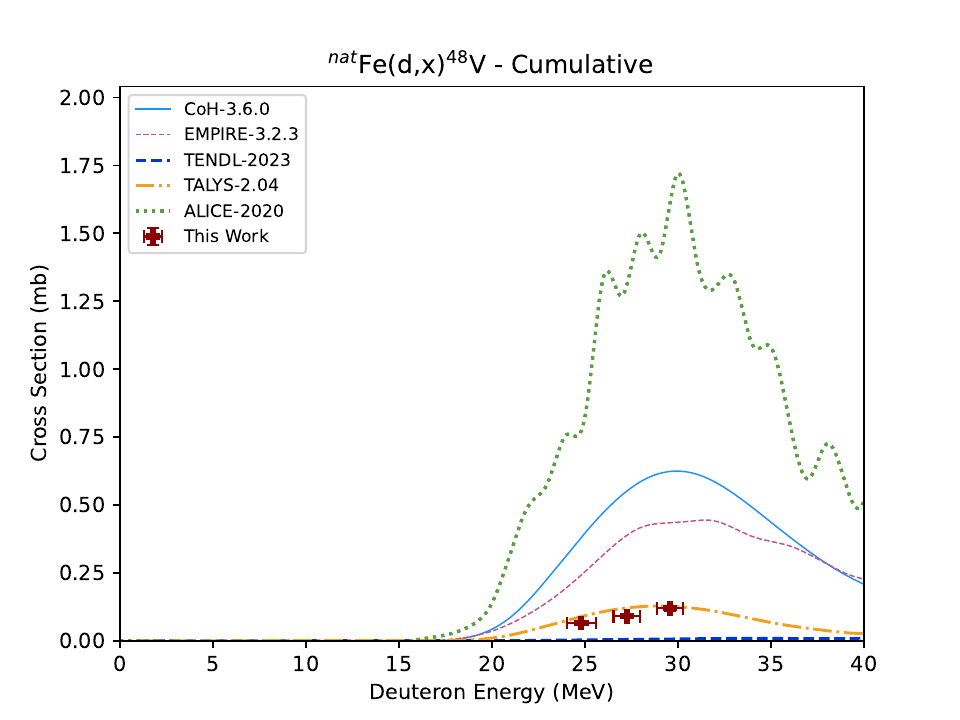}
    % \caption{}
    % \label{fig:Fe_48V_c}
  \end{minipage} 
  \begin{minipage}[b]{0.6\textwidth}
    \includegraphics[width=\textwidth]{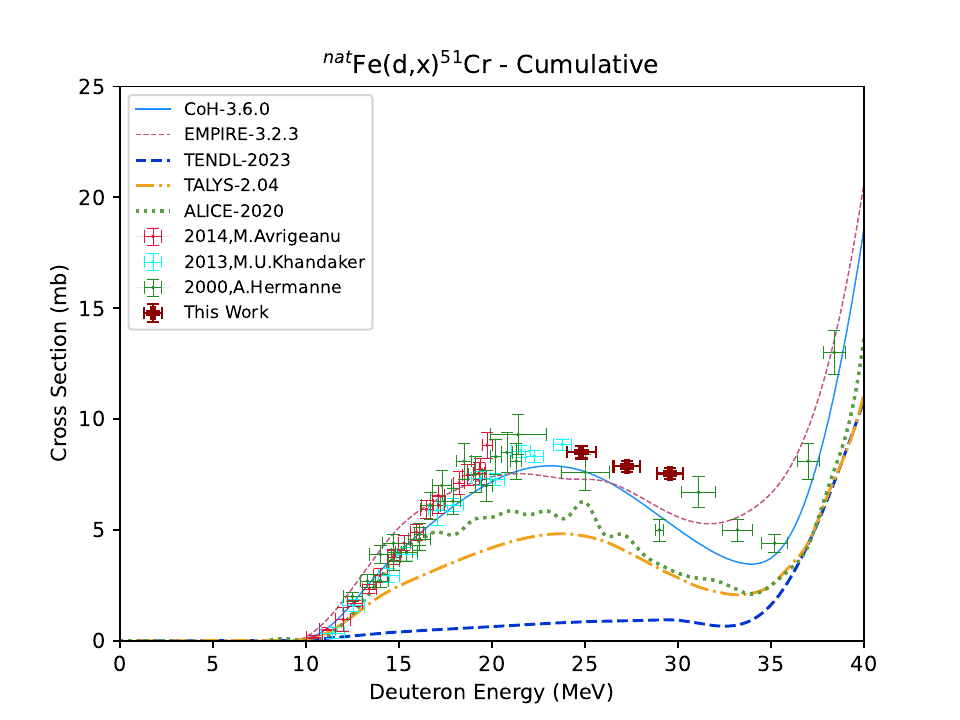}
    % \caption{}
    % \label{fig:Fe_51Cr_c}
  \end{minipage} 
  \begin{minipage}[b]{0.6\textwidth}
    \includegraphics[width=\textwidth]{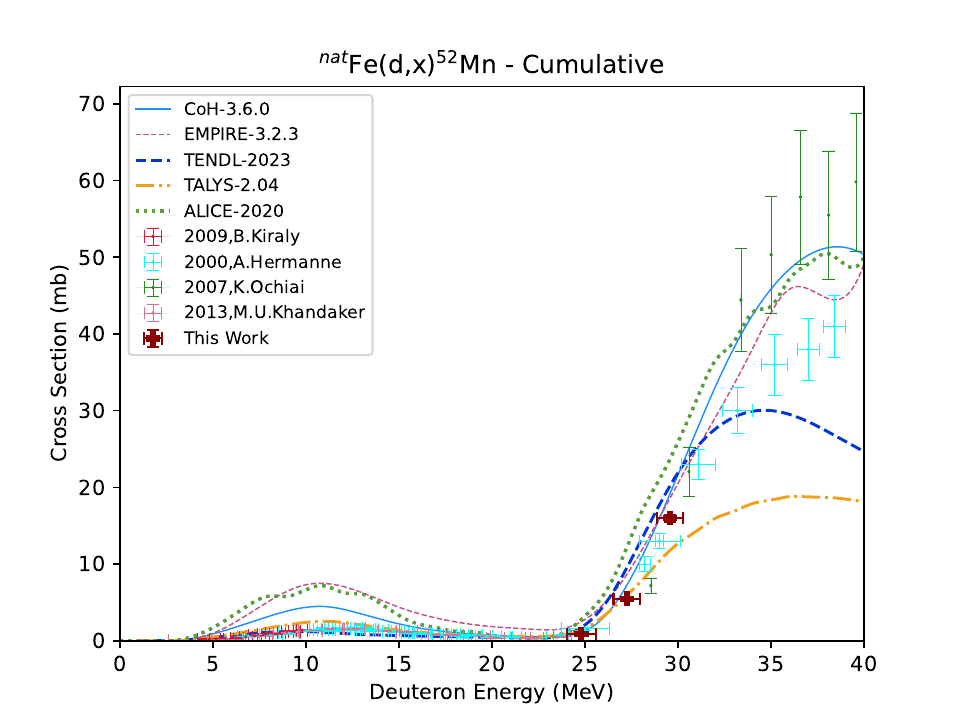}
    % \caption{}
    % \label{fig:Ni_52Mn_c}
  \end{minipage}%\hspace*{-0.2em}
  \hfill
  %a\hskip 0pt plus .35 fill b\hskip 0pt plus .65 fill 
  \begin{minipage}[b]{0.6\textwidth}
    \includegraphics[width=\textwidth]{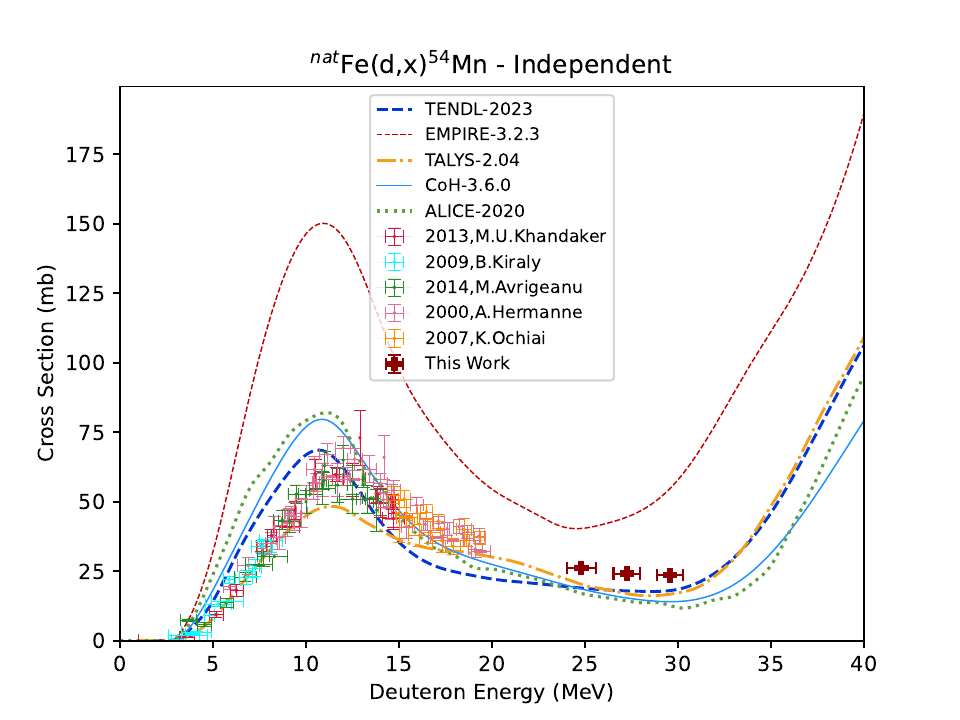}
    % \caption{}
    % \label{fig:Ni_64Mn_c}
  \end{minipage}
  \vfill
  \begin{minipage}[b]{0.6\textwidth}
    \includegraphics[width=\textwidth]{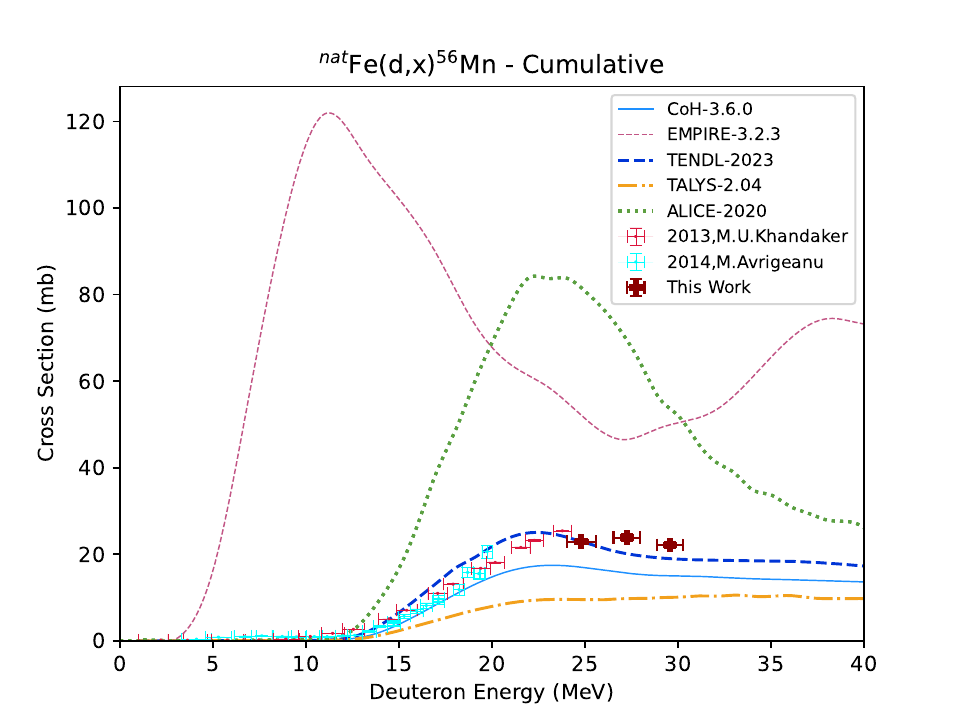}
    % \caption{}
    % \label{fig:Ni_56Mn_c}
  \end{minipage}  
  \caption{Excitation functions for reactions on natural iron. }
  \label{fig:iron1}
 \end{adjustwidth}
\end{figure}

\begin{figure}[h!!]
    \begin{adjustwidth}{-1.5cm}{}
  \begin{minipage}[b]{0.6\textwidth}
    \includegraphics[width=\textwidth]{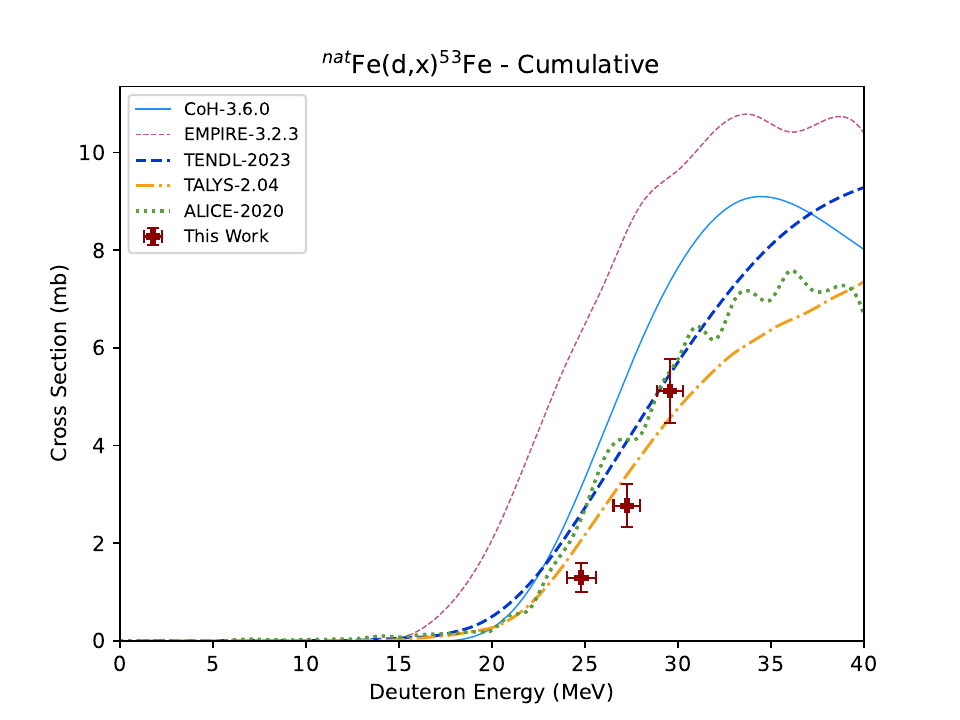}
    % \caption{}
    \label{fig:Fe_53Fe_c}
  \end{minipage} 
  \begin{minipage}[b]{0.6\textwidth}
    \includegraphics[width=\textwidth]{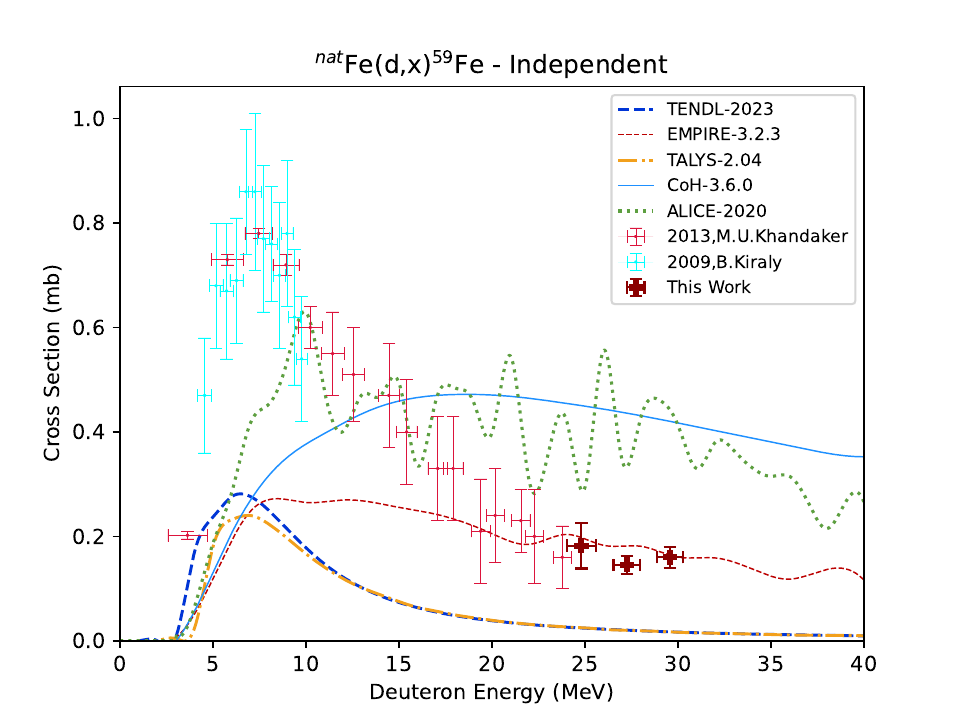}
    % \caption{}
    % \label{fig:Fe_59Fe_i}
  \end{minipage} 
  \begin{minipage}[b]{0.6\textwidth}
    \includegraphics[width=\textwidth]{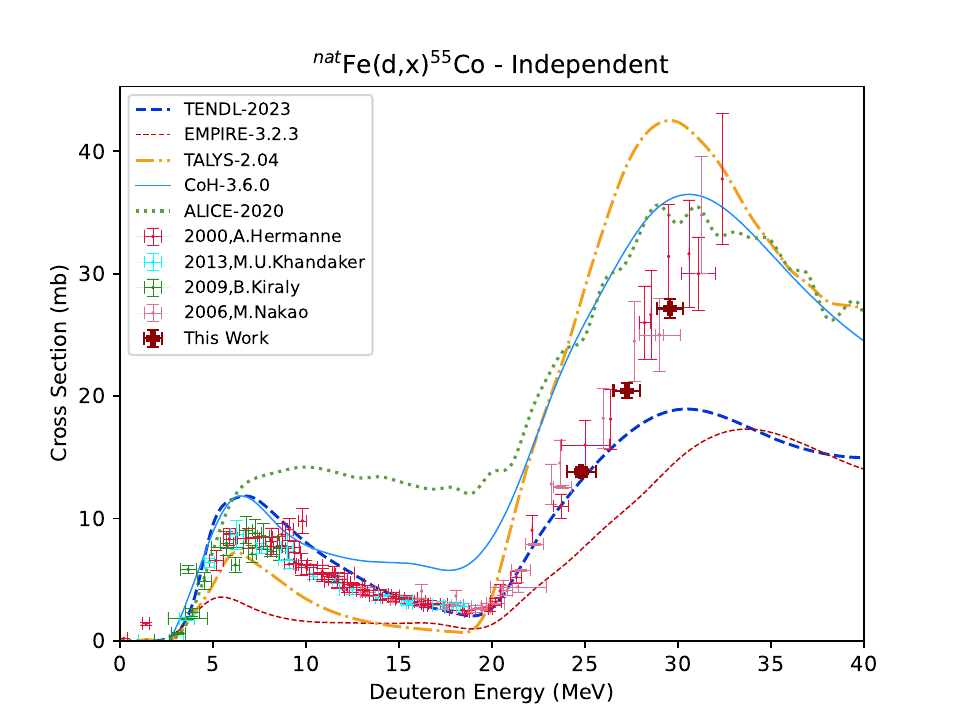}
    % \caption{}
    % \label{fig:Fe_55Co_i}
  \end{minipage}%\hspace*{-0.2em}
  \hfill
  % a\hskip 0pt plus .35 fill b\hskip 0pt plus .65 fill 
  \begin{minipage}[b]{0.6\textwidth}
    \includegraphics[width=\textwidth]{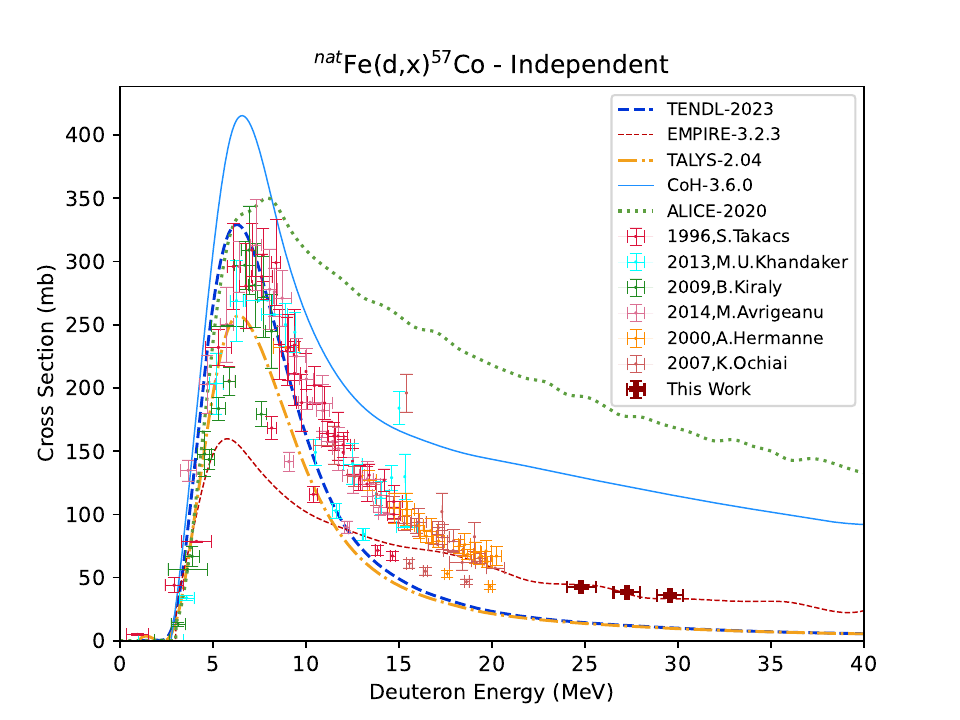}
    % \caption{}
    % \label{fig:Fe_57Co_i}
  \end{minipage}
  % \vfill
  \begin{minipage}[b]{0.6\textwidth}
    \includegraphics[width=\textwidth]{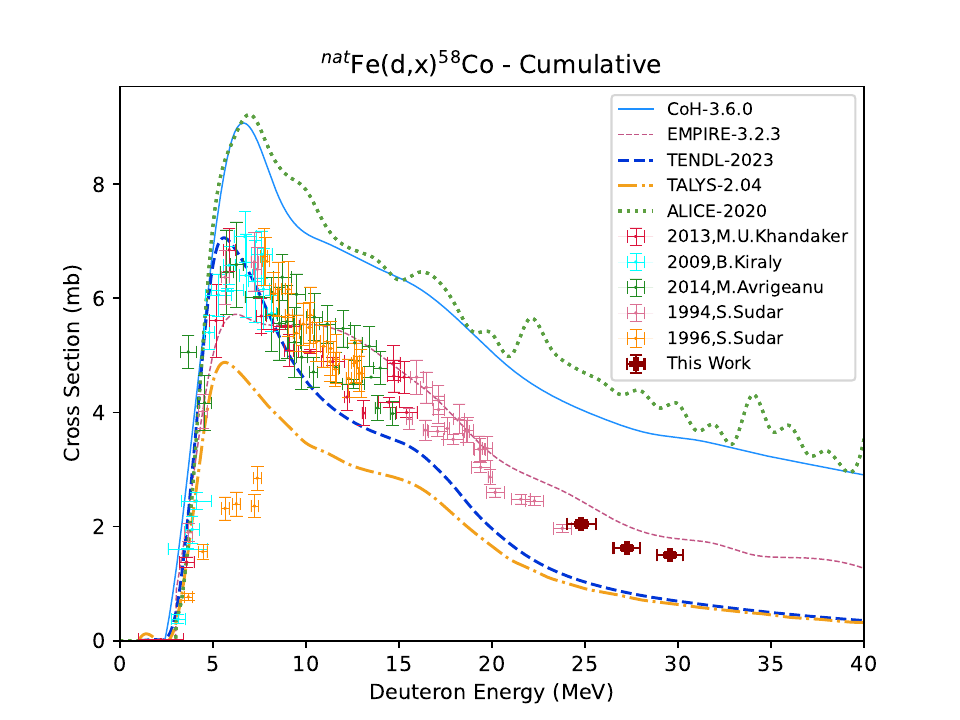}
    % \caption{}
    % \label{fig:Fe_58Co_c}
  \end{minipage}  
  \caption{Excitation functions for reactions on natural iron. }
  \label{fig:iron2}
 \end{adjustwidth}
\end{figure}

\begin{figure}[h!!]
    \begin{adjustwidth}{-1.5cm}{}
  \begin{minipage}[b]{0.6\textwidth}
    \includegraphics[width=\textwidth]{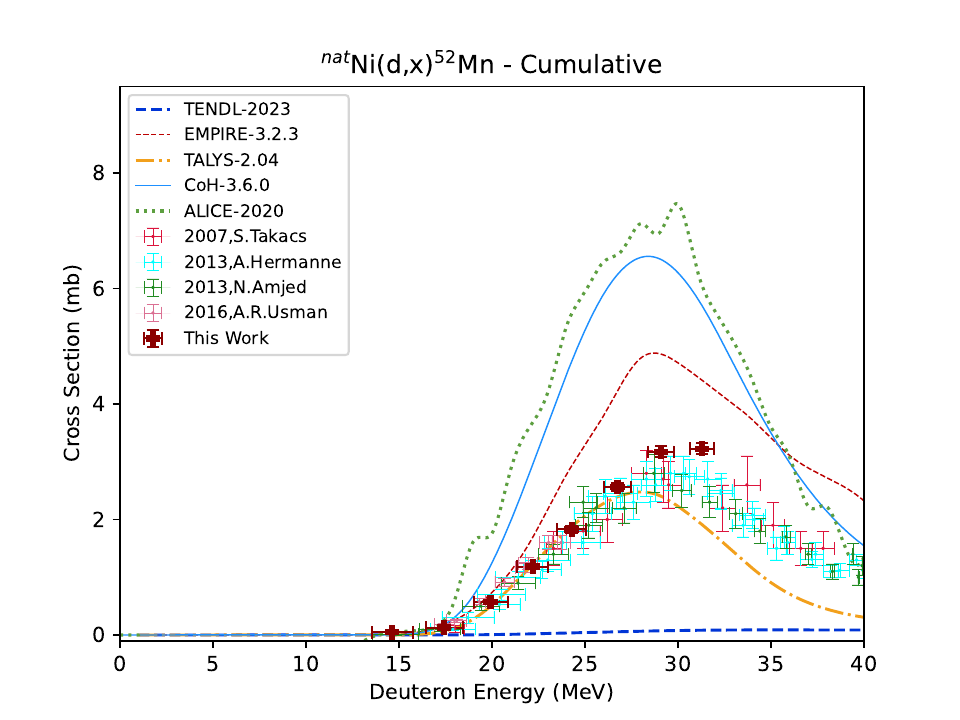}
    % \caption{}
    % \label{fig:Ni_52Mn_c}
  \end{minipage} 
  \begin{minipage}[b]{0.6\textwidth}
    \includegraphics[width=\textwidth]{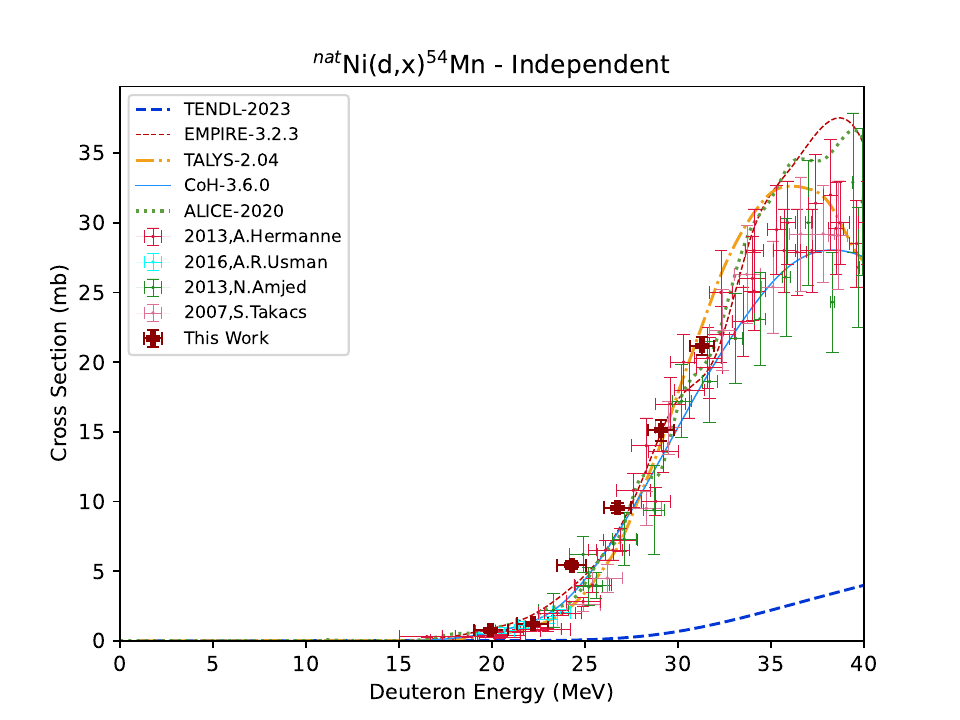}
    % \caption{}
    % \label{fig:Ni_54Mn_i}
  \end{minipage} 
  % \begin{minipage}[b]{0.6\textwidth}
  %   \includegraphics[width=\textwidth]{Ni_56Mn_c.png}
  %   % \caption{}
  %   % \label{fig:Ni_56Mn_c}
  % \end{minipage}%\hspace*{-0.2em}
  % \hfill
  %a\hskip 0pt plus .35 fill b\hskip 0pt plus .65 fill 
  \begin{minipage}[b]{0.6\textwidth}
    \includegraphics[width=\textwidth]{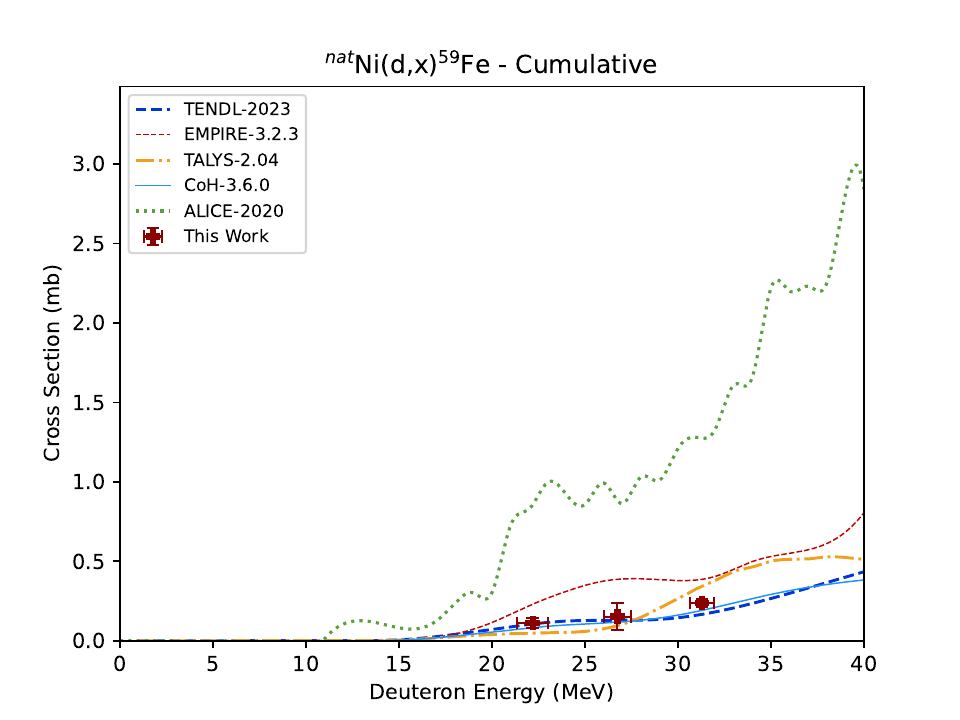}
    % \caption{}
    % \label{fig:Ni_59Fe_c}
  \end{minipage}
  \hfill
  \begin{minipage}[b]{0.6\textwidth}
    \includegraphics[width=\textwidth]{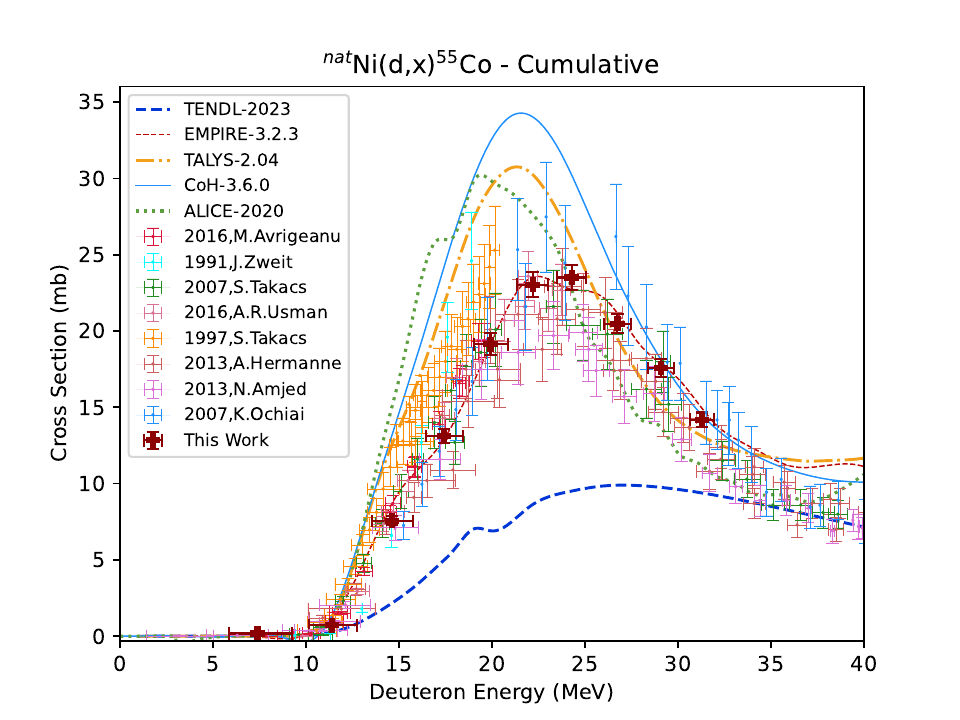}
    % \caption{}
    % \label{fig:Ni_55Co_c}
  \end{minipage}  
  \begin{minipage}[b]{0.6\textwidth}
    \includegraphics[width=\textwidth]{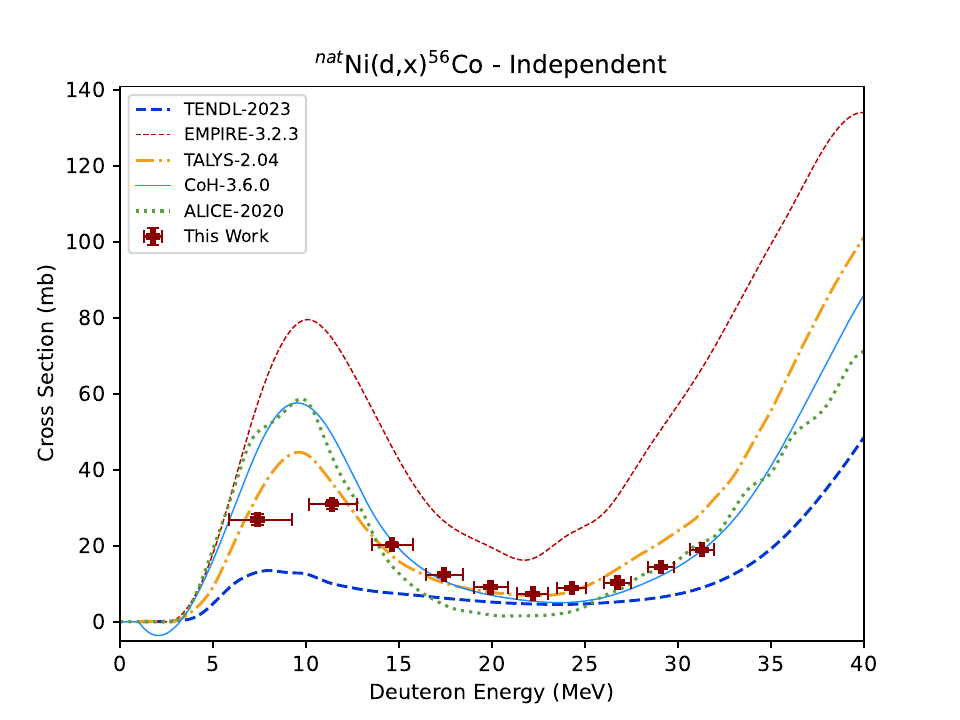}
    % \caption{}
    % \label{fig:Ni_56Co_i}
  \end{minipage}  
  \caption{Excitation functions for reactions on natural nickel. }
  \label{fig:nickel1}
 \end{adjustwidth}
\end{figure}

\begin{figure}[h!!]
    \begin{adjustwidth}{-1.5cm}{}
  \begin{minipage}[b]{0.6\textwidth}
    \includegraphics[width=\textwidth]{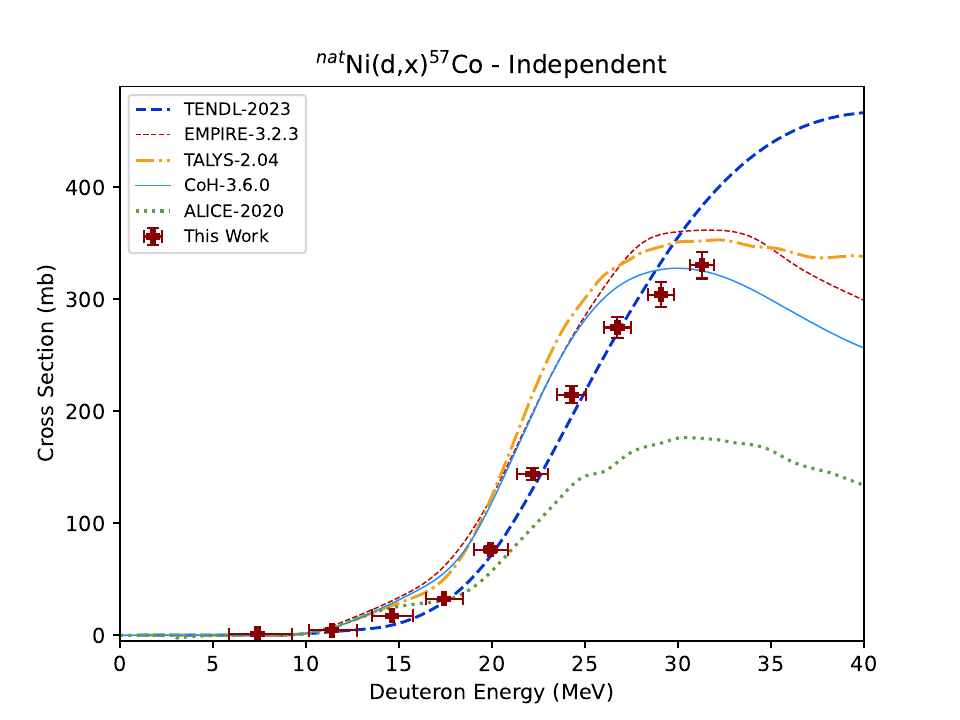}
    % \caption{}
    % \label{fig:Ni_57Co_i}
  \end{minipage} 
  \begin{minipage}[b]{0.6\textwidth}
    \includegraphics[width=\textwidth]{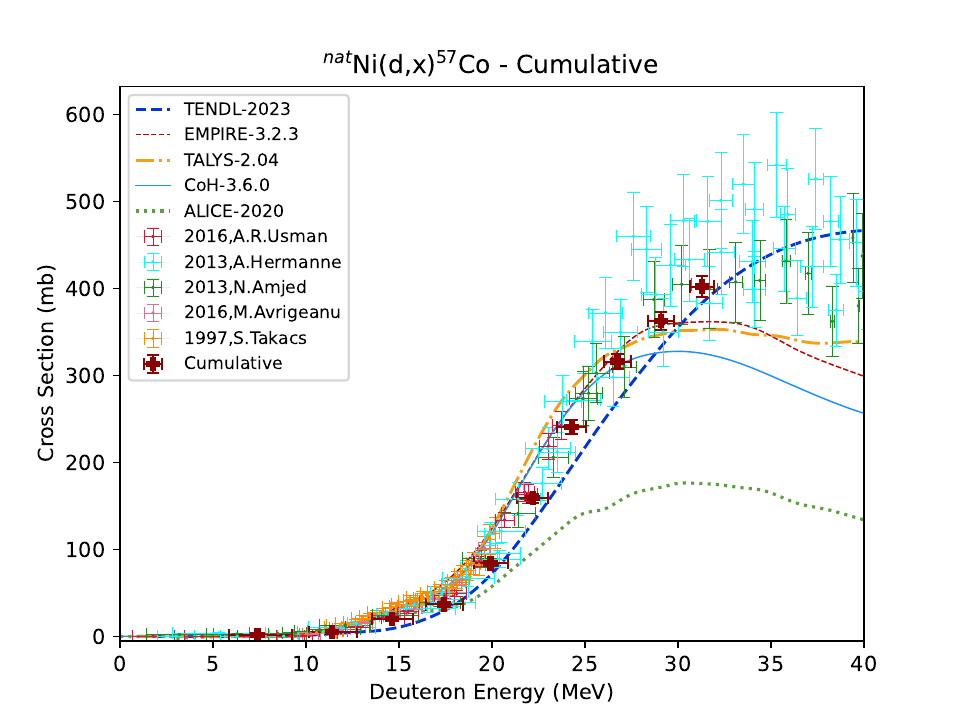}
    % \caption{}
    % \label{fig:Ni_57Co_c}
  \end{minipage} 
  \begin{minipage}[b]{0.6\textwidth}
    \includegraphics[width=\textwidth]{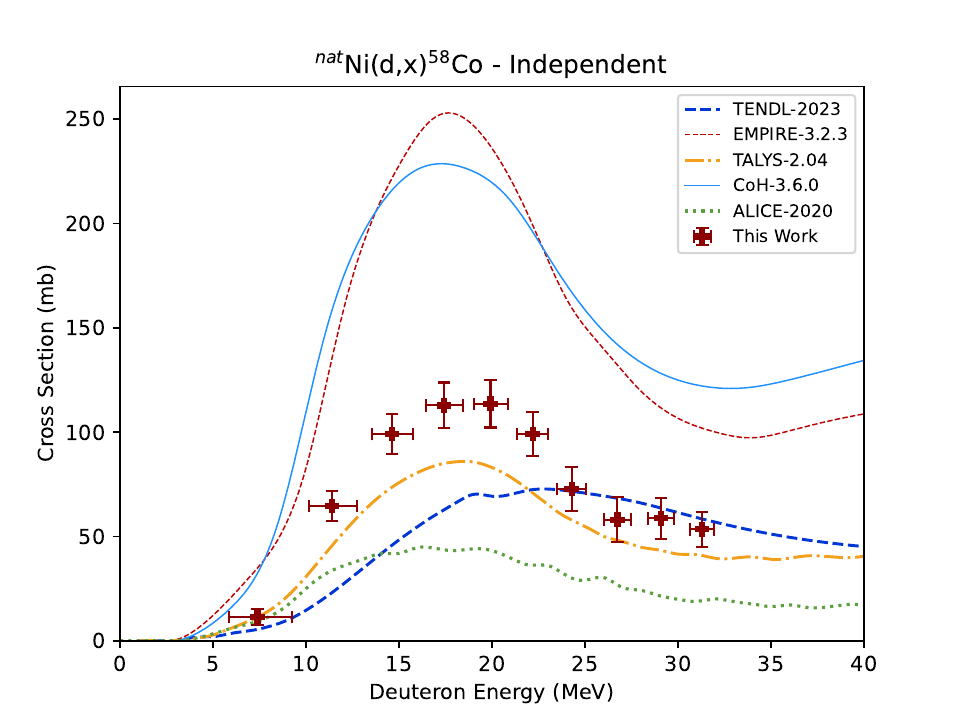}
    % \caption{}
    % \label{fig:Ni_58Co_i}
  \end{minipage}%\hspace*{-0.2em}
  \hfill
  %a\hskip 0pt plus .35 fill b\hskip 0pt plus .65 fill 
  \begin{minipage}[b]{0.6\textwidth}
    \includegraphics[width=\textwidth]{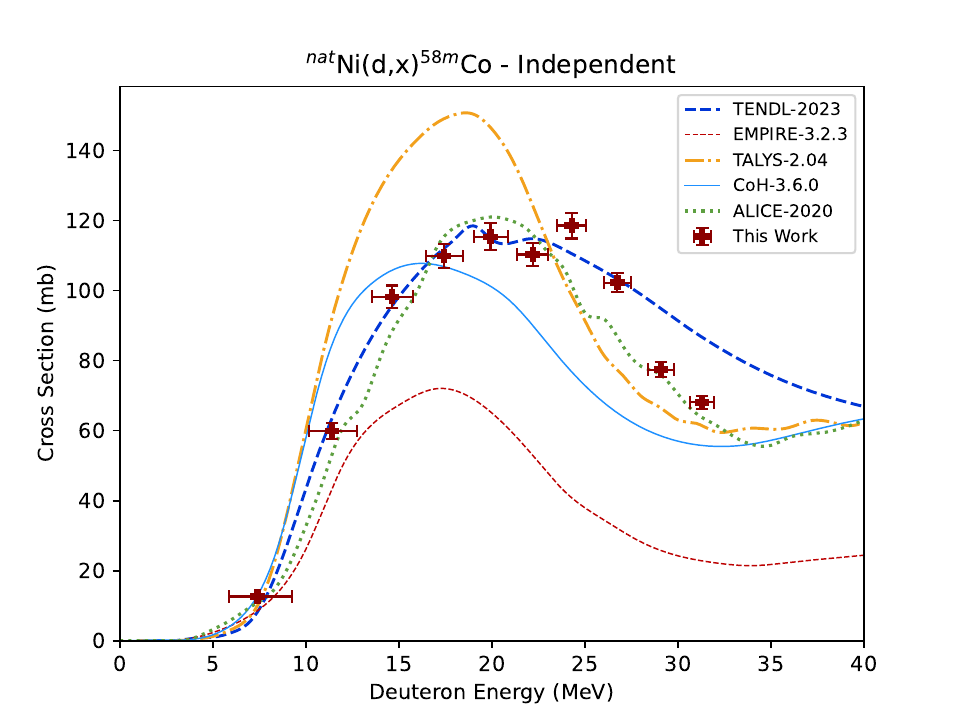}
    % \caption{}
    % \label{fig:Ni_58mCo_i}
  \end{minipage}
  \vfill
  \begin{minipage}[b]{0.6\textwidth}
    \includegraphics[width=\textwidth]{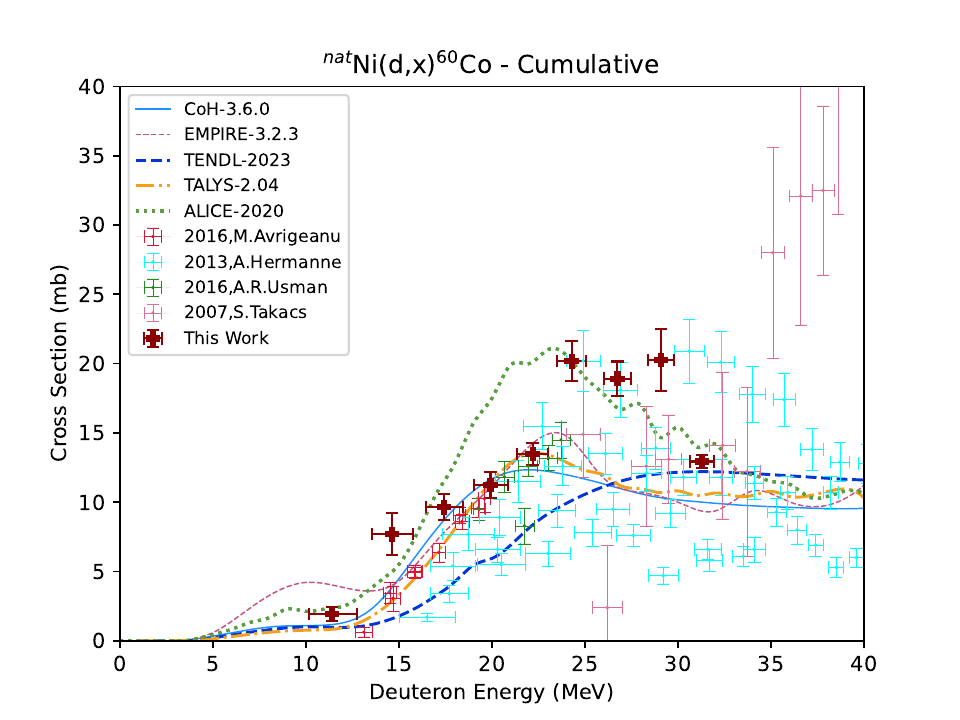}
    % \caption{}
    % \label{fig:Ni_60Co_c}
  \end{minipage}  
  \caption{Excitation functions for reactions on natural nickel. }
  \label{fig:nickel2}
 \end{adjustwidth}
\end{figure}

\begin{figure}[h!!]
    \begin{adjustwidth}{-1.5cm}{}
  \begin{minipage}[b]{0.6\textwidth}
    \includegraphics[width=\textwidth]{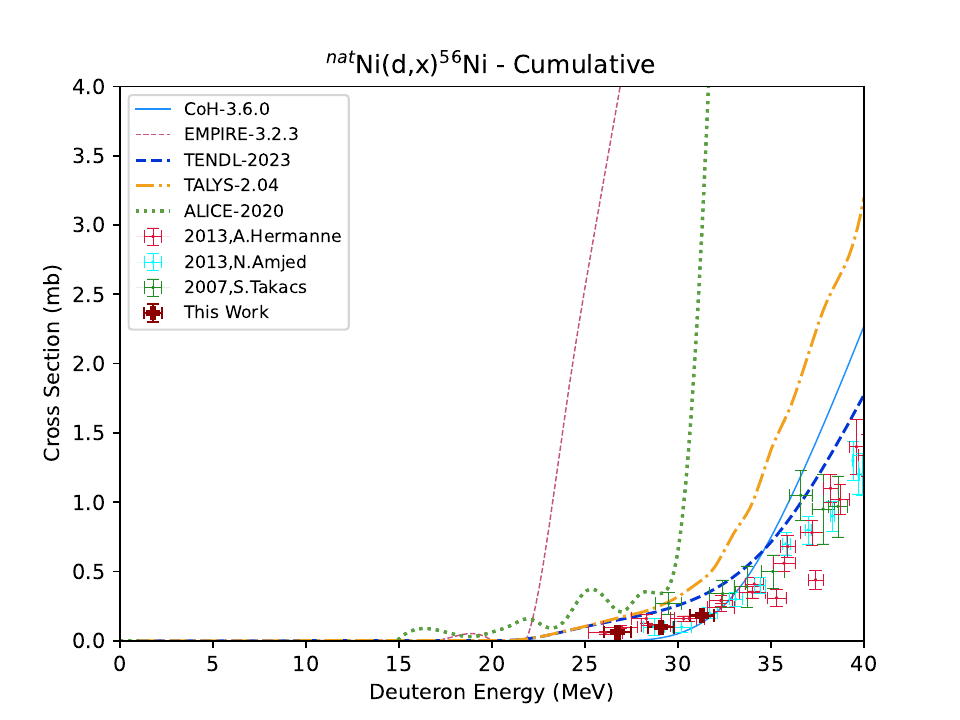}
    % \caption{}
    % \label{fig:Ni_56Ni_c}
  \end{minipage} 
  \begin{minipage}[b]{0.6\textwidth}
    \includegraphics[width=\textwidth]{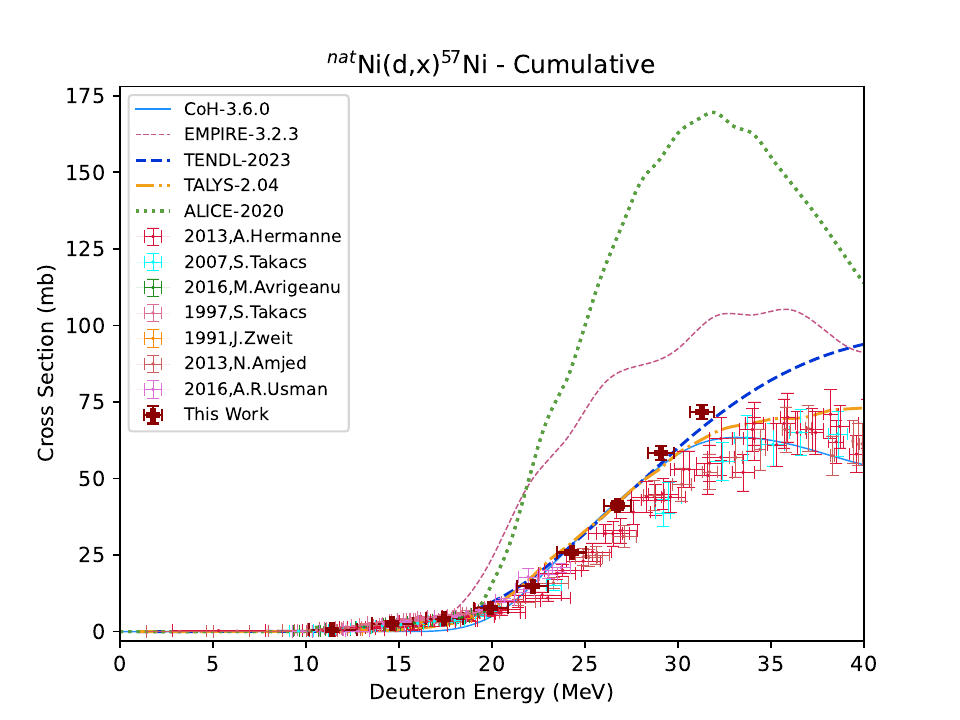}
    % \caption{}
    % \label{fig:Ni_57Ni_c}
  \end{minipage} 
  \begin{minipage}[b]{0.6\textwidth}
    \includegraphics[width=\textwidth]{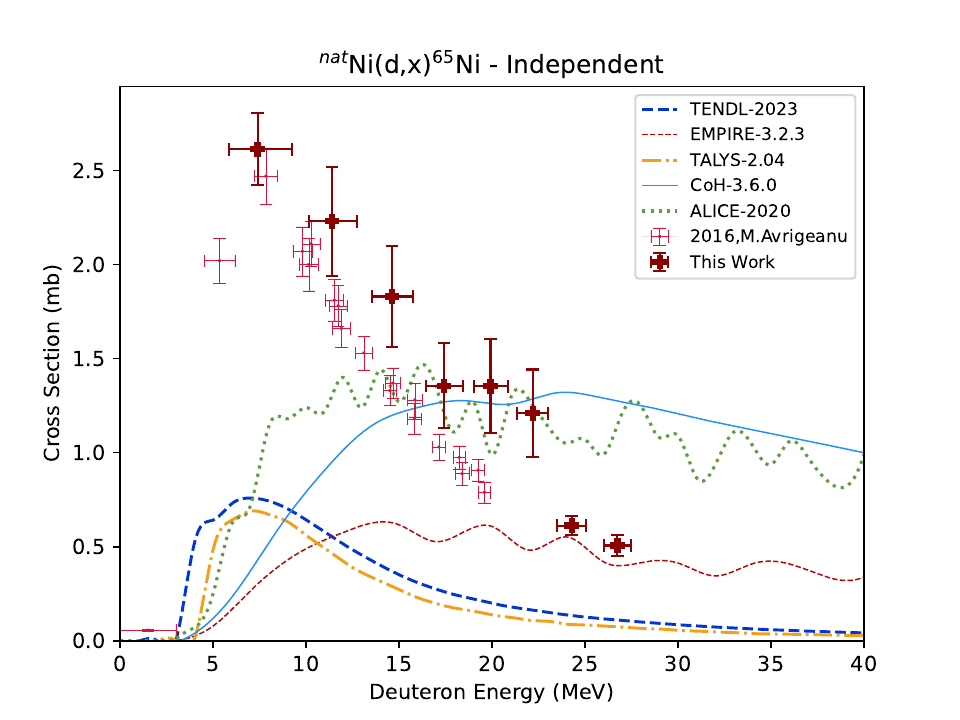}
    % \caption{}
    % \label{fig:Ni_65Ni_i}
  \end{minipage}%\hspace*{-0.2em}
  \hfill
  %a\hskip 0pt plus .35 fill b\hskip 0pt plus .65 fill 
  \begin{minipage}[b]{0.6\textwidth}
    \includegraphics[width=\textwidth]{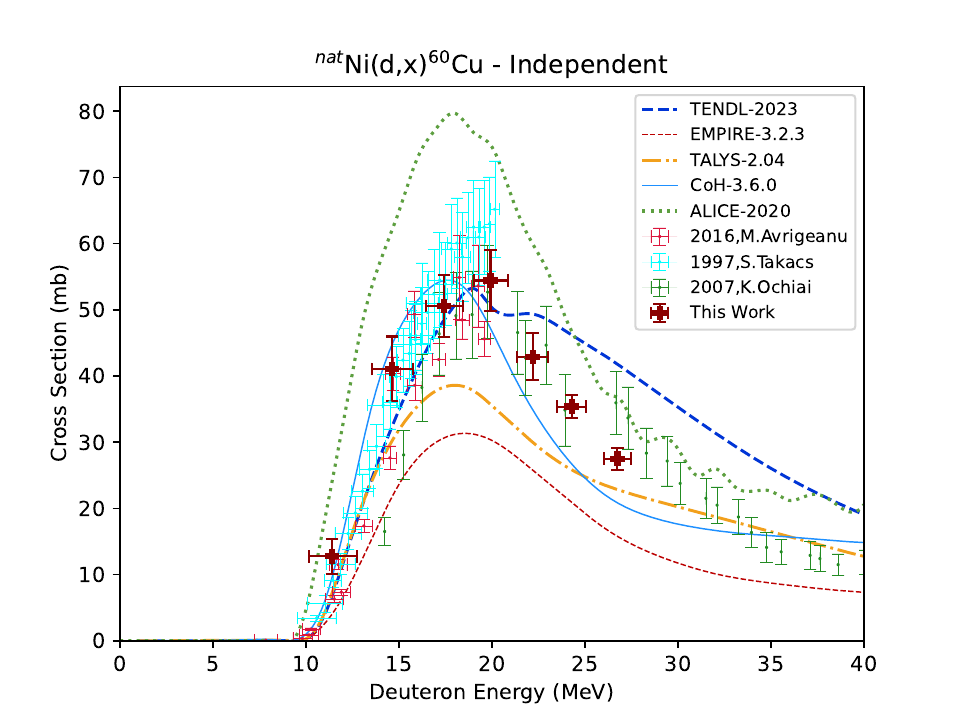}
    % \caption{}
    % \label{fig:Ni_60Cu_i}
  \end{minipage}
  \vfill
  \begin{minipage}[b]{0.6\textwidth}
    \includegraphics[width=\textwidth]{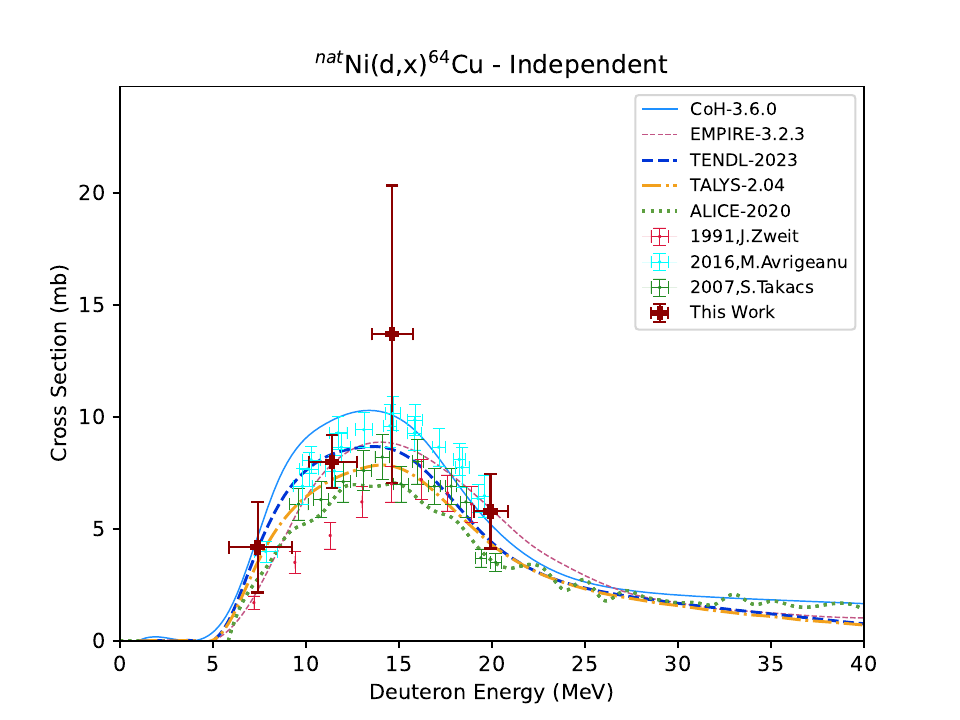}
    % \caption{}
    % \label{fig:Ni_64Cu_i}
  \end{minipage}  
  \caption{Excitation functions for reactions on natural nickel. }
  \label{fig:nickel3}
 \end{adjustwidth}
\end{figure}

\begin{figure}[h!!]
    \begin{adjustwidth}{-1.5cm}{}
  \begin{minipage}[b]{0.6\textwidth}
    \includegraphics[width=\textwidth]{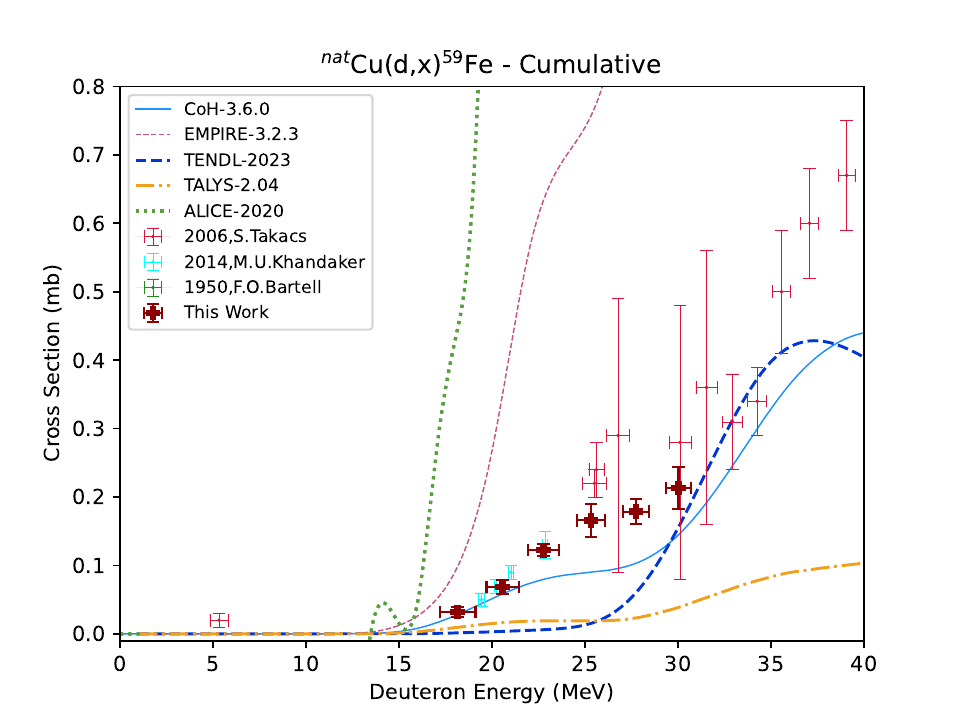}
    % \caption{}
    % \label{fig:Cu_60Co_c}
  \end{minipage} 
  \begin{minipage}[b]{0.6\textwidth}
    \includegraphics[width=\textwidth]{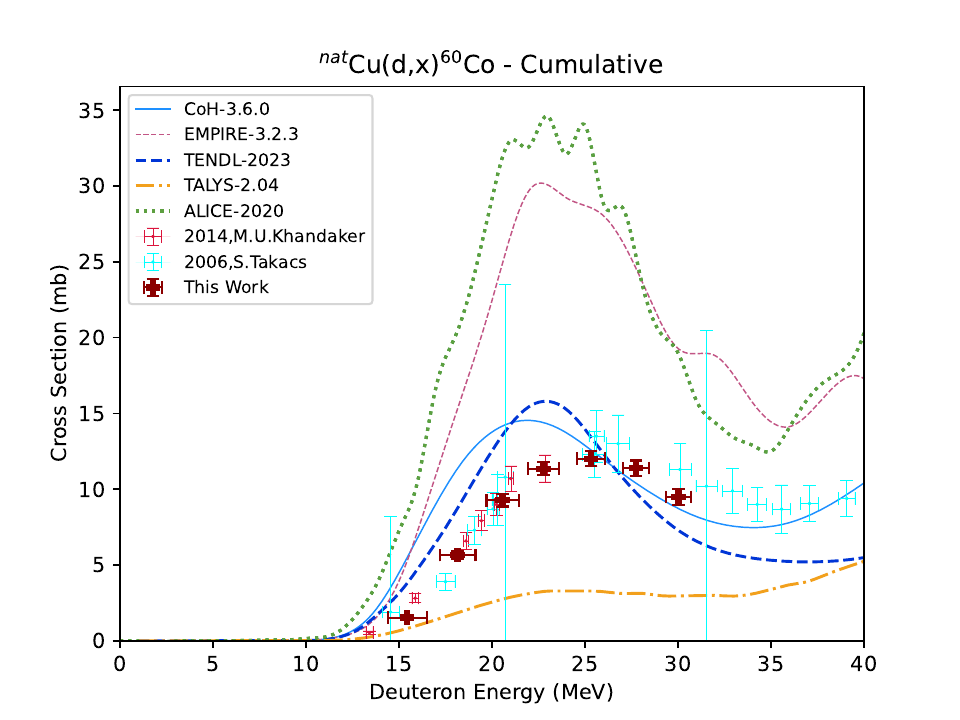}
    % \caption{}
    % \label{fig:Cu_60Co_c}
  \end{minipage} 
  \begin{minipage}[b]{0.6\textwidth}
    \includegraphics[width=\textwidth]{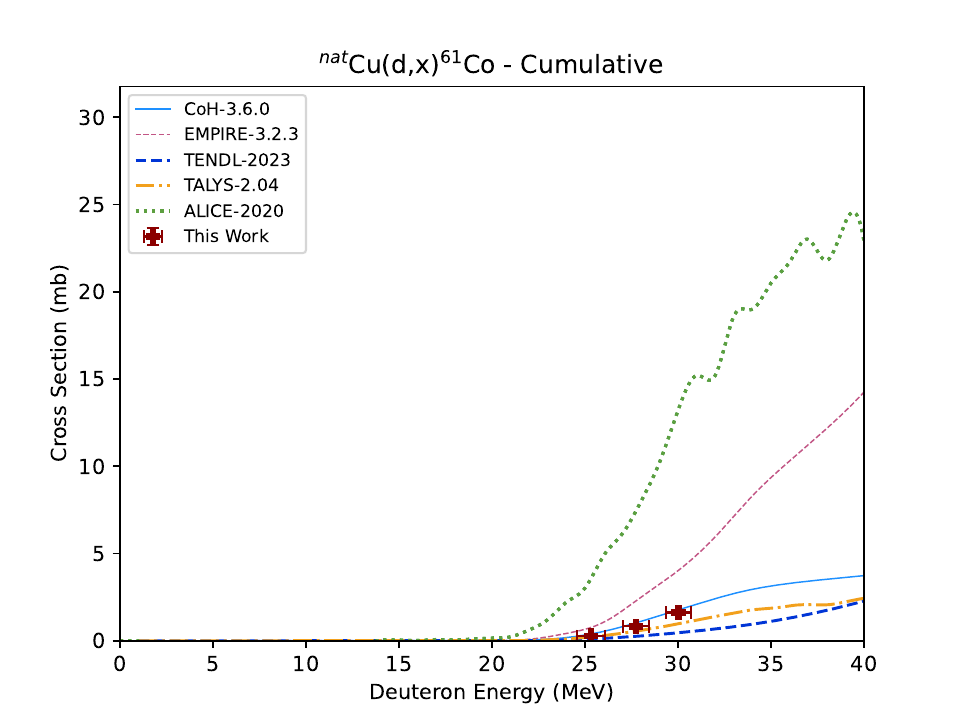}
    % \caption{}
    % \label{fig:Cu_60Co_c}
  \end{minipage} 
  \begin{minipage}[b]{0.6\textwidth}
    \includegraphics[width=\textwidth]{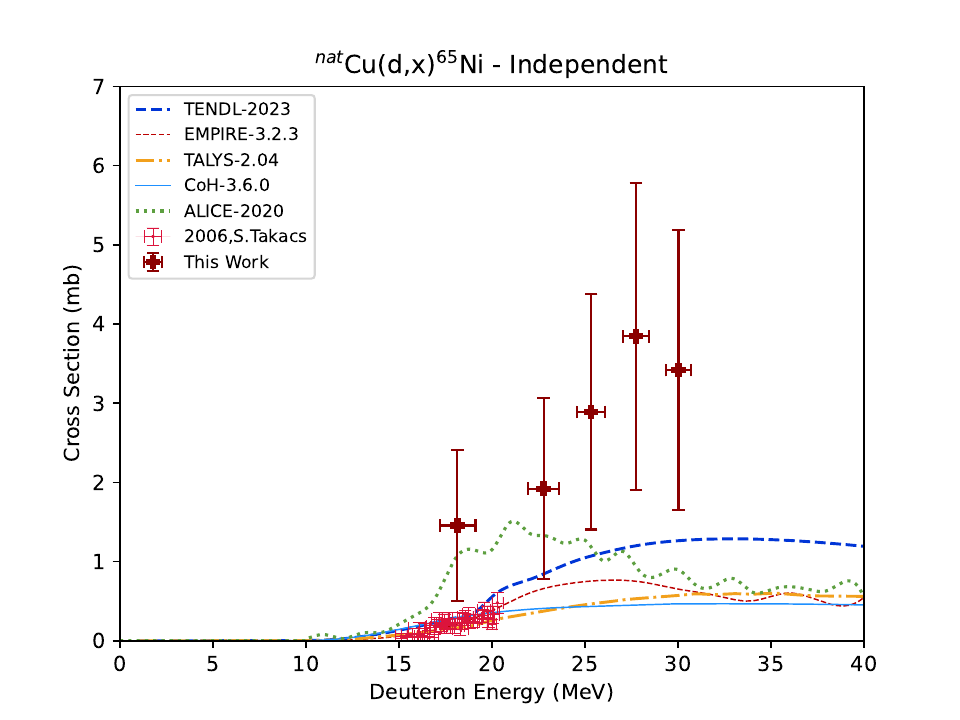}
    % \caption{}
    % \label{fig:Cu_65Ni_i}
  \end{minipage} 
  \begin{minipage}[b]{0.6\textwidth}
    \includegraphics[width=\textwidth]{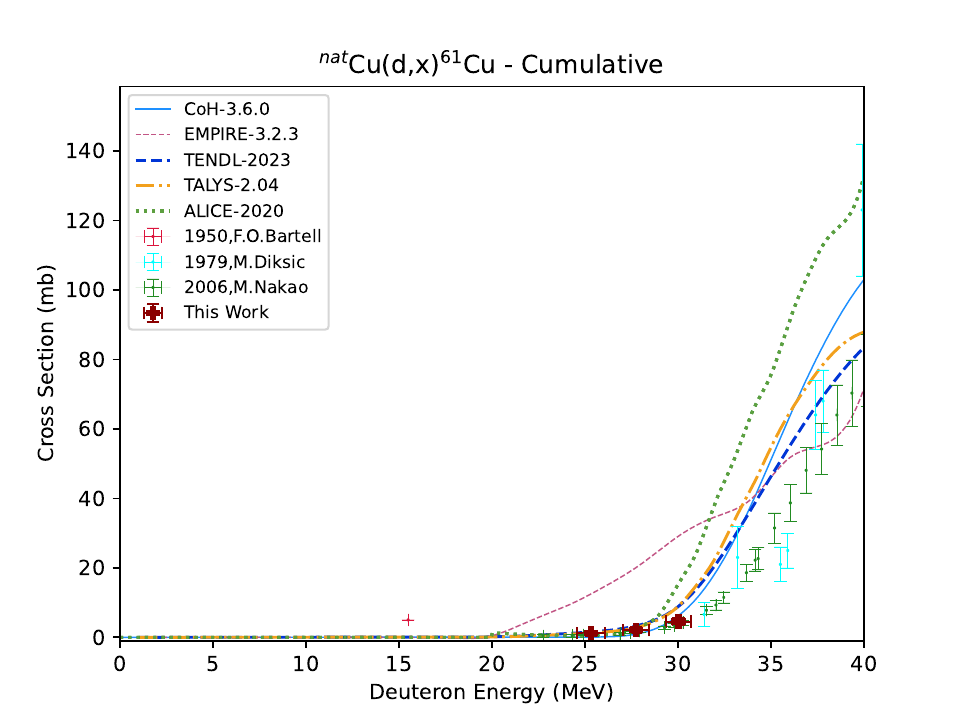}
    % \caption{}
    % \label{fig:Cu_61Cu_c}
  \end{minipage}%\hspace*{-0.2em}
  \hfill
  %a\hskip 0pt plus .35 fill b\hskip 0pt plus .65 fill 
  \begin{minipage}[b]{0.6\textwidth}
    \includegraphics[width=\textwidth]{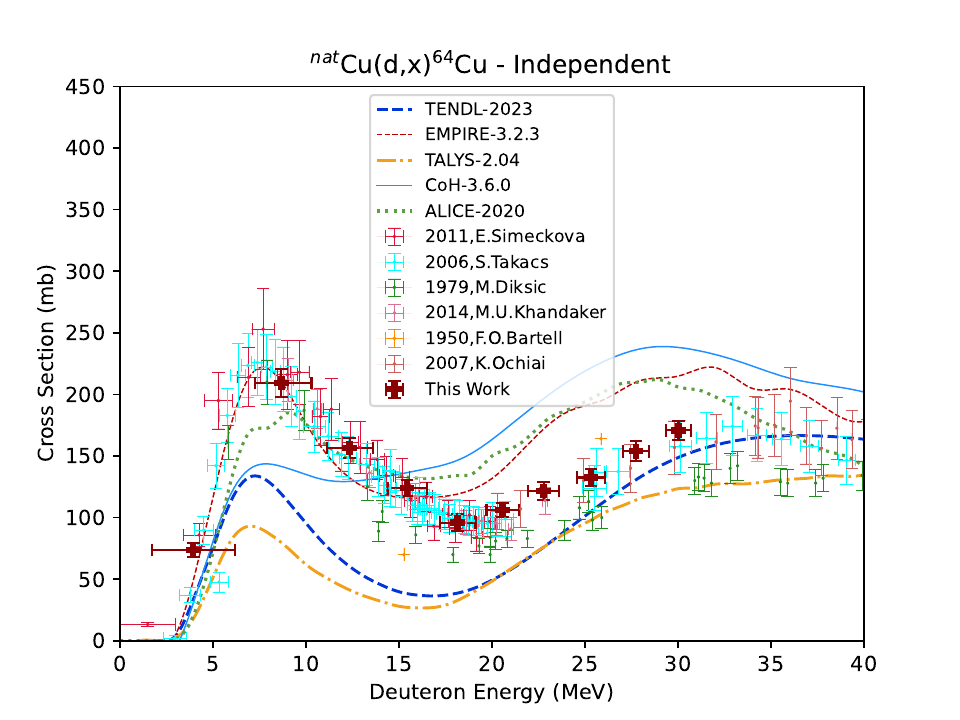}
    % \caption{}
    % \label{fig:Cu_64Cu_i}
  \end{minipage}
  \caption{Excitation functions for reactions on natural copper. }
  \label{fig:cupper1}
 \end{adjustwidth}
\end{figure}

Excitation functions for the monitor reactions \natFe(d,x)\ce{^{56}Co}, \natNi(d,x)\ce{^{56,58}Co},\ce{^{61}Cu} and \natCu(d,x)\ce{^{62,65}Zn} are presented in \autoref{fig:MonitorReactions}, and \ce{^{63}Zn} is presented in \autoref{fig:Cu_63Zn_i}. Excitation functions for reactions induced in iron are presented in \autoref{fig:iron1} and \autoref{fig:iron2}. Excitation functions for reactions induced in nickel are presented in \autoref{fig:nickel1}, \autoref{fig:nickel2}, and \autoref{fig:nickel3}. Excitation functions for reactions induced in copper are presented in \autoref{fig:cupper1}. All excitation functions are compared to experimental data retrieved from the EXFOR database \cite{Khandaker2013, Takacs1996, Avrigeanu2014, Kiraly2009, Clark1969, Takacs2001, Ochiai2007, Avrigeanu2016, Takacs1997, Hermanne2013, Zweit1991, Takacs2007, Amjed2013, Usman2016a, Ochiai2007, Avrigeanu2016, Carzaniga2019, Herman2007, Takacs2006a, Khandaker2014, Bartell1950, Nakao2006, Weissman2015, Hermanne2000a, sudar1994, Sudar1996a, Diksic1979}.

\section{Tabulated cross sections}  \label{chapter:cross_sections_tabulated}
The tabulated measured cross sections obtained in this work are presented in Tables \ref{tab:Iridium_Cross_sections}, \ref{tab:Iron_Cross_sections}, \ref{tab:Nickel_crossSections} and \ref{tab:Copper_Cross_sections} for reactions in iridium, iron, nickel and copper, respectively. 

\setlength{\tabcolsep}{2pt} % Default value: 6pt
\renewcommand{\arraystretch}{1.5} % Default value: 1
\begin{sidewaystable}
\caption{The measured cross section for \natIr(d,x) reactions. Subscript \textit{i} indicates that the measurement is independent while subscript \textit{c} indicates that the measurement is cumulative. }
    \footnotesize
    \centering
    \begin{tabular}{lllllllllll}
    \hline
    \multicolumn{11}{c} {Production cross sections (mb)}\\
    \hline
     \makecell{E$_\textnormal{d}$ (MeV)} & \makecell{30.65$^{+0.76}_{-0.75}$} & \makecell{28.40$^{+0.80}_{-0.79}$} & \makecell{26.03$^{+0.82}_{-0.82}$} & \makecell{23.54$^{+0.88}_{-0.87}$} & \makecell{21.38$^{+0.94}_{-0.92}$} & \makecell{19.03$^{+1.00}_{-0.99}$} & \makecell{16.43$^{+1.11}_{-1.08}$} & \makecell{13.51$^{+1.28}_{-1.22}$} & \makecell{10.09$^{+1.55}_{-1.41}$} & \makecell{5.63$^{+2.21}_{-1.83}$} \\
     
    \Xhline{2\arrayrulewidth}
     \makecell{\ce{^{188}Pt}$_\textnormal{i}$} & \makecell{0.95 $\pm$ 0.13} & \makecell{0.30$\pm$0.09} & \makecell{0.17$\pm$0.05} & \makecell{-} & \makecell{-} & \makecell{-} & \makecell{-} & \makecell{-} & \makecell{-} & \makecell{-} \\
    
     \makecell{\ce{^{189}Pt}$_\textnormal{i}$} & \makecell{486.47$\pm$21.86} & \makecell{341.24$\pm$16.64} & \makecell{172.11$\pm$8.03} & \makecell{30.72$\pm$1.48} & \makecell{1.04$\pm$0.07} & \makecell{0.09$\pm$0.02} & \makecell{-} & \makecell{-} & \makecell{-} & \makecell{-} \\
    \makecell{\ce{^{191}Pt}$_\textnormal{i}$} & \makecell{597.10$\pm$16.55} & \makecell{483.60$\pm$13.79} & \makecell{353.99$\pm$9.67} & \makecell{165.12$\pm$5.15} & \makecell{71.05$\pm$2.19} & \makecell{77.53$\pm$2.57} & \makecell{128.24$\pm$4.03} & \makecell{137.37$\pm$4.42} & \makecell{53.45$\pm$3.23} & \makecell{1.05$\pm$0.06} \\
      
     \makecell{\ce{^{193m}Pt}$_\textnormal{i}$} & \makecell{48.11$\pm$6.33} & \makecell{46.78$\pm$2.18} & \makecell{55.68$\pm$2.17} & \makecell{51.79$\pm$2.12} & \makecell{58.31$\pm$1.96} & \makecell{77.98$\pm$2.89} & \makecell{115.33$\pm$4.09} & \makecell{148.98$\pm$5.54} & \makecell{56.18$\pm$2.85} & \makecell{1.55$\pm$0.12} \\
     
    \makecell{\ce{^{188}Ir}$_\textnormal{c}$} & \makecell{1.37$\pm$0.01} & \makecell{0.47$\pm$0.07} & \makecell{0.34$\pm$0.08} & \makecell{-} & \makecell{-} & \makecell{-} & \makecell{-} & \makecell{-} & \makecell{-} & \makecell{-} \\ 
    
    \makecell{$^{188m1+g}$Ir$_\textnormal{c}$} & \makecell{0.42$\pm$0.16} & \makecell{0.17$\pm$0.11} & \makecell{0.17$\pm$0.09} & \makecell{-} & \makecell{-} & \makecell{-} & \makecell{-} & \makecell{-} & \makecell{-} & \makecell{-} \\ 
    
    \makecell{\ce{^{189}Ir}$_\textnormal{c}$} & \makecell{334.77$\pm$22.25} & \makecell{235.95$\pm$16.33} & \makecell{84.03$\pm$6.03} & \makecell{16.81$\pm$5.17} & \makecell{-} & \makecell{-} & \makecell{-} & \makecell{-} & \makecell{-} & \makecell{-} \\     
    
    \makecell{\ce{^{190m2}Ir}$_\textnormal{i}$} & \makecell{8.40$\pm$0.25} & \makecell{4.69$\pm$0.14} & \makecell{2.67$\pm$0.08} & \makecell{1.09$\pm$0.04} & \makecell{0.44$\pm$0.01} & \makecell{0.17$\pm$0.01} & \makecell{0.07$\pm$0.01} & \makecell{-} & \makecell{-} & \makecell{-} \\  
    
    \makecell{\ce{^{190}Ir}$_\textnormal{c}$} & \makecell{86.65$\pm$2.89} & \makecell{62.80$\pm$2.14} & \makecell{44.26$\pm$1.47} & \makecell{27.29$\pm$1.02} & \makecell{18.73$\pm$0.71} & \makecell{14.02$\pm$0.55} & \makecell{12.40$\pm$0.51} & \makecell{8.26$\pm$0.43} & \makecell{-} & \makecell{-} \\

    \makecell{$^{190m1+g}$Ir$_\textnormal{c}$} & \makecell{85.92$\pm$2.89} & \makecell{62.39$\pm$2.14} & \makecell{44.03$\pm$1.47} & \makecell{27.19$\pm$1.02} & \makecell{18.69$\pm$0.71} & \makecell{14.01$\pm$0.55} & \makecell{12.39$\pm$0.51} & \makecell{8.26$\pm$0.43} & \makecell{-} & \makecell{-} \\
    
    \makecell{\ce{^{192}Ir}$_\textnormal{c}$} & \makecell{183.35$\pm$5.18} & \makecell{147.35$\pm$4.28} & \makecell{120.82$\pm$3.34} & \makecell{95.47$\pm$3.02} & \makecell{86.86$\pm$2.70} & \makecell{87.91$\pm$2.93} & \makecell{96.17$\pm$3.05} & \makecell{103.00$\pm$3.35} & \makecell{60.91$\pm$2.44} & \makecell{5.54$\pm$0.32} \\ 
    
    \makecell{\ce{^{194}Ir}$_\textnormal{c}$} & \makecell{50.94$\pm$2.28} & \makecell{51.41$\pm$2.29} & \makecell{61.39$\pm$2.39} & \makecell{69.45$\pm$2.76} & \makecell{82.71$\pm$3.21} & \makecell{96.64$\pm$4.15} & \makecell{119.84$\pm$4.66} & \makecell{142.26$\pm$5.72} & \makecell{89.76$\pm$4.22} & \makecell{6.35$\pm$0.43} \\     
     
    \end{tabular}
    \label{tab:Iridium_Cross_sections}
\end{sidewaystable}

\setlength{\tabcolsep}{6pt} % Default value: 6pt
\renewcommand{\arraystretch}{1.5} % Default value: 1

\begin{table}
    \caption{The measured cross section for \natFe(d,x) reactions. Subscript \textit{i} indicates that the measurement is independent while subscript \textit{c} indicates that the measurement is cumulative.}
    \small
    \centering
    \begin{tabular}{llll}
    \hline
    \multicolumn{4}{c} {Production cross sections (mb)}\\
    \hline
     \makecell{E$_\textnormal{d}$ (MeV)} & \makecell{29.57$^{+0.68}_{-0.67}$} & \makecell{27.25$^{+0.73}_{-0.72}$} & \makecell{24.80$^{+0.77}_{-0.76}$} \\
    \hline
    \makecell{$^{58}$Co$_\textnormal{c}$} & \makecell{1.50$\pm$0.05} & \makecell{1.62$\pm$0.05} & \makecell{2.05$\pm$0.07} \\
    \makecell{$^{57}$Co$_\textnormal{i}$} & \makecell{35.91$\pm$1.06} & \makecell{38.37$\pm$1.13} & \makecell{42.63$\pm$1.27} \\
    \makecell{$^{55}$Co$_\textnormal{i}$} & \makecell{27.15$\pm$0.80} & \makecell{20.44$\pm$0.60} & \makecell{13.82$\pm$0.40} \\
    \makecell{$^{59}$Fe$_\textnormal{i}$} & \makecell{0.16$\pm$0.02} & \makecell{0.15$\pm$0.02} & \makecell{0.18$\pm$0.04} \\
    \makecell{$^{53}$Fe$_\textnormal{c}$} & \makecell{5.11$\pm$0.65} & \makecell{2.77$\pm$0.44} & \makecell{1.29$\pm$0.30} \\
    \makecell{$^{56}$Mn$_\textnormal{c}$} & \makecell{22.14$\pm$0.65} & \makecell{23.86$\pm$0.80} & \makecell{22.91$\pm$0.65} \\
    \makecell{$^{54}$Mn$_\textnormal{i}$} & \makecell{23.58$\pm$0.70} & \makecell{24.18$\pm$0.72} & \makecell{26.12$\pm$0.79} \\
    \makecell{$^{52}$Mn$_\textnormal{c}$} & \makecell{16.00$\pm$0.46} & \makecell{5.48$\pm$0.16} & \makecell{0.91$\pm$0.03} \\
    \makecell{$^{51}$Cr$_\textnormal{c}$} & \makecell{7.54$\pm$0.23} & \makecell{7.86$\pm$0.25} & \makecell{8.51$\pm$0.29} \\
    
    \makecell{$^{48}$V$_\textnormal{c}$} & \makecell{0.12$\pm$0.01} & \makecell{0.09$\pm$0.01} & \makecell{0.07$\pm$0.01} \\
    \multicolumn{4}{c} {\underline{Monitor reaction}} \\
    \makecell{$^{56}$Co$_\textnormal{i}$} & \makecell{115.74$\pm$3.55} & \makecell{143.76$\pm$4.22} & \makecell{183.75$\pm$5.41}\\
    
    \hline
    \end{tabular}
    \label{tab:Iron_Cross_sections}
\end{table}

\setlength{\tabcolsep}{2pt} % Default value: 6pt
\renewcommand{\arraystretch}{1.5} % Default value: 1
\begin{sidewaystable}[]
\caption{The measured cross section for \natNi(d,x) reactions. Sub letter \textit{i} indicates that the measurement is independent while sub letter \textit{c} indicates that the measurement is cumulative. }
    \scriptsize
    \centering
    \begin{tabular}{lllllllllll}
    \hline
    \multicolumn{11}{c} {Production cross sections (mb)}\\
    \hline
     \makecell{E$_\textnormal{d}$ (MeV)} & \makecell{31.29$^{+0.66}_{-0.66}$} & \makecell{29.08$^{+0.70}_{-0.69}$} & \makecell{26.74$^{+0.74}_{-0.73}$} & \makecell{24.28$^{+0.79}_{-0.77}$} & \makecell{22.19$^{+0.84}_{-0.83}$} & \makecell{19.92$^{+0.92}_{-0.90}$} & \makecell{17.42$^{+1.00}_{-0.97}$} & \makecell{14.63$^{+1.14}_{-1.08}$} & \makecell{11.41$^{+1.35}_{-1.25}$} & \makecell{7.40$^{+1.84}_{-1.55}$} \\
    \Xhline{2\arrayrulewidth}
         
    \hline
    
    \makecell{$^{64}$Cu$_\textnormal{i}$} & \makecell{-} & \makecell{-} & \makecell{-} & \makecell{-} & \makecell{-} & \makecell{5.78$\pm$1.66} & \makecell{-} & \makecell{13.69$\pm$6.64} & \makecell{7.99$\pm$1.19} &\makecell{4.17$\pm$2.02}  \\
    
    \makecell{$^{60}$Cu$_\textnormal{i}$} & \makecell{-} & \makecell{-} & \makecell{27.51$\pm$1.68} & \makecell{35.38$\pm$1.71} & \makecell{42.90$\pm$3.55} & \makecell{54.43$\pm$4.62} & \makecell{50.61$\pm$4.68} & \makecell{41.09$\pm$4.89} & \makecell{12.76$\pm$2.64} &\makecell{-}  \\  
    
    \makecell{$^{65}$Ni$_\textnormal{i}$} & \makecell{-} & \makecell{-} & \makecell{0.51$\pm$0.06} & \makecell{0.61$\pm$0.05} & \makecell{1.21$\pm$0.23} & \makecell{1.35$\pm$0.25} & \makecell{1.35$\pm$0.23} & \makecell{1.83$\pm$0.27} & \makecell{2.23$\pm$0.29} &\makecell{2.61$\pm$0.19}  \\ 
    
    \makecell{$^{57}$Ni$_\textnormal{c}$} & \makecell{71.67$\pm$2.37} & \makecell{58.38$\pm$2.21} & \makecell{41.16$\pm$1.38} & \makecell{25.85$\pm$0.95} & \makecell{14.79$\pm$0.62} & \makecell{7.66$\pm$0.36} & \makecell{4.22$\pm$0.18} & \makecell{2.61$\pm$0.15} & \makecell{0.65$\pm$0.05} &\makecell{-}  \\ 
    
    \makecell{$^{56}$Ni$_\textnormal{c}$} & \makecell{0.18$\pm$0.01} & \makecell{0.10$\pm$0.01} & \makecell{0.06$\pm$0.01} & \makecell{-} & \makecell{-} & \makecell{-} & \makecell{-} & \makecell{-} & \makecell{-} &\makecell{-}  \\ 
    
    \makecell{$^{60}$Co$_\textnormal{c}$} & \makecell{12.94$\pm$0.48} & \makecell{20.26$\pm$2.23} & \makecell{18.91$\pm$1.24} & \makecell{20.17$\pm$1.44} & \makecell{13.48$\pm$0.77} & \makecell{11.25$\pm$0.95} & \makecell{9.64$\pm$0.95} & \makecell{7.69$\pm$1.52} & \makecell{1.94$\pm$0.51} &\makecell{-}  \\  
    
    \makecell{$^{58m}$Co$_\textnormal{i}$} & \makecell{68.06$\pm$1.92} & \makecell{77.42$\pm$2.20} & \makecell{102.27$\pm$2.79} & \makecell{118.56$\pm$3.70} & \makecell{110.28$\pm$3.40} & \makecell{115.36$\pm$3.86} & \makecell{109.86$\pm$3.46} & \makecell{98.24$\pm$3.19} & \makecell{59.93$\pm$2.38} &\makecell{12.66$\pm$0.69}  \\
    
    \makecell{$^{58g}$Co$_\textnormal{i}$} & \makecell{53.45$\pm$8.36} & \makecell{58.67$\pm$9.72} & \makecell{58.02$\pm$10.71} & \makecell{72.68$\pm$10.59} & \makecell{99.10$\pm$10.52} & \makecell{113.57$\pm$11.36} & \makecell{112.78$\pm$10.98} & \makecell{99.08$\pm$9.48} & \makecell{64.66$\pm$7.30} &\makecell{11.36$\pm$3.74}  \\
    
    \makecell{$^{57}$Co$_\textnormal{c}$} & \makecell{402.12$\pm$11.70} & \makecell{362.58$\pm$10.85} & \makecell{315.94$\pm$9.06} & \makecell{240.63$\pm$7.74} & \makecell{158.92$\pm$5.17} & \makecell{84.14$\pm$3.08} & \makecell{36.78$\pm$1.23} & \makecell{20.01$\pm$0.75} & \makecell{5.03$\pm$0.23} &\makecell{1.29$\pm$0.29}  \\
    
    \makecell{$^{57}$Co$_\textnormal{i}$} & \makecell{330.46$\pm$11.94} & \makecell{304.20$\pm$11.08} & \makecell{274.78$\pm$9.16} & \makecell{214.79$\pm$7.79} & \makecell{144.12$\pm$5.21} & \makecell{76.48$\pm$3.10} & \makecell{32.56$\pm$1.25} & \makecell{17.40$\pm$0.77} & \makecell{4.38$\pm$0.24} &\makecell{1.29$\pm$0.29}  \\
    
    \makecell{$^{56}$Co$_\textnormal{i}$} & \makecell{18.97$\pm$0.56} & \makecell{14.38$\pm$0.46} & \makecell{10.33$\pm$0.31} & \makecell{8.88$\pm$0.30} & \makecell{7.40$\pm$0.25} & \makecell{9.06$\pm$0.33} & \makecell{12.43$\pm$0.41} & \makecell{20.30$\pm$0.69} & \makecell{30.91$\pm$1.24} &\makecell{26.89$\pm$1.50}  \\

    \makecell{$^{55}$Co$_\textnormal{c}$} & \makecell{14.19$\pm$0.48} & \makecell{17.59$\pm$0.59} & \makecell{20.47$\pm$0.63} & \makecell{23.50$\pm$0.81} & \makecell{23.03$\pm$0.82} & \makecell{19.13$\pm$0.72} & \makecell{13.11$\pm$0.47} & \makecell{7.57$\pm$0.30} & \makecell{0.71$\pm$0.04} &\makecell{0.18$\pm$0.03}  \\
    
    \makecell{$^{59}$Fe$_\textnormal{c}$} & \makecell{0.24$\pm$0.03} & \makecell{-} & \makecell{0.15$\pm$0.09} & \makecell{-} & \makecell{0.11$\pm$0.03} & \makecell{-} & \makecell{-} & \makecell{-} & \makecell{-} &\makecell{-}  \\
    
    \makecell{$^{54}$Mn$_\textnormal{i}$} & \makecell{21.17$\pm$0.64} & \makecell{15.11$\pm$0.74} & \makecell{9.55$\pm$0.35} & \makecell{5.42$\pm$0.29} & \makecell{1.23$\pm$0.08} & \makecell{0.76$\pm$0.17} & \makecell{-} & \makecell{-} & \makecell{-} &\makecell{-}  \\
    
    \makecell{$^{52}$Mn$_\textnormal{c}$} & \makecell{3.22$\pm$0.09} & \makecell{3.17$\pm$0.10} & \makecell{2.56$\pm$0.07} & \makecell{1.83$\pm$0.06} & \makecell{1.19$\pm$0.04} &  \makecell{0.57$\pm$0.02} & \makecell{0.13$\pm$0.01} & \makecell{0.05$\pm$0.02} & \makecell{-} &\makecell{-}  \\
    
    \multicolumn{11}{c} {\underline{Monitor reactions}}\\
    
    \makecell{$^{56}$Co$_\textnormal{c}$} & \makecell{18.39$\pm$1.02} & \makecell{14.16$\pm$2.10} & \makecell{10.08$\pm$1.12} & \makecell{8.83$\pm$0.27} & \makecell{7.67$\pm$0.24} & \makecell{9.57$\pm$0.31} & \makecell{13.22$\pm$0.41} & \makecell{19.73$\pm$0.60} & \makecell{31.65$\pm$0.91} &\makecell{27.78$\pm$0.96}  \\ 
    \makecell{$^{58}$Co$_\textnormal{c}$} & \makecell{116.66$\pm$18.54} & \makecell{133.05$\pm$22.35} & \makecell{155.58$\pm$29.03} & \makecell{190.24$\pm$28.10} & \makecell{217.14$\pm$23.76} & \makecell{241.76$\pm$24.98} & \makecell{236.83$\pm$23.97} & \makecell{191.83$\pm$18.91} & \makecell{127.59$\pm$14.39} &\makecell{24.81$\pm$8.13}  \\ 
    \makecell{$^{61}$Cu$_\textnormal{i}$} & \makecell{15.25$\pm$1.41} & \makecell{15.44$\pm$1.41} & \makecell{15.20$\pm$0.88} & \makecell{14.67$\pm$0.88} & \makecell{14.49$\pm$1.19} & \makecell{15.39$\pm$1.23} & \makecell{17.98$\pm$1.44} & \makecell{25.45$\pm$1.93} & \makecell{36.70$\pm$2.48} &\makecell{62.33$\pm$4.18}  \\   
    \end{tabular}
    \label{tab:Nickel_crossSections}
\end{sidewaystable}

\setlength{\tabcolsep}{2pt} % Default value: 6pt
\renewcommand{\arraystretch}{1.5} % Default value: 1
\begin{sidewaystable}[]
\caption{The measured cross section for \natCu(d,x) reactions. Sub letter \textit{i} indicates that the measurement is independent while sub letter \textit{c} indicates that the measurement is cumulative. }
    \scriptsize
    \centering
    \begin{tabular}{lllllllllll}
    \hline
    \multicolumn{11}{c} {Production cross sections (mb)}\\
    \hline
     \makecell{E$_\textnormal{d}$ (MeV)} & \makecell{30.03$^{+0.67}_{-0.67}$} & \makecell{27.74$^{+0.72}_{-0.71}$} & \makecell{25.32$^{+0.77}_{-0.76}$} & \makecell{22.77$^{+0.83}_{-0.82}$} & \makecell{20.57$^{+0.89}_{-0.87}$} & \makecell{18.14$^{+0.97}_{-0.94}$} & \makecell{15.43$^{+1.08}_{-1.04}$} & \makecell{12.34$^{+1.27}_{-1.20}$} & \makecell{8.68$^{+1.62}_{-1.43}$} & \makecell{3.94$^{+2.25}_{-2.22}$} \\
    \Xhline{2\arrayrulewidth}

    \makecell{$^{64}$Cu$_\textnormal{i}$} & \makecell{170.76$\pm$7.76} & \makecell{153.78$\pm$8.20} & \makecell{132.56$\pm$6.93} & \makecell{121.54$\pm$7.12} & \makecell{106.07$\pm$5.81} & \makecell{95.92$\pm$7.14} & \makecell{123.79$\pm$6.62} & \makecell{156.65$\pm$8.20} & \makecell{209.38$\pm$11.27} &\makecell{73.54$\pm$5.70}  \\ 
    
    \makecell{$^{61}$Cu$_\textnormal{c}$} & \makecell{4.54$\pm$0.87} & \makecell{2.07$\pm$0.95} & \makecell{1.18$\pm$0.72} & \makecell{-} & \makecell{-} & \makecell{-} & \makecell{-} & \makecell{-} & \makecell{-} &\makecell{-}  \\ 
    
    \makecell{$^{65}$Ni$_\textnormal{i}$} & \makecell{3.42 $\pm$ 1.76} & \makecell{3.84$\pm$ 1.94} & \makecell{2.89$\pm$1.48} & \makecell{1.92$\pm$1.14} & \makecell{-} & \makecell{-} & \makecell{-} & \makecell{-} & \makecell{-} &\makecell{-}  \\ 
    
    \makecell{$^{61}$Co$_\textnormal{c}$} & \makecell{1.62 $\pm$ 0.09} & \makecell{0.82 $\pm$ 0.07} & \makecell{0.29 $\pm$ 0.05} & \makecell{-} & \makecell{-} & \makecell{-} & \makecell{-} & \makecell{-} & \makecell{-} &\makecell{-}  \\ 
    
    \makecell{$^{60}$Co$_\textnormal{c}$} & \makecell{9.49$\pm$0.52} & \makecell{11.38$\pm$0.51} & \makecell{12.02$\pm$0.51} & \makecell{11.36$\pm$0.43} & \makecell{9.27$\pm$0.41} & \makecell{5.65$\pm$0.26} & \makecell{1.53$\pm$0.12} & \makecell{-} & \makecell{-} & \makecell{-}  \\ 
    
    \makecell{$^{59}$Fe$_\textnormal{c}$} & \makecell{0.21$\pm$0.03} & \makecell{0.18$\pm$0.02} & \makecell{0.17$\pm$0.02} & \makecell{0.12$\pm$0.01} & \makecell{0.07$\pm$0.01} & \makecell{0.03$\pm$0.01} & \makecell{-} & \makecell{-} & \makecell{-} &\makecell{-  }  \\ 
    
    \multicolumn{11}{c} {\underline{Monitor reactions}} \\
    \makecell{$^{62}$Zn$_\textnormal{i}$} & \makecell{32.51$\pm$1.24} & \makecell{26.56$\pm$1.09} & \makecell{17.24$\pm$0.83} & \makecell{9.37$\pm$0.43} & \makecell{3.26$\pm$0.17} & \makecell{0.32$\pm$0.13} & \makecell{-} & \makecell{-} & \makecell{-} &\makecell{-}  \\
    
    \makecell{$^{63}$Zn$_\textnormal{i}$} & \makecell{112.37$\pm$9.38} & \makecell{137.19$\pm$11.41} & \makecell{172.92$\pm$13.89} & \makecell{226.46$\pm$16.14} & \makecell{273.59$\pm$19.27} & \makecell{322.03$\pm$22.76} & \makecell{300.45$\pm$21.21} & \makecell{170.56$\pm$12.23} & \makecell{60.07$\pm$3.19} &\makecell{-}  \\
    
    \makecell{$^{65}$Zn$_\textnormal{i}$} & \makecell{62.65$\pm$1.90} & \makecell{76.11$\pm$2.47} & \makecell{98.93$\pm$2.87} & \makecell{135.19$\pm$4.09} & \makecell{177.91$\pm$4.97} & \makecell{224.29$\pm$8.55} & \makecell{254.98$\pm$7.25} & \makecell{249.63$\pm$7.83} & \makecell{149.24$\pm$8.45} &\makecell{14.79$\pm$5.49}  \\
    \end{tabular}
    \label{tab:Copper_Cross_sections}
\end{sidewaystable}

% 
%     
% \pagebreak
% 
% \onecolumn
% 
% %% The Appendices part is started with the command \appendix;
% %% appendix sections are then done as normal sections
% %% \appendix
% 
% %% \section{}
% %% \label{}
% 
% %% References
% %%
% %% Following citation commands can be used in the body text:
% %% Usage of \cite is as follows:
% %%   \cite{key}         ==>>  [#]
% %%   \cite[chap. 2]{key} ==>> [#, chap. 2]
%%

% \twocolumn


% ``Overfull \hbox in .bbl'' message fixed by commenting out ''write.url'' in /usr/share/texlive/texmf-dist/bibtex/bst/elsarticle/elsarticle-num.bst for affected entry types (likely, 'article' and 'book', all but 'phdthesis')

% Note: replace '../../library' with the path to the library.bib file that Mendeley generates for you!

%\IfFileExists{../../library.bib}{\bibliography{../../library}}{\bibliography{library}}
% \thispagestyle{fancyTOC}

\bibliography{library.bib, library_kevin.bib} 
\bibliographystyle{elsarticle-num}

\end{document}